\documentclass[twocolumn,superscriptaddress,floatfix,preprintnumbers,prd ,nofootinbib,hyperref]{revtex4-2}
\pdfoutput=1
\usepackage[colorlinks=true,breaklinks=true]{hyperref}
\usepackage[normalem]{ulem}
\usepackage[utf8]{inputenc}
\hypersetup{allcolors=[rgb]{0.0 0.0 0.6},linkcolor=[rgb]{0.75 0.05 0.05}}
\usepackage{amsmath,amssymb}
\usepackage{mathtools}
\usepackage{cancel}
\usepackage{tensor}
\usepackage{amsmath, amssymb}

\usepackage{dcolumn}
\usepackage{multirow}
\usepackage[dvipsnames]{xcolor}
\usepackage{float}
\usepackage{bbold}
\usepackage{letltxmacro}
\LetLtxMacro{\oldcite}{\cite}
\renewcommand{\cite}[1]{\mbox{\oldcite{#1}}}
\long\def\exclude#1{}

\newcommand{\gagg}{g_{a\gamma \gamma}}

\DeclareMathOperator{\eV}{eV}

\DeclareMathOperator{\keV}{keV}

\DeclareMathOperator{\ms}{ms}

\DeclareMathOperator{\diag}{diag}

\DeclareMathOperator{\km}{km}
\DeclareMathOperator{\m}{m}
\DeclareMathOperator{\cm}{cm}
\DeclareMathOperator{\mm}{mm}

\DeclareMathOperator{\Tesla}{T}

\newcommand{\beq}{\begin{equation}}
\newcommand{\eeq}{\end{equation}}

\def\ga{\,\,\raise0.14em\hbox{$>$}\kern-0.76em\lower0.28em\hbox
{$\sim$}\,\,}

\newcommand{\ie}{\emph{i.e.~}}
\newcommand{\eg}{\emph{e.g.~}}

\long\def\exclude#1{}

\newcommand{\OO}[1]{\mathcal O \left( #1 \right)}

\newcommand{\EE}{\mathcal{E}}
\newcommand{\RR}{\mathcal{R}}
\renewcommand{\AA}{\mathcal{A}}
\newcommand{\MMsq}{\mathcal{M}^2}
\newcommand{\abs}[1]{\bigl| #1 \bigr|}
\newcommand{\PsiVector}[2]{
\begin{pmatrix}
#1 \\
#2
\end{pmatrix}
}

\newcommand{\spatial}[1]{\prescript{(3)}{}{#1}}
\newcommand{\UU}{\mathcal{U}}
\newcommand{\nel}{n_{\scriptscriptstyle{\rm EL}}}
\newcommand{\tder}{\partial_t}
\newcommand{\xder}{\partial_x}
\newcommand{\yder}{\partial_y}

\allowdisplaybreaks

\bibpunct{[}{]}{,}{n}{}{,}

\hypersetup{
    colorlinks=true,
    citecolor=[rgb]{.1, .7, .6},
    linkcolor=[rgb]{.2, .55, .95},
    filecolor=magenta,
    urlcolor=[rgb]{.1, .7, .6},
}

\begin{document}
    
\preprint{DESY-26-046}
\preprint{CERN-TH-2026-069}

\title{Simulating Axion Electrodynamics in Magnetized Plasmas: \\[6pt]  {\large \rm \emph{Energy transfer in the inhomogeneous and strongly varying limit}}}

\author{Fabrizio Corelli} \email{fabrizio.corelli@aei.mpg.de}
\affiliation{Dipartimento di Fisica, ``Sapienza'' Universit\`a di Roma \& Sezione INFN Roma1, Piazzale Aldo Moro
5, 00185, Roma, Italy}
\affiliation{Max Planck Institute for Gravitational Physics (Albert Einstein Institute), Am Mühlenberg 1, 14476 Potsdam, Germany}

\author{Estanis Utrilla Gin\'{e}s}
\affiliation{Instituto de F\'{i}sica Corpuscular (IFIC), University of Valencia-CSIC,
Parc Cient\'{i}fic UV, c/ Catedr\'{a}tico Jos\'{e} Beltr\'{a}n 2, E-46980 Paterna, Spain}

\author{Enrico Cannizzaro}\email{enrico.cannizzaro@tecnico.ulisboa.pt} 
\affiliation{CENTRA, Departamento de Fisica, Instituto Superior T\'ecnico – IST, Universidade de Lisboa – UL, Avenida Rovisco Pais 1, 1049 Lisboa, Portugal}
\author{Andrea Caputo}\email{andrea.caputo@cern.ch}
\affiliation{Dipartimento di Fisica, ``Sapienza'' Universit\`a di Roma \& Sezione INFN Roma1, Piazzale Aldo Moro
5, 00185, Roma, Italy}
\affiliation{Department of Theoretical Physics, CERN, Esplanade des Particules 1, P.O. Box 1211, Geneva 23, Switzerland}
\affiliation{Department of Particle Physics and Astrophysics, Weizmann Institute of Science, Rehovot 7610001, Israel}

\author{Samuel J. Witte}\email{Samuel.Witte@physics.ox.ac.uk}
\affiliation{Rudolf Peierls Centre for Theoretical Physics, University of Oxford, UK}
\affiliation{Deutsches Elektronen-Synchrotron DESY, Notkestraße 85, 22607 Hamburg, Germany}
\affiliation{
II. Institute of Theoretical Physics, Universität Hamburg, 22761, Hamburg, Germany}

%==========================

\begin{abstract}
In this work we study the electromagnetic response induced by axions in a magnetized plasma, focusing specifically on characterizing energy transfer and energy losses from the ambient axion field in highly inhomogeneous and strongly varying backgrounds.  Using a suite of both frequency-domain and time-domain simulations, we solve for: the efficiency of photon excitation in a rapidly varying background, the indirect excitation of Alfvén modes, occurring when a Langmuir-Ordinary (LO) mode is resonantly excited near a combined cutoff-resonance of the dispersion relations of the LO and Alfvén modes,  and the excitation of electric fields in small localized plasma under-densities. We identify a particularly interesting regime in which energy can be transferred into sub-luminal plasma modes ($\omega < k$) with an efficiency greater than that of super-luminal modes ($\omega > k$). Our results highlight a variety of less conventional ways in which axions (and other light degrees of freedom that mix with electromagnetism, such as dark photons or gravitons) can interact in extreme astrophysical environments. 
\end{abstract}

\maketitle

\section{Introduction}

Axions are extremely well-motivated candidates for new fundamental physics beyond the Standard Model. Many of the searches for axions, both in the laboratory and in astrophysics, exploit their coupling to electromagnetism (see e.g.~\cite{graham2015experimental,Irastorza:2018dyq,adams2022axion,Arza:2026rsl, Caputo:2024oqc, OHare:2024nmr} for various reviews), which arises via the dimension-five coupling $\mathcal{L} \supset (g_{a\gamma\gamma} / 4) \,  a \,  F_{\mu \nu} \tilde{F}^{\mu\nu} $, where $g_{a\gamma\gamma}$ is the axion-photon coupling constant (in units of inverse energy), and $F$ and $\tilde{F}$ are the electromagnetic field strength tensor and its dual. This interaction allows for: axions to undergo a two-photon decay (see \eg~\cite{ressell1991limits,Caputo:2018ljp,Caputo:2018vmy,Ghosh:2020hgd,Buen-Abad:2021qvj,Sun:2021oqp,Dev:2023ijb,Sun:2023gic,Roach:2022lgo,Foster:2021ngm,Roy:2023omw,Janish:2023kvi,Calore:2022pks}), axions to be produced from non-orthogonal electric and magnetic fields (see \eg~\cite{Prabhu:2021zve,Noordhuis:2022ljw,Noordhuis:2023wid,Caputo:2023cpv}), a rotation in the polarization of electromagnetic waves as they traverse a varying axion background (see \eg~\cite{Raffelt:1987im,rosenberg2000searches,asztalos2006searches,liu2019searching, Castillo:2022zfl, Caputo:2019tms, Ivanov:2018byi, Sigl:2018fba, Fedderke:2019ajk}), and the mixing of axions and photons in the presence of background magnetic fields (see \eg~\cite{Bibber1987,Raffelt:1987im}). Each of these processes are efficient in different regimes, allowing for complementary searches for axions across a broad range of masses and in a variety of environments. For example: the former process is typically only efficient for heavy axions; axion production requires large time-varying $\vec{E} \cdot \vec{B}$ fields, which are  screened in most astrophysical plasmas (although not always -- see \cite{Prabhu:2021zve,Noordhuis:2022ljw,Noordhuis:2023wid}); and birefringence induces a polarization shift proportional to the difference of $\delta(g_{a\gamma\gamma}a)$ at emission and detection, such that generating large observable effects typically necessitate axion field values $a \sim f_a$ (with $f_a$ being their decay constant) at some location in the Universe. This often leaves axion-photon mixing as the most promising process across a much broader range of axion parameter space (as is evidenced by the fact that this process drives most limits on the axion-photon coupling for axion masses below $m_a \lesssim 10^{-4} \,{\rm eV}$).

Axion-photon mixing is typically heavily suppressed in the non-relativistic limit (see \eg~\cite{Raffelt:1987im,Marsh:2021ajy,Gines:2024ekm,Smarra:2024vzg}) -- this is merely a consequence of the fact that the axion and photon dispersion relations become increasingly discrepant in this limit (stemming from the fact that axions have a mass, $m_a$, while photons do not). There are well-known ways, however, of evading this suppression. One idea which has served as the basis of a variety of experimental~\cite{arik2009probing,lawson2019tunable,MADMAX:2019pub} and indirect astrophysical probes~\cite{Pshirkov:2007st,Huang:2018lxq,Hook:2018iia,Safdi:2018oeu,Battye:2019aco,Leroy:2019ghm,Foster:2020pgt,Prabhu:2021zve,Buckley2021,Edwards:2020afl,Witte:2021arp,Battye2021,battye2021robust,Nurmi:2021xds,Foster:2022fxn,Witte:2022cjj,Battye:2023oac,mcdonald2023generalized,Xue:2023ejt,Tjemsland:2023vvc,Ruz:2024gkl,Roy:2025mqw} of  axions is to exploit the fact that background media can alter the photon dispersion relation, allowing for resonant axion-photon transitions (similar effects also occur for dark photons, see e.g.~\cite{Arias:2012az,Jaeckel:2008fi,McDermott:2019lch,Corelli:2024lvc, Caputo:2020bdy, Caputo:2020rnx, Arsenadze:2024ywr, Witte:2020rvb, Trost:2024ciu, Bolton:2022hpt, Mirizzi:2009iz}). As a simple example, one can consider axions propagating through a dilute non-thermal electron plasma with density $n_e$; here, the plasma induces an `effective mass' (or a gap in the dispersion relation) for transverse photons (\ie, the dispersion relation is given by $\omega^2 = k^2 + \omega_p^2$) set by the plasma frequency of the medium $\omega_p \simeq \sqrt{e^2 n_e / m_e}$, with $e$ and $m_e$ being the charge and mass of the electron. For a spatially varying background plasma density, there may exist a point at which $\omega_p \simeq m_a$; here, the dispersion relations become degenerate, and resonant axion-photon transitions can occur. For low- or high-density (meaning far away from resonances) uniform plasmas, the efficiency of axion-photon conversion can be derived analytically, using e.g. perturbative approaches~\cite{Raffelt:1987im,Smarra:2024vzg} or Green's functions~\cite{Gines:2024ekm}. For slowly varying resonances, \ie when the background changes slowly with respect to the inverse momentum scale, and for certain classes of dielectric tensors, one can instead use the WKB approximation to derive an effective (Landau-Zener) conversion probability, see \eg~\cite{Gines:2024ekm} for a discussion. While the aforementioned cover a vast range of possibilities, there are a variety of astrophysical scenarios in which the traditional approaches break down, and when no well-defined separation of scales exist. This is the regime of interest here.

\begin{figure}
    \centering
    \includegraphics[width=\linewidth]{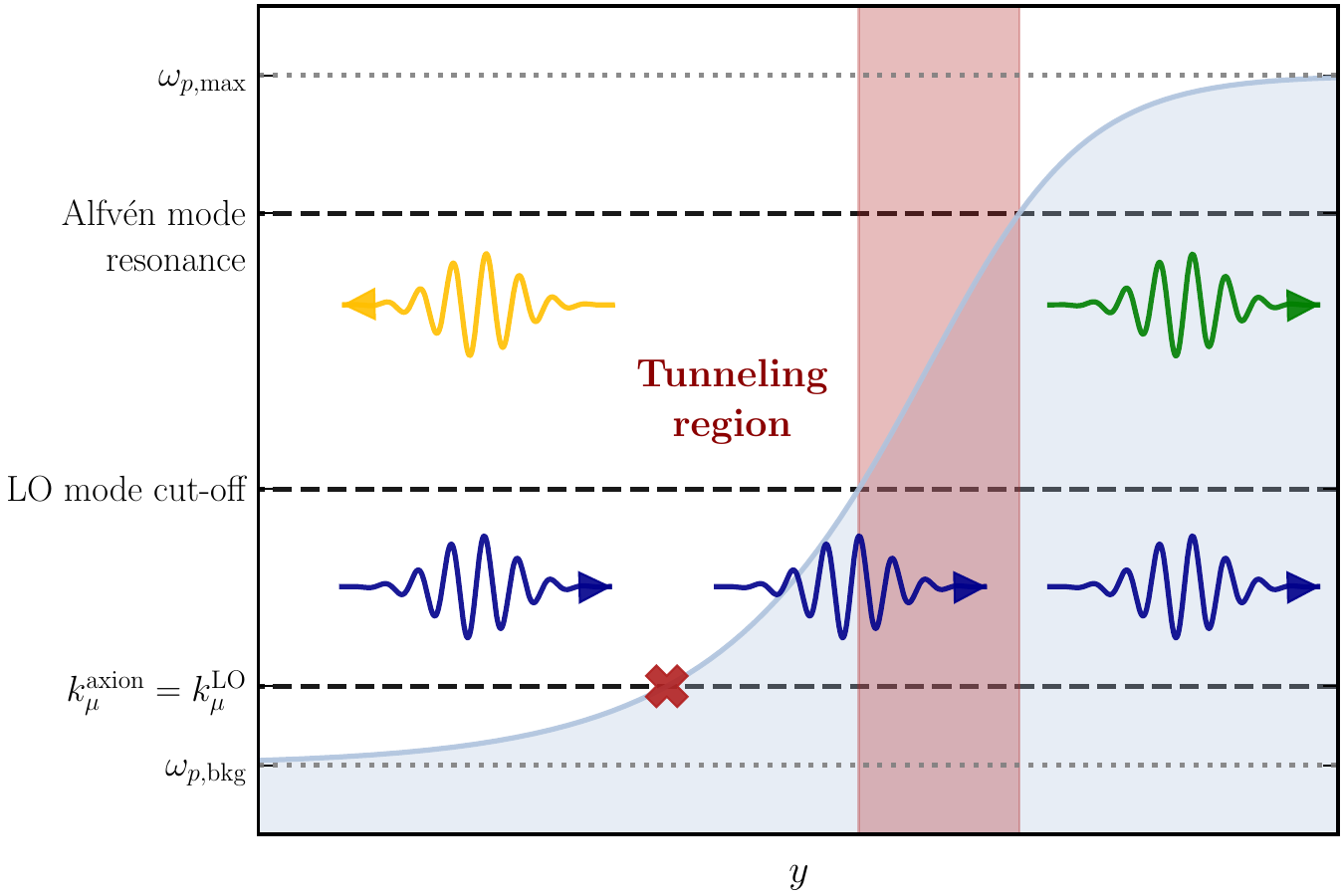}
    \caption{Illustration of how axions can lead to the indirect excitation of Alfv\'en modes. Incoming axion wave (dark blue) travels from a low density background into a high density background (plasma density profile shown with light blue shaded region). At $k_\mu^{\rm axion} = k_\mu^{\rm LO}$ the axion resonantly excites an LO mode (red cross); the LO modes travels a small distance before being reflected at the cut-off (yellow). Note that Alfv\'en mode only exists for plasma densities above the resonance point. For sufficiently large gradients, the excited LO mode can tunnel through the barrier and excite the Alfv\'en mode (green). For non-relativistic/semi-relativistic axions and large gradients, the WKB approximation is strongly violated across all processes. Numerical simulations are developed in this work in order to understand the efficiency of this process.  }
    \label{fig:alfven_cart}
\end{figure}

Let us briefly digress to discuss a few scenarios in which the effects discussed are realized, and what sort of interesting questions arise from the presence of these features in the context of axion physics. These examples are by no means exhaustive, and are simply intended to motivate the need for a broader understanding of axion electrodynamics in complex astrophysical environments. 
With that being said, consider the following:
\begin{itemize}
    \item Many plasmas found in astrophysical environments contain highly inhomogeneous distributions of plasma on small scales. This can stem directly from turbulent structure of the electromagnetic fields present in the system (as occurs \eg near black holes~\cite{hawley1995local}, in molecular clouds~\cite{crutcher2012magnetic}, in supernovae remnants~\cite{reynolds2012magnetic}, or in filamentary structures in the Central Molecular Zone (CMZ) of the galaxy~\cite{federrath2016link}), it can arise from small-scale structure in the plasma (as \eg in the case of plasmoids~\cite{philippov2019pulsar,yuan2020plasmoid,nathanail2022magnetic}, or as in current-sheets appearing near neutron stars and black holes~\cite{cerutti2015particle,east2018magnetosphere,selvi2024current} ), or it can result from inhomogeneities in plasma production mechanisms (as occurs \eg in the case of magnetars~\cite{beloborodov2007corona,thompson2020pair}). As a concrete example, one can consider the open field lines of pulsars, which host pair discharges that serve to supply the currents that in turn support the structure of the magnetosphere~\cite{Sturrock1971a, RudermanSutherland1975,Arons:1979bc} (in the case of magnetars, inhomogeneous plasma loading may occur also in the closed zones -- see e.g. ~\cite{Roy:2025mqw} for a discussion of leading models). The open field line bundle near the surface of the star spans a characteristic distance $r_{\rm pc} \sim R_{\rm NS} \sqrt{R_{\rm NS} \Omega} \sim  \mathcal{O}(100 \, {\rm m})$ (where $R_{\rm NS}$ and $\Omega$ are the radius of the neutron star and rotational frequency, respectively), with inhomogeneities in the phase space distribution arising on much smaller scales. This of course implies very small variations in the dispersion relations of the plasma, which can alter the mixing with light axions. Natural questions that one might want to ask include: how can axion-photon mixing be enhanced or suppressed in the presence strongly varying inhomogeneous backgrounds, and can high density axion field configurations (\eg, `axion clouds' formed around pulsars~\cite{Noordhuis:2023wid}, or quasi-bound states formed via black hole superradiance~\cite{Arvanitaki:2009fg,Baryakhtar:2020gao,Witte:2024drg}) dissipate energy into small-scale local inhomogeneities (see e.g.~\cite{Noordhuis:2023wid,Caputo:2023cpv})?
    \item Typically, axions are assumed to mix directly with a single, super-luminal ($\omega > k$) plasma mode. This is because energy-momentum conservation requires any mis-match in the 4-momentum to be balanced by the spatial-temporal inhomogeneity of magnetic field, which is often relatively smooth and slowly varying. Super-luminal modes, however, can mix with sub-luminal modes. This happens, for example, when there exists a pole, or neighboring poles and cut-offs, in the photon dispersion relations -- at such a point, energy can tunnel from one mode to another. For non-relativistic (and semi-relativistic) axions traversing magnetized plasmas, the point of peak resonant axion-photon mixing occurs exactly near a pole-cut-off pair, implying that mode-mixing can occur {\emph{while}} axions are sourcing electromagnetic modes. This is illustrated schematically in Fig.~\ref{fig:alfven_cart}, and we return to a detailed discussion of this phenomenon later in Sec.~\ref{sec:results}.  In the context of axion physics, this opens a number of interesting questions, including: how is axion mixing altered near poles and cut-offs, and can axions indirectly mix with sub-luminal plasma modes?  
\end{itemize}

Let us reiterate that the goal of this work is to study the electrodynamic response induced by, and energy transfer from, axions in scenarios where conventional analytic approximations break down. We have provided a number of examples above which demonstrate that such scenarios are generic and unavoidable in astrophysics -- this is particularly true when one studies low-energy axions near extreme astrophysical objects, such as near neutron stars and black holes. For this reason we focus specifically on the case of magnetized plasmas, but our general approach can be straightforwardly applied to other scenarios.  In order to approach this problem, we develop a suite of both frequency and time domain simulations that allow us to solve for the evolution and energy transfer in these systems. These simulations are built upon preceding work, which studied axion-photon mixing in magnetized plasma~\cite{Gines:2024ekm} and dark photon-photon mixing in the case where there are multiple nearby level-crossings~\cite{Corelli:2024lvc}.

This manuscript is divided as follows. We begin with an overview of axion electrodynamics (Sec.~\ref{sec:axelec}); this is mostly a review in order to familiarize the reader with the fundamental concepts and notation. Since the frameworks behind both our time-domain and frequency-domain simulations were outlined in~\cite{Corelli:2024lvc} and ~\cite{Gines:2024ekm}, we briefly review them in Sec.~\ref{subsec:numerics_sims} and defer the details to  Appendix~\ref{sec:Simulations}, before moving directly to their applications. Here, we begin (Sec.~\ref{subsec:wkb}) by demonstrating how our simulations can easily recover the expected mixing behavior in various asymptotic limits when a strongly varying, but smooth, background ignites resonant transitions over short distance scales (such that the WKB approximation is strongly violated). We then consider the case where the resonant excitation of a plasma mode is very close to point of mode-mixing (which we argue is natural for axions in magnetized media), allowing for axions to indirectly excite Alfvén modes (Sec.~\ref{sec:modeexcite}); here, we show that the efficiency with which the Alfvén mode is excited can actually, in some limiting circumstances, exceed the excitation efficiency of the Langmuir-Ordinary mode (which is the primary mode with which the axion mixes). We then analyze how small-scale inhomogeneities in the effective plasma frequency can alter the locally induced electric fields (Sec.~\ref{sec:inhomo}), demonstrating how one can potentially overcome the strong parametric suppression of the axion-induced electric in a dense plasma (typically scaling as $\sim (\omega_p/m_a)^2$). This work serves as the basis for broader studies of axion electrodynamics in extreme environments, and has interesting implications for situations in which large axion field values around extreme compact objects are naturally realized, such as in the case of axion clouds near pulsars or axion superradiance. 

\section{Axion Electrodynamics}\label{sec:axelec}
Let us start with a general discussion of axion electrodynamics, with the specific goal of highlighting various limiting scenarios which drive energy losses from a background axion field. This section does not contain any novel material, but rather serves to highlight the regimes in which analytic approaches break down, why numerical simulations can serve to move beyond that, and the strengths and weaknesses of working with time- and/or frequency domain simulations. The expert reader may wish to skip this section, as various aspects are already summarized in the literature (see e.g. ~\cite{Raffelt:1987im,Marsh:2021ajy,Gines:2024ekm,Smarra:2024vzg}). 

Let us start from the general Lagrangian describing an axion coupled to electromagnetism
\begin{align}
    \mathcal{L} &= -\frac{1}{4} F_{\mu\nu}F^{\mu\nu} - A_\mu J^\mu - \frac{1}{2} \bigl(\nabla_\mu a \bigr)\bigl(\nabla^\mu a \bigr) \notag \\
                &- \frac{m_a^2}{2} a^2 - \frac{1}{4}\gagg F_{\mu\nu} \tilde{F}^{\mu\nu} a,
    \label{eq:AxionPhotonLagrangian}
\end{align}
where $a$ is the axion and $m_a$ is its mass; $A_\mu$ is the electromagnetic 4-potential, $F_{\mu\nu}$ its field strength and $J^\mu$ is the electromagnetic 4-current; $\gagg$ is a constant encoding the strength of the coupling between the axion and the photon, and $\tilde{F}^{\mu\nu}$ is the dual of $F_{\mu\nu}$:
\begin{equation}
    \tilde{F}^{\mu\nu} = - \frac{1}{2} \epsilon^{\mu\nu\rho\sigma} F_{\rho\sigma},
    \label{eq:Fdual}
\end{equation}
where $\epsilon^{\mu\nu\rho\sigma}$ is the Levi-Civita tensor, with $\epsilon^{0123} = -1 / \sqrt{-g}$. The field equations for the axion and the electromagnetic field obtained from \ref{eq:AxionPhotonLagrangian} then read
\begin{align}
    (\Box - m_a^2) a = \frac{1}{4} \gagg F_{\mu\nu} \tilde{F}^{\mu\nu}, \label{eq:AxionFieldEquation} \\
    \nabla_\mu F^{\mu\nu} = J^\nu - \gagg \tilde{F}^{\mu\nu} \nabla_\mu a. \label{eq:EMFieldEquation}
\end{align}
For the plasma, we will assume ions to be at rest, and we will describe electrons with a cold fluid model, characterized by the equations
\begin{align}
    u^\nu \nabla_\nu u^\mu &= \frac{e}{m_e} F^{\mu\nu} u_\nu, \\
    \nabla_{\mu} (n_e u^\mu) &= 0,
    \label{eq:ColdPlasmaEquation}
\end{align}
where $u^\mu$ is the 4-velocity of the fluid, $n_e$ is the density of electrons, while $e$ is the elementary charge and $m_e$ is the electron mass. 

Here, we have maintained the possibility of using a generic metric $g_{\mu\nu}$. In what follows, we will restrict our attention to flat space, and adopt a Minkowski metric. Written in terms of electric $\vec{E}$ and magnetic fields $\vec{B}$, Eqns.~\ref{eq:AxionFieldEquation} and \ref{eq:EMFieldEquation} reduce to 
\begin{eqnarray}
 (\Box - m_a^2) a &=& - \gagg  (\vec{E} \cdot \vec{B}) \label{eq:aom_eb} \\
 \nabla \cdot \vec{E} &=& - \rho - \gagg \vec{B} \cdot \nabla a  \label{eq:nabladotEGauss}\\ 
 \nabla \times \vec{B} - \partial_t \vec{E} &=& - \vec{J} + \gagg (\vec{B} \partial_t a - \vec{E} \times \nabla a) \label{eq:maxEB} \, ,
\end{eqnarray}
with $J^\nu = (\rho, \vec{J})$. Note that the new contributions to Maxwells equations appear as effective charge densities $\rho_{ae} =  \gagg \vec{B} \cdot \nabla a$ and current densities $\vec{J}_{ae} = - \gagg \dot{a} \vec{B} + \gagg \vec{E} \times \nabla a$, meaning their gradients will naturally source perturbations in the electromagnetic fields.
The goal of this work effectively amounts to studying the solutions to Eqns.~\ref{eq:aom_eb}-\ref{eq:maxEB} and Eq.~\ref{eq:ColdPlasmaEquation} in various limiting regimes. In particular, we will focus on the case where the initial amplitude of the axion field is large (justifying dropping the source term in Eq.~\ref{eq:aom_eb}), and in the limit where the background magnetic field can be treated as strong, allowing for the perturbative expansion $\vec{B} = \vec{B}_0(\vec{x}) + \delta\vec{B}(\vec{x},t)$, $\vec{E}(\vec{x},t) = \delta \vec{E}(\vec{x},t)$, where $|\delta\vec{B}|, |\delta\vec{E}| \sim \mathcal{O}(g_{a\gamma\gamma}) \ll |\vec{B}_0|$.

Before discussing the scenarios of interest, let us briefly review axion electrodynamics in various well-known limits, highlighting when and where the underlying assumptions break down. Much of this discussion follows that of \cite{Gines:2024ekm}, however we will highlight a number of points which are particularly relevant for this study. 

\subsection{Vacuum mixing}

We will start with the simplest limit, corresponding to weak mixing (\ie the limit where the excited electromagnetic fields remain small relative to the background field) in vacuum. Setting $J^\nu = 0$, and dropping terms of order $\mathcal{O}(g_{a\gamma\gamma}^2)$, 
Eqns.~\ref{eq:aom_eb}-\ref{eq:maxEB} reduce to
\begin{eqnarray}
 (\Box - m_a^2) a &=& 0 \\
 \nabla \cdot \delta \vec{E} &=& - \gagg \vec{B}_0 \cdot \nabla a  \\ 
 \nabla \times (\vec{B}+ \delta \vec{B}) - \partial_t \delta \vec{E} &=& \gagg \vec{B}_0 \partial_t a \label{eq:maxEB_vac1} \, .
\end{eqnarray}
In order to shift this into a more compact form, we will take the time derivative of Amperes law, use  the Maxwell–Faraday equation $\nabla \times \vec{E} = -\partial_t \vec{B}$ to re-write the magnetic field in terms of the electric field; the result is given by
\begin{eqnarray}\label{eq:wave_vac}
    \nabla^2 \delta \vec{E} - \nabla(\nabla \cdot \delta \vec{E}) - \partial_t^2 \delta \vec{E} = \gagg \vec{B}_0 \partial_t^2 a \, ,
\end{eqnarray}
where in the second line we have used vector identities to re-write the first terms. Note also that the axion equation of motion has simply reduced to that of a free scalar field, which motivates adopting a plane wave Ansatz, $a = a_0 e^{-i\omega t + i k_a x} $. Without loss of generality, we will assume the axion plane wave travels in the x-direction, and that the background magnetic field is uniform over some spatial region and oriented in the x-y plane. This motivates an Ansatz for the electric field $\delta \vec{E}(\vec{x}) = \vec{\mathcal{E}}(x) e^{-i\omega t} $, which  yields a solution 
\begin{eqnarray}\label{eq:vac_1}
     \mathcal{E}_x(x) &=& - \gagg B_x(x) a_0 e^{ i k_a x}  \\
    \mathcal{E}_y(x) &=& -\frac{i\omega \gagg a_0}{2}\int dx^\prime e^{i \omega |x- x^\prime| + i k_a x^\prime} B_y(x^\prime)  \\ 
    \mathcal{E}_z(x) &=& 0 \, . \label{eq:vac_3}
\end{eqnarray}
This is the well-known result which shows that: $i)$ a longitudinal electric field is excited in the magnetic field region which oscillates with a phase  consistent with that of the axion, and $ii)$ a transverse electric field is excited, with an amplitude at asymptotically large $x$ which is set by the Fourier transform of $B_y$ evaluated at $k = (\omega - k_{a})$. This example is rather trivial, but highlights two important aspects -- energy dissipation from the axion field can proceed either via the direct excitation of a mode, or indirectly by exciting an electric field which subsequently perturbs a background plasma \footnote{We also briefly comment below on a third possibility: the direct excitation of electromagnetic modes from an inhomogeneous axion distribution. As we demonstrate, however, this effect is heavily suppressed in most systems, and is thus unlikely to be a driving force of energy transfer. }.

\subsection{Isotropic Plasma}

\subsubsection{Non-resonant mixing}
\label{sec:NonResonantConversion}
The derivation above has thrown away the response of any background plasma. In general, one must simultaneously solve the equation of motion for the charged particles alongside Maxwells equations.  Here, we provide a simple example for the case of a cold isotropic plasma. 

We will assume that the background plasma has no relevant thermal contribution, is non-relativistic, and single-species (electron only) -- here, the relevant equation of motion is merely
\begin{eqnarray}
    m_e \partial_t \vec{v}_e = - e (\vec{E} + \vec{v}\times \vec{B}) \, .
\end{eqnarray}
Decomposing the velocity into components parallel to and perpendicular to the magnetic field, and once again assuming $\delta E, \delta B \propto e^{-i \omega t}$,  one finds 
\begin{eqnarray}
    v_{||} &=& \frac{-i e}{m_e \omega }E_{||} \label{eq:v_par}\\
    v_{\perp, \pm} &=& \frac{-ie}{m_e(\omega \pm \omega_c)} E_\pm \, ,
\end{eqnarray}
where $v_\pm = v_{\perp, 1} \pm i v_{\perp, 2}$, and $\omega_c = e B_0 / m_e$ is the cyclotron frequency. Defining the current as $\vec{J} = -e n v$, and taking the limit where $\omega_c \ll \omega$ (\ie the isotropic plasma limit), one can generalize the vacuum wave equation as \
\begin{equation}\label{eq:waveeq_iso}
    \nabla^2 \delta \vec{E} - \nabla(\nabla \cdot \delta \vec{E}) - \partial_t^2 \delta \vec{E} = \frac{i \omega_p^2}{\omega} \partial_t \delta E +  \gagg \vec{B}_0 \partial_t^2 a \, \,  .
\end{equation}
with $\omega_p^2 \equiv e^2 n / m_e$ being the plasma frequency.
Under the same assumptions as before, this reduces to the two equations 
\begin{eqnarray}\label{eq:diffE_iso}
    \mathcal{E}_x &=& -e^{i k_a x}\frac{\omega^2 a_0 \gagg B_x(x)}{\omega^2 - \omega_p^2} \\
    (\partial_x^2 + \omega^2 - \omega_p^2) \mathcal{E}_y &=& -\omega^2 \gagg B_y(x) a_0 e^{ik_a x}\, .
\end{eqnarray}
As before, the first equation describes the longitudinal electric field sourced by the axion. Note that in the limit $\omega_p \rightarrow 0$, this smoothly reproduces the result in the proceeding section, while in the limit that $\omega \ll \omega_p$, one finds the well-known in-medium suppression which scales as $\omega^2 / \omega_p^2$. Thus, indirect energy transfer via the electric field is highly inefficient for light axions.

In the limit where the plasma frequency is uniform, Eq.~\ref{eq:diffE_iso} can be solved using Green's functions (as we did above in the vacuum case), with the solution given by 
\begin{eqnarray}\label{eq:gf_uniisopl}
    \mathcal{E}_y(x) &=& - \frac{i\omega^2 \gagg a_0}{2 \sqrt{\omega^2 - \omega_p^2}} \nonumber \\ &\times & \int dx^\prime e^{i k_\gamma |x- x^\prime|} B_y(x^\prime)  \, .
\end{eqnarray}
with $k_\gamma^2 = \omega^2-m_a^2 +i \epsilon$. 
In the limit where $\omega_p$ is small, this reproduces the vacuum result, while in the limit where $\omega_p$ is large, there is a large suppression in the production of the asymptotic mode (coming both from the pre-factor, and the requirement that the Fourier transform of $B_y$ be evaluated at increasingly high $k$).  One can now define a conversion probability, $p_{a\rightarrow \gamma}$, which captures the ratio of out-flowing electromagnetic energy to in-flowing axion energy (typically computed in the asymptotic limit):
\begin{eqnarray}\label{eq:pc_0def}
    p_{a\rightarrow \gamma} = \frac{\left<S \right>}{T^{20}} = \frac{\frac{1}{2}k_\gamma \omega |A_y|^2}{\frac{1}{2}k_a \omega a_0^2} \, ,
\end{eqnarray}
where $a_0$ is the axion amplitude, $S$ the Poynting flux, and $A_y = E_y / (- i \omega)$.

Let us briefly highlight an alternative, but equivalent, approach, which instead treats the problem via basis transformations. Here, we can start from the wave equation in \ref{eq:waveeq_iso} and the axion equation of motion; working directly with the vector potential (rather than the electric field), and assuming adopting an Ansatz for all fields of the form $f(t, x) = \tilde f(x) e^{i \omega t}$, the equations for the axion and $A_\parallel$ (\ie the component of $\vec{A}$ parallel to $\vec{B}$), read
\begin{equation}
    (\omega^2 + \partial_x^2)
    \begin{pmatrix}
        \tilde A_\parallel \\
        \tilde a 
    \end{pmatrix}
     = 
    \begin{pmatrix}
        \omega_p^2 & -i \omega \gagg  B_0 \\
        i \omega \gagg B_0 & m_a^2
    \end{pmatrix}
    \begin{pmatrix}
        \tilde A_\parallel \\
        \tilde a
    \end{pmatrix},
\end{equation}
where $\omega_p = \sqrt{\frac{n_e e^2}{m_e}}$ is the plasma frequency. Note that we avoided writing explicitly the dependence on the $x$ coordinate on $\tilde A_\parallel$ and $\tilde a$. By performing a phase redefinition of the axion field $\bar a = i \tilde a$, we arrive to the well-known equation \cite{Raffelt:1987im, Smarra:2024vzg}
\begin{equation}
    (\omega^2 + \partial_x^2)
    \begin{pmatrix}
        \tilde A_\parallel \\
        \bar a 
    \end{pmatrix}
     = 
    \begin{pmatrix}
        \omega_p^2 & - \omega \gagg  B_0 \\
        - \omega \gagg B_0 & m_a^2
    \end{pmatrix}
    \begin{pmatrix}
        \tilde A_\parallel \\
        \bar a
    \end{pmatrix},
    \label{eq:AxionLinearSystem}
\end{equation}
which can be written in the form $(\omega^2 + \partial_x^2) \tilde \Psi = \MMsq \tilde \Psi$, where 
\begin{equation}
    \Psi = 
    \begin{pmatrix}
        \tilde A_\parallel \\
        \bar a 
    \end{pmatrix},
\end{equation}
and
\begin{equation}
    \MMsq = 
    \begin{pmatrix}
        \omega_p^2 & - \omega \gagg  B_0 \\
        - \omega \gagg B_0 & m_a^2
    \end{pmatrix}
\end{equation}
is the real and symmetric mass matrix. We now introduce $\epsilon = \frac{\omega \gagg B_0}{m_a^2}$, which encodes the strength of the coupling. The mass matrix then reads
\begin{equation}
    \MMsq = 
    \begin{pmatrix}
        \omega_p^2 & - \epsilon m_a^2 \\
        - \epsilon m_a^2 & m_a^2
    \end{pmatrix}.
\end{equation}
Assuming a constant plasma density, the system can be diagonalized by performing a basis rotation of an angle \cite{Raffelt:1987im}
\begin{equation}
    \theta = \frac{1}{2} \arctan{\biggl(\frac{2 ~ \epsilon ~ m_a^2}{m_a^2 - \omega_p^2}} \biggr),
    \label{eq:MixingAngle}
\end{equation}
so that
\begin{align}
    \tilde \Psi_D &= R(\theta) \tilde \Psi \label{eq:DiagonalSystem}\\
    \MMsq_D &= R(\theta) \MMsq R(\theta)^T = 
    \begin{pmatrix}
        \mu_-^2 & 0 \\
        0& \mu_+^2
    \end{pmatrix},
    \label{eq:DiagonalMassMatrix}
\end{align}
where
\begin{equation}
    R(\theta) = 
    \begin{pmatrix}
        \cos \theta & \sin \theta \\
        -\sin \theta & \cos \theta
    \end{pmatrix}
    \label{eq:RotationMatrix}
\end{equation}
is the rotation matrix, and
\begin{equation}
    \mu_\pm^2 = \frac{m_a^2 + \omega_p^2 \pm \sqrt{(m_a^2 - \omega_p^2)^2 + 4 m_a^4 \epsilon^2}}{2},
    \label{eq:MassMatrixEigenvalues}
\end{equation}
are the eigenvalues of the mass matrix, which depend on the frequency $\omega$ through the parameter $\epsilon$. Following \cite{Smarra:2024vzg}, we will call the basis where $\MMsq$ is diagonal the \textit{unmixed basis}.
Here we will consider $\epsilon \ll 1$, and we get
\begin{equation}
    \mu_\pm^2 = m_a^2 + \OO{\epsilon^2}, \quad \mu_\mp^2 = \omega_p^2 + \OO{\epsilon^2},
    \label{eq:MassEigenvaluesExpansion}
\end{equation}
where the upper and lower signs refer, respectively, to the cases in which $m_a > \omega_p$ and $m_a < \omega_p$. Note, also, that when $m_a > \omega_p$ then $0 \le \theta < \pi / 4$, while when $m_a < \omega_p$ then $- \pi / 4 < \theta \le 0$.

A solution to the system of equations can be written as
\begin{equation}
    \tilde \Psi_D = 
    \begin{pmatrix}
        A_A e^{-i k_A x} \\
        A_a e^{-i k_a x}
    \end{pmatrix},
\end{equation}
where the wave numbers are given by
\begin{equation}
    k_A^2 = \omega^2 - \mu_\mp^2, \quad k_a^2 = \omega^2 - \mu_\pm^2,
    \label{eq:DispersionRelationUnmixed}
\end{equation}
when $m_a \gtrless \omega_p$. Given Eq.~\ref{eq:MassEigenvaluesExpansion}, we see that the component $(1, 0)$ has a dispersion relation closer to a photon, while $(0,1)$ has a dispersion relation closer to an axion; for this reason, we will refer to them as the photon and axion in the unmixed basis, despite in the original basis they have both a component of $\tilde A$ and $\bar a$. 

In order to illustrate how this approach yields the equivalent of Eq.~\ref{eq:gf_uniisopl}, we adopt an  Ansatz \cite{An:2020jmf, An:2023mvf}
\begin{equation}
    \tilde A_\parallel(y) = \hat A_\parallel(x) e^{-i k_\gamma x}, \quad \bar a(y) = \hat a(x) e^{-i k_a x},
\end{equation}
where $k_a = \sqrt{\omega^2 - m_a^2}$ and $k_\gamma = \sqrt{\omega^2 - \omega_p^2}$. Assuming a constant plasma frequency, applying the WKB approximation,  $\abs{\partial_x \hat A_\parallel} \ll k_x \abs{\hat A_\parallel}$, $\abs{\partial^2_x \hat A_\parallel} \ll k_x \abs{\partial_x \hat A_\parallel}$, and assuming the mixing is in the inefficient regime (meaning the axion amplitude can be treated as constant), one finds  
\begin{eqnarray}
    A_{||} = -\frac{\gagg a_0 \omega}{2\sqrt{\omega^2 - \omega_p^2}} \int dx^\prime e^{-i (k_\gamma - k_a) x^\prime} B(x^\prime) \, ,
\end{eqnarray}
which reproduces the previously found solution (after relating the vector potential with the electric field). One can then use Eq.~\ref{eq:pc_0def} to directly compute the conversion probability.

There is an alternative way to obtain the result of non-resonant mixing. For the sake of simplicity, let us consider a  plasma background with a barrier-like shape, with the plasma frequency given by the expression
\begin{equation}
    \omega_p(x) = 
    \begin{cases}
        \omega_{p, \rm{bkg}} & \, \text{if} \quad x < x_0 \\
        \omega_{p, \rm{max}} & \, \text{if} \quad x \ge x_0
    \end{cases},
\end{equation}
where $x_0$ is the position of the boundary of the barrier. The plasma frequencies are chosen in such a way that $\omega_{p, \rm{bkg}} < m_a < \omega_{p, \rm{max}}$, and the transition between them is continuous, so that the resonance region, where $\omega_p \sim m_a$ is infinitely small, and one can safely ignore resonant conversions.

Consider a system  initialized with a wave packet of the axion in the unmixed basis, so that it propagates towards the barrier without undergoing vacuum mixing. When it crosses $x_0$ the mixing angle changes, so that the state has to be re-projected on the new eigenstates, which are subject to different dispersion relations and propagate with different velocities. The conversion probability can then be estimated from the energy flux of the axion $F_a$, outside and inside the barrier, computed in the mixed basis:
\begin{equation}
    p_{a \rightarrow \gamma} = \frac{F_{a, \rm{bkg}} - F_{a, \rm{max}}}{F_{a, \rm{bkg}}} = 1 - \frac{F_{a, \rm{max}}}{F_{a, \rm{bkg}}}.
    \label{eq:NonResonantConversionProbabilityDef}
\end{equation}
If the initial state is monochromatic, the two waves inside the barrier will interfere, since they both have a component of the axion field, producing oscillations in the conversion probability as a result. However, in the case of a wave packet, after some time the two components inside the barrier will separate, having different group velocities. The axion energy flux will thus converge to the sum the contributions from the two eigenstates, and $p_{a \rightarrow \gamma}$ will approach a constant value. We will assume the wave packet to be well localized in frequency, so that it can be assumed to be monochromatic when performing the change of basis from the background to the barrier region, but we will neglect the interference of the waves when computing the energy-flux, in order to obtain the conversion probability in the asymptotic state, when the wave packets have already separated. 

\begin{figure}
    \centering
    \includegraphics[width=\columnwidth]{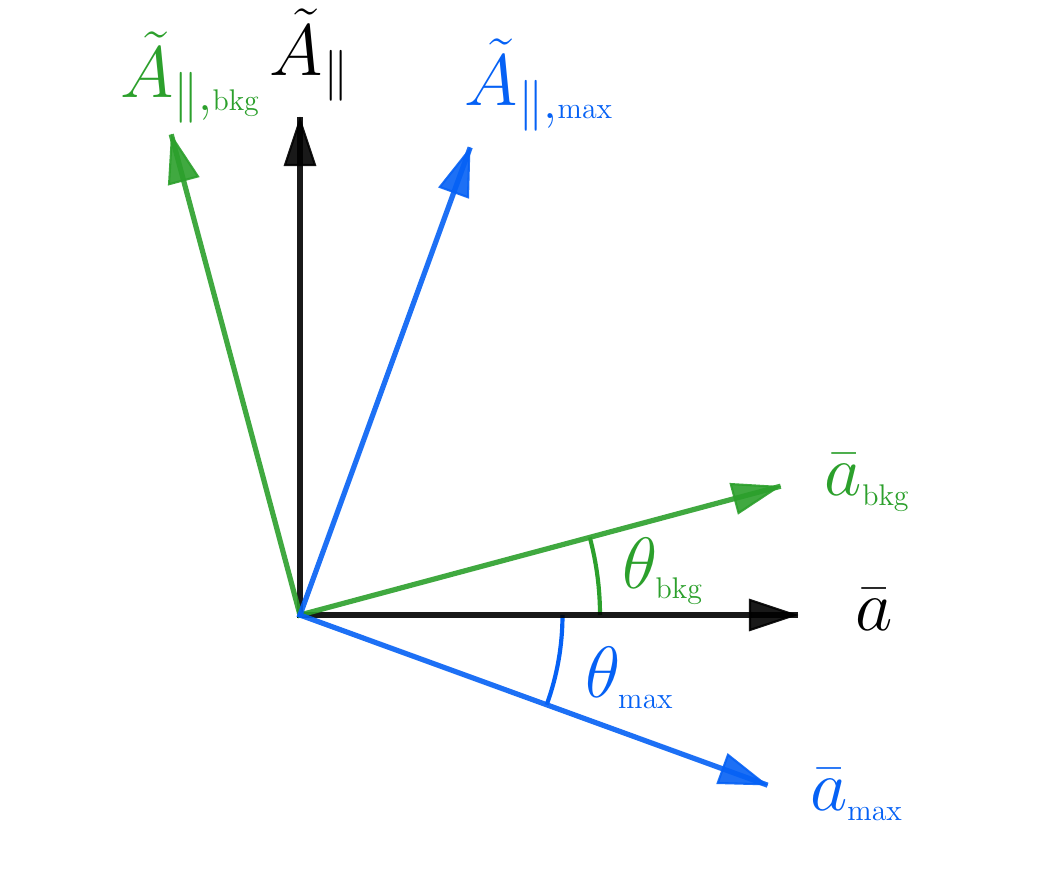}
    \caption{ Schematic representation of the rotation of the unmixed basis, inside (in blue) and outside (in green) the plasma barrier, compared to the original (mixed) one (shown in black). The mixing angles on the background, $\theta_{\rm{bkg}}$, and at the top of the barrier, $\theta_{\rm{max}}$, have positive and negative sign, respectively, due to the fact that the plasma frequencies satisfy $\omega_{p, \rm{bkg}} < m_a < \omega_{p, \rm{max}}$.}
    \label{fig:NonResonantConversionBases}
\end{figure}

We start from an initial state which is defined by the axion field in the unmixed basis, which can be written as
\begin{equation}
    \Psi_{D, \rm{bkg}} = \PsiVector{0}{A_a} e^{i(\omega t - k_{a, \rm{bkg}} x)} = \hat \Psi_{D, \rm{bkg}} e^{i(\omega t - k_{a, \rm{bkg}} x)},
    \label{eq:NRCUnmixedBkg}
\end{equation}
where $\hat \Psi_{D, \rm{bkg}} = (0, A_a)$, $k_{a, \rm{bkg}} = \sqrt{\omega^2 - \mu_{+, \rm{bkg}}^2}$, and $\mu_{+, \rm{bkg}}^2$ is the eigenvalue in Eq.~\ref{eq:MassMatrixEigenvalues} computed at $\omega_p = \omega_{p, \rm{bkg}}$. In the mixed basis, the state reads
\begin{align}
    \Psi_{\rm{bkg}} &= R(\theta_{\rm{bkg}})^T \Psi_{D, \rm{bkg}} = R(-\theta_{\rm{bkg}}) \Psi_{D, \rm{bkg}} \notag  \\
                    &=\PsiVector{-\sin \theta_{\rm{bkg}}}{\cos \theta_{\rm{bkg}}} A_a e^{i(\omega t - k_{a, \rm{bkg}} x)} \notag \\
                    &= \hat \Psi_{\rm{bkg}} e^{i(\omega t - k_{a, \rm{bkg}} x)}.
    \label{eq:NRCMixedBkg}
\end{align}
Inside the barrier the state has to be rewritten in terms of the eigenstates of the unmixed basis at $\omega_p = \omega_{p, \rm{max}}$. The amplitudes of these components, encoded in $\hat \Psi_{D, \rm{max}}$ can be obtained from $\hat \Psi_{\rm{bkg}}$ as
\begin{align}
    \hat \Psi_{D, \rm{max}} &= R(\theta_{\rm{max}}) \hat \Psi_{\rm{bkg}} = R(\theta_{\rm{max}}) R(\theta_{\rm{bkg}})^T \hat \Psi_{D, \rm{bkg}} \notag \\
                         &= R(\theta_{\rm{max}}) R(-\theta_{\rm{bkg}}) \hat \Psi_{D, \rm{bkg}} \notag \\
                         &= R(\theta_{\rm{max}} - \theta_{\rm{bkg}}) \hat \Psi_{D, \rm{bkg}} \notag \\
                         &= R(\theta_{\rm{bkg}} - \theta_{\rm{max}})^T \hat \Psi_{D, \rm{bkg}} \notag \\
                         &= \PsiVector{-\sin{(\theta_{\rm{bkg}} - \theta_{\rm{max}})}}{\cos{(\theta_{\rm{bkg}} - \theta_{\rm{max}})}} A_a
\end{align}
Note that the fact that $\hat \Psi_{D, \rm{max}}$ can be obtained by applying a rotation of an angle $\theta_{\rm{bkg}} - \theta_{\rm{max}}$ to $\Psi_{D, \rm{bkg}}$ is also visible from Fig.~\ref{fig:NonResonantConversionBases}. The two components will then be subject to different dispersion relations, and inside the barrier we will have
\begin{equation}
    \Psi_{D, \rm{max}} =  A_a \PsiVector{-\sin{(\theta_{\rm{bkg}} - \theta_{\rm{max}})} e^{i(\omega t - k_{A, \rm{max}} x)}}{\cos{(\theta_{\rm{bkg}} - \theta_{\rm{max}})} e^{i(\omega t - k_{a, \rm{max}} x)}},
    \label{eq:NRCUnmixedMax}
\end{equation}
where now $k_{A, \rm{max}} = \sqrt{\omega^2 - \mu_{+, \rm{max}}^2}$, and $k_{a, \rm{max}} = \sqrt{\omega^2 - \mu_{-, \rm{max}}^2}$. Here the eigenvalues of the mass matrix have been computed at $\omega_p = \omega_{p, \rm{max}}$, and have been assigned to the photon and axion components according to the fact that $\omega_{p, \rm{max}} > m_a$.

We can now transform this state back to the mixed basis, by applying the rotation $R(\theta_{\rm{\max}})^T$ to $\Psi_{D, \rm{max}}$. Since we will compute the axion energy flux independently on the two eigenstates, and we write them separately as
\begin{align}
    \Psi_{\rm{max}} &= - A_a \sin{(\theta_{\rm{bkg}} - \theta_{\rm{max}})} \PsiVector{\cos \theta_{\rm{max}}}{\sin \theta_{\rm{max}}} e^{i(\omega t - k_{A, \rm{max}} x)} \notag \\
                    &+ A_a \cos{(\theta_{\rm{bkg}} - \theta_{\rm{max}})} \PsiVector{-\sin \theta_{\rm{max}}}{\cos \theta_{\rm{max}}} e^{i(\omega t - k_{a, \rm{max}} x)}.
    \label{eq:NRCMixedMax}
\end{align}
For a generic eigenstate of the mass matrix 
\begin{equation}
    \Psi = \PsiVector{A_A}{A_a} e^{i(\omega t - k x)},
\end{equation}
with $\omega$ and $k$ satisfying the appropriate dispersion relation, the axion energy flux $F_a = \frac{d \EE}{dS dt}$ can be computed by averaging  the energy per unit surface over a period $T = 2 \pi / \omega$. This is the same as the energy per unit surface contained in a segment of length $\lambda = \frac{2\pi}{k} = v_p T$, normalized by $T$:
\begin{equation}
    F_a = \frac{1}{T} \int_{x}^{x + v_p T} dx' \, \rho_a (x');
    \label{eq:AxionEnergyFluxDef}
\end{equation}
where $\rho_a(x)$ is the axion energy density
\begin{equation}
    \rho_a(x) = \frac{1}{2} \Bigl[(\tder a)^2 + (\xder a)^2 + m_a^2 a^2 \Bigr],
\end{equation}
computed using the original axion field appearing in Eq.~\ref{eq:AxionPhotonLagrangian}. Keeping therefore into consideration the redefinition $\bar a = i \tilde a$ performed before Eq.~\ref{eq:AxionLinearSystem} we obtain that
\begin{equation}
    F_a = \frac{A_a^2 (\omega^2 + k^2 + m_a^2)}{4} v_p.
    \label{eq:AxionEnergyFlux}
\end{equation}
With this we can now compute the axion energy flux of the initial state from Eq.~\ref{eq:NRCMixedBkg} as
\begin{equation}
    F_{a, \rm{bkg}} = A_a^2 \cos^2 \theta_{\rm{bkg}} \frac{\omega^2 + k_{a, \rm{bkg}}^2 + m_a^2}{4} \frac{\omega}{k_{a, \rm{bkg}}}.
\end{equation}
$F_{a, \rm{max}}$ is instead computed summing over the two eigenstates in Eq.~\ref{eq:NRCMixedMax}:
\begin{align}
    F_{a, \rm{max}} &= A_a^2 \sin^2 \theta_{\rm{max}} \sin^2{(\theta_{\rm{bkg}} - \theta_{\rm{max}})} \notag \\
                    &\times \frac{\omega^2 + k_{A, \rm{max}}^2 + m_a^2}{4} \frac{\omega}{k_{A, \rm{max}}} \notag \\
                    &+A_a^2 \cos^2 \theta_{\rm{max}} \cos^2{(\theta_{\rm{bkg}} - \theta_{\rm{max}})} \notag \\
                    &\times \frac{\omega^2 + k_{a, \rm{max}}^2 + m_a^2}{4} \frac{\omega}{k_{a, \rm{max}}}.
\end{align}
Since when computing the conversion probability the factors $\frac{A_a^2}{4}$ cancel, we can define $\mathcal F_a = 4 F_a / A_a^2$, and compute the probability as
\begin{equation}
    p_{a \rightarrow \gamma} = 1 - \frac{\mathcal{F}_{a, \rm{max}}}{\mathcal{F}_{a, \rm{bkg}}}.
\end{equation}
Notice that in the limit of weak mixing (and a uniform magnetic field), this procedure reproduces the identical conversion probability as would be obtained by combining Eq.~\ref{eq:gf_uniisopl} with Eq.~\ref{eq:pc_0def}.

\subsubsection{Resonant transitions}

The other limit of interest is the one in which the background has a varying density profile, $\omega_p(x)$. In this case, one cannot use the Green's functions approach; instead, it is more natural to assume that the background density varies slowly, and apply a WKB approximation to an initial Ansatz $\mathcal{E}_y(x) = \tilde{\mathcal{E}}_y(x) e^{i k_a x}$, where $k_a \tilde{\mathcal{E}}_y \gg \partial_x \tilde{\mathcal{E}}_y$, and $\partial_x^2 \tilde{\mathcal{E}}_y \ll k_a \partial_x \tilde{\mathcal{E}}_y$ (note the second condition follows from the former so long as the envelope does not have any sharp features). 

Under this approximation, one can write 
\begin{eqnarray}
    2 i k_a \partial_x \tilde{\mathcal{E}}_y \simeq - (m_a^2 - \omega_p^2) \tilde{\mathcal{E}}_y - \omega^2 \gagg B_y(x) a_0 
\end{eqnarray}
which can be integrated to give
\begin{eqnarray}\label{eq:E_iso_res}
    \tilde{\mathcal{E}}_y \simeq \frac{i \omega^2 \gagg a_0}{2 k_a}  e^{-i\phi(x)} \, \int_{-\infty}^x d x^\prime B_y(x^\prime) e^{i \phi(x^\prime)}
\end{eqnarray}
with 
\begin{eqnarray}
    \phi(x) = \int_0^x \, dx^\prime \frac{m_a^2 - \omega_p^2(x^\prime)}{2 k_a} \, .
\end{eqnarray}
Here, one has a resonant enhancement of the mixing process, with the resonance being triggered when the dispersion relation of the axion matches that of the photon. One can use the stationary phase approximation to compute Eq.~\ref{eq:E_iso_res} asymptotically far from the location of the resonance itself 
\begin{equation}\label{eq:iso_res_ey}
  \tilde{\mathcal{E}}_y = \frac{i \omega^2 \gagg a_0}{2 k_a} B_y(x_c) e^{i \, {\rm sgn}(\phi(x_c)) + i \pi/4} \sqrt{\frac{2\pi}{|\partial_x^2 \phi\big|_{x_c}|}} \, ,
\end{equation}
where $x_c$ denotes the resonance point (here, corresponding to where $\omega_p = m_a$). Notice that a key assumption in this derivation is the validity of the WKB approximation, which follows from the constraint that the background quantities are sufficiently slowly varying. 

It is worth pointing out that in this limit, the dielectric tensor is diagonal, ${\bf{\epsilon}} = {\rm diag}(1 - \omega_p^2/\omega^2,1 - \omega_p^2/\omega^2,1 - \omega_p^2/\omega^2)$. The standard dispersion relations can be derived by plugging in a plane solution to the wave equation in Eq.~\ref{eq:waveeq_iso} with $a \rightarrow 0$, and solving for the eigenmodes. This effectively amounts to solving for the eigenmodes of the matrix ${\bf{\mathcal{C}}}\equiv n^2 \delta_{ij} - n_i n_j - \epsilon_{ij}$. Doing this procedure, one finds 2 transverse propagating modes with dispersion relations $\omega^2 = k^2 + \omega_p^2$, and one longitudinal mode with $\omega^2 = \omega_p^2$. The axion purely excites the transverse mode, and the longitudinal mode is entirely de-coupled from the transverse mode (since the polarizations are orthogonal). This is not generically true for arbitrary media, however, and as we will see below, the axion can in some circumstances indirectly excite other modes.

Turning to the secondary matrix approach, and working in the original (mixed) basis, we can write an Ansatz for the solution of Eq.~\ref{eq:AxionLinearSystem} as \cite{An:2020jmf, An:2023mvf}
\begin{equation}
    \tilde A_\parallel(x) = \hat A_\parallel(y) e^{-i k_x x}, \quad \bar a(x) = \hat a(x) e^{-i k_x x},
\end{equation}
where $k_x = \sqrt{\omega^2 - m_a^2}$. If the scales over which the variation of the plasma density occur are large enough that we can consider $\abs{\partial_x \hat A_\parallel} \ll k_x \abs{\hat A_\parallel}$, $\abs{\partial^2_x \hat A_\parallel} \ll k_x \abs{\partial_x \hat A_\parallel}$, and the same for $\hat a$, then we can apply a WKB approximation and write the system in a Schr\"odinger-like form in terms of the $x$ coordinate. By working in the interaction picture, the equations can be solved from a starting point $x_i$ to a generic $x$, and the probability of conversion from axion to photon reads \cite{Raffelt:1987im, An:2020jmf, An:2023mvf}
\begin{equation}
    p_{a \to \gamma} = \epsilon^2 \left| \int_{x_i}^{x} dx' \, \Delta_a e^{i \phi(x')} \right|^2
    \label{eq:ConversionProbabilityWKB}
\end{equation}
where $\phi(x) = \int_{x_i}^x dx' \, \bigl( \Delta(x') - \Delta_a \bigr)$, while $\Delta = - \frac{\omega_p(x)^2}{2 k_x}$ and $\Delta_a = -\frac{m_a^2}{2 k_x}$. The integrand in Eq.~\ref{eq:ConversionProbabilityWKB} is highly oscillating, so that $p_{a \to \gamma}$ mainly receives contribution from the stationary points where $\phi'(x) = 0$, which corresponds to points for which $m_a \simeq \omega_p$. So if the plasma frequency crosses the value of the axion mass, then we have a resonant conversion process; in the case the stationary points $x_n$ are well-separated (meaning the relative distance between adjacent resonances is larger than the width of the resonance $\sim \sqrt{\frac{2}{|\phi''(x)|}}$), the probability can be estimated by means of the stationary-phase approximation, which yields the well-known Landau-Zener (LZ) formula:
\begin{equation}
    p_{\scriptscriptstyle{LZ}} = \sum_{x_n} \epsilon^2 \frac{m_a^4}{2 k_x^2} \frac{\pi}{\abs{\Delta'(x_n)}}.
    \label{eq:LandauZener}
\end{equation}
Comparing this expression with that of Eq.~\ref{eq:iso_res_ey}, one can see that these approaches are equivalent for a single isolated resonance.

\subsection{Anisotropic Plasma}
We now turn our attention to magnetized plasmas, in which $\omega_c \gg \omega,\omega_p$ -- these plasmas support diverse modes, making their dynamics intrinsically more complex than that of the isotropic example given above. For simplicity, we will work in the infinite magnetized limit -- for the interests here, this distinction is not important, but we will highlight below where this approximation enters\footnote{Note that this limit is directly imposed in our frequency domain simulations, as here we can directly input the dielectric tensor, while it is not imposed in our time domain simulations, since the code makes use of the full profile of the magnetic field. Even in the the case of the latter, however, we impose $\omega_c \gg \omega, \omega_p$.}. 

The wave equation can, in general, be expressed as
\begin{eqnarray}\label{eq:waveeq_ani}
    -\nabla^2 \delta \vec{E} + \nabla(\nabla \cdot \delta \vec{E}) - \omega^2 {\bf{\epsilon}} \delta \vec{E}  =   \gagg \vec{B}_0 \partial_t^2 a \, \, ,
\end{eqnarray}
where the dielectric tensor under the assumption that the medium linearly responds to an external field follows from $\vec{J} = i \omega (\mathbb{1} - {\bf{\epsilon}}) \cdot \vec{E}$, where $\vec{J} = J_{||}$, and $J_{||}$ is inferred from Eq.~\ref{eq:v_par}

One can directly obtain the plasma modes present in this media using the same procedure as before; here, three modes are obtained, with dispersion relations  
\begin{align}
    \omega_{\rm A}^2 &= \frac{1}{2}(k^2 + \omega_p^2 - \sqrt{k^4 + \omega_p^4 - 2 k^2 \omega_p^2 \cos(2\theta_B)}) \label{eq:AlfvenDispersionRelation} \\
    \omega_{\rm LO}^2 &= \frac{1}{2}(k^2 + \omega_p^2 + \sqrt{k^4 + \omega_p^4 - 2 k^2 \omega_p^2 \cos(2\theta_B)}) \\
    \omega_m^2 &= k^2 c_m^2 
\end{align}
corresponding to the Alfv\'{e}n, the Langmuir-Ordinary, and the magnetosonic modes~\cite{gedalin1998long}. Here, $\theta_B$ is the angle between $\vec{k}$ and $\vec{B}$. In the infinity magnetized limit, $c_m \rightarrow 1$; for non-thermal systems, it can more generally be approximated as $c_m \sim v_A / (1 + v_A)$, with $v_A = B / \sqrt{\rho_m}$ the Alfv\'{e}n velocity (with $\rho_m$ the plasma energy density). The polarizations corresponding to each mode (in the basis with $\vec{k} \propto \hat{x}$ and $\vec{B}$ in the x-y plane), are given by 
\begin{eqnarray}
\boldsymbol{\epsilon_A}(\omega^2) &\propto& \left( \frac{\omega_p^2 \sin\theta \cos\theta}{\omega_A^2 - \omega_p^2 \cos^2\theta},\ 1 \, , 0\right)\, , \label{eq:AlfenPolarizationVector} \\
\boldsymbol{\epsilon_{\rm LO}}(\omega^2) &\propto &\left( \frac{\omega_p^2 \sin\theta \cos\theta}{\omega_{\rm LO}^2 - \omega_p^2 \cos^2\theta}, \, 1 ,\ 0\right)\, \\
\boldsymbol{\epsilon_{\rm m}} &=& (0, 0 , 1)
\end{eqnarray}
where the proportionality signs of the Alfv\'en and LO mode indicate that neither have been normalized.

We can start by noting that axion dark matter is typically only expected to strongly mix with the LO mode. This statement stems from the fact that the Alfv\'{e}n mode is sub-luminal, $\omega < k$, while the axion is always super-luminal $\omega > k$ (thus, in order for energy and momentum to be conserved, there must exist a strong inhomogeneity in the background). We will later demonstrate how and when this intuition can break down, but for now, we will focus merely on understanding LO mode excitation via resonant axion mixing.

\begin{figure}
    \includegraphics[width=0.49\textwidth]{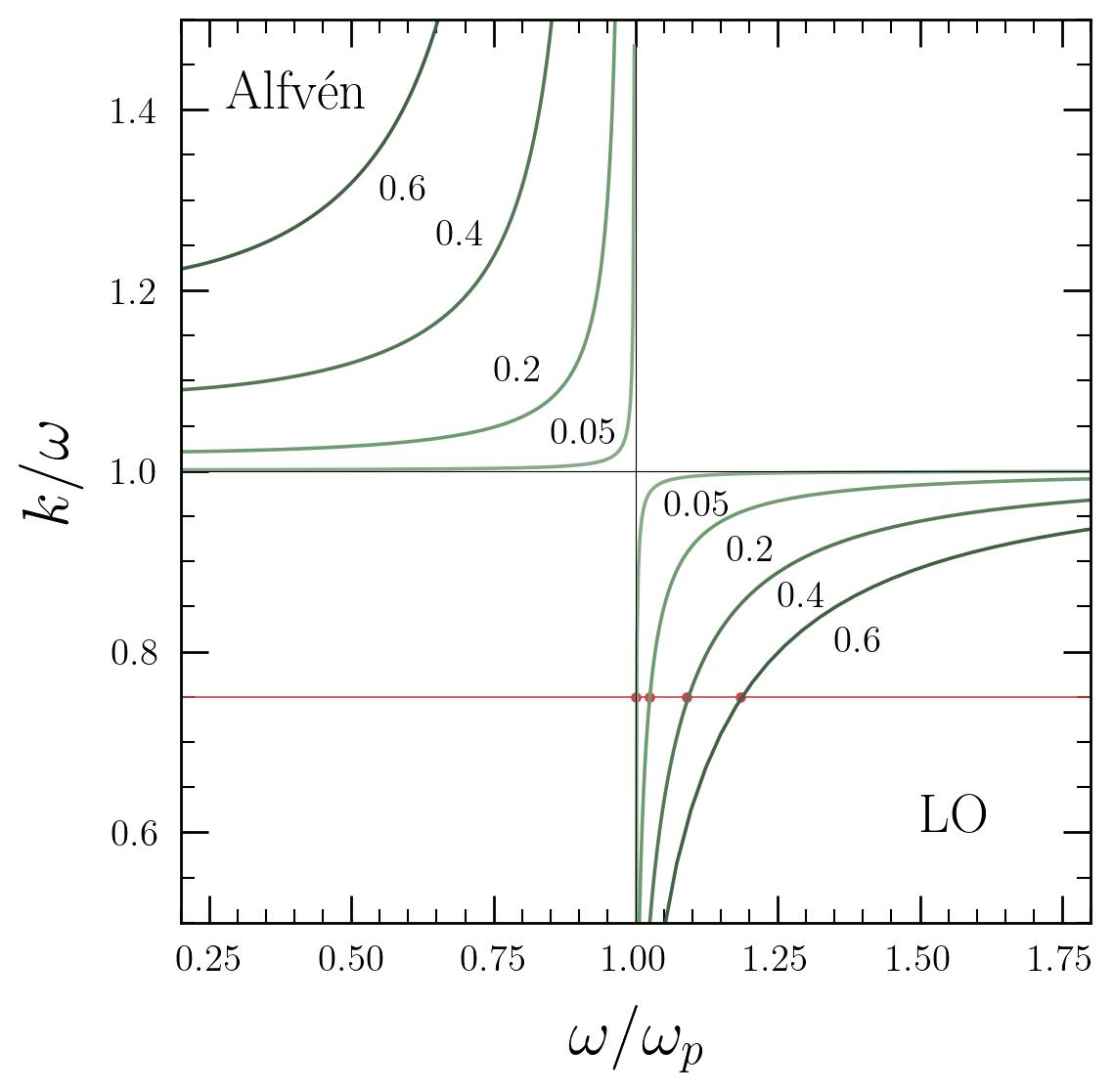}
    \caption{\label{fig:mode_comp} Comparison of the LO (bottom right quadrant) and Alfv\'{e}n (top left quadrant) modes, plotted as the ratio of $k/\omega$ vs the ratio of $\omega / \omega_p$, for various values of $\theta_B$ (relevant values, in radians, are written next to each curve). For reference, we also provide an illustrative example for the axion in red, highlighting the level crossings with the LO mode using small red points.  Despite one mode being super-luminal (LO) and the other sub-luminal (Alfv\'en), one can see the cut-off (occurring for the LO mode at $\omega/\omega_p = 1$) in the dispersion relation become increasingly degenerate with the resonance of the Alfv\'en mode (occurring at $\omega = \omega_p |\cos\theta_B|$) in the limit that $\theta_B \rightarrow 0$. Should the background vary sufficiently quickly near the cut-off and resonance, one will expect a non-zero tunneling probability of any incident wave (see sections below). }
\end{figure}

Expanding the wave equation in Eq.~\ref{eq:waveeq_ani}, and assuming the background gradients are confined to the x-y plane such that the z-coordinate decouples, yields 
\begin{eqnarray}
    \partial_x^2 \delta E_y - \partial_x\partial_y \delta E_x + (\omega^2 - \omega_p^2 \sin^2\theta_B) \delta E_y \nonumber \\ - \omega_p^2 \sin\theta_B \cos\theta_B \delta E_x = -\omega^2 \gagg B_y a \\
    \partial_y^2 \delta E_x - \partial_x\partial_y \delta E_y + (\omega^2 - \omega_p^2 \cos^2\theta_B) \delta E_x \nonumber \\ - \omega_p^2 \sin\theta_B \cos\theta_B \delta E_y = -\omega^2 \gagg B_x a \, .
\end{eqnarray}
Previously, one could simplify this expression by assuming that $E_y$ and $E_x$ were functions merely of the x-coordinate; however, in an anisotropic media, waves with $k \ll \omega$ undergo refraction, and can thus pick up additional functional dependencies on the y-coordinate. Non-resonant mixing is thus complicated by the non-degeneracy of the axion and photon world lines. In the case of resonant mixing, one can either adopt a density matrix formalism~\cite{McDonald:2023ohd}, or adopt the Eikonal approximation $\vec{E} = A \hat{\epsilon} e^{i \theta(\vec{x})}$, and derive analytic expressions for the excitation~\cite{McDonald:2024uuh} -- here, we briefly outline the latter approach, following the derivation of ~\cite{McDonald:2024uuh}.  

Starting from the wave equation in Eq.~\ref{eq:waveeq_ani}, one can make the standard Eikonial Ansatz for the electric field $\delta \vec{E} = \mathcal{A} \hat{\epsilon} e^{i \Theta(\vec{x})}$ to yield the following equation:
\begin{eqnarray}
    \mathcal{D}_{ij}(\mathcal{A} \hat{\epsilon}_j) + i \frac{\partial \mathcal{D}_{ij}}{\partial k_l} \nabla_l (\mathcal{A} \hat{\epsilon}_j) + i \frac{(\mathcal{A} \hat{\epsilon}_j)}{2} \frac{\partial \mathcal{D}_{ij}}{\partial k_l \partial k_{l^\prime}} \nabla_l k_{l^\prime}  \nonumber \\ +
    \frac{1}{2}\frac{\partial \mathcal{D}_{ij}}{\partial k_l \partial k_{l^\prime}}\nabla_l \nabla_{l^\prime} (\mathcal{A}\epsilon_j) = \gagg\omega^2 a_0 B_i e^{i (\vec{k}_a \cdot \vec{x} - \Theta)} 
\end{eqnarray}
where we have defined the wave operator~\cite{McDonald:2024uuh}
\begin{eqnarray}
    \mathcal{D}_{ij} = - |k|^2 \delta_{ij} + k_i k_j + \omega^2 \epsilon_{ij}(\omega, \vec{x}) \, .
\end{eqnarray}
Projecting this differential equation along the polarization vector from the left yields a sourced transport equation
\begin{eqnarray}
    [\epsilon^* \cdot (\partial_{\vec{k}}\mathcal{D}) \cdot \epsilon] \cdot \nabla \mathcal{A} +  \\  \left[ \epsilon^* \cdot (\partial_{\vec{k}}\mathcal{D}) \cdot \nabla \epsilon + \frac{1}{2} \epsilon_i^* (\partial_{k_l}\partial_{k_{l^\prime}} \mathcal{D}_{ij}) \epsilon_j \nabla_l k_{l^\prime} \right] \mathcal{A} \nonumber \\ = \gagg \omega^2 a_0 (\vec{B} \cdot \epsilon^*) e^{i(k_a \cdot \vec{x} - \Theta)} \, .
\end{eqnarray}
Here, one can identify the group velocity 
\begin{eqnarray}
    \vec{v}_g \equiv \partial \omega / \partial \vec{k} = \frac{\epsilon^* \cdot (\partial_{\vec{k}} \mathcal{D}) \cdot \epsilon}{\partial_\omega \mathcal{H}}
\end{eqnarray}
where $\mathcal{H}$ is the matrix defining the eigenvalue problem $\mathcal{H}\epsilon = 0$ (the eigenmodes defining the relevant dispersion relations). This can be recast as a simple transport equation of the form
\begin{eqnarray}
    \vec{v}_g \cdot \nabla U_\gamma + (\nabla \cdot \vec{v}_g)U_\gamma  \nonumber \\ = \frac{1}{4}\gagg \omega^2 a_0 (\vec{B} \cdot \epsilon^*) e^{i(k_a \cdot \vec{x} - \Theta)} A^* + {\rm h.c.}
\end{eqnarray}
where 
\begin{eqnarray}
    U_\gamma = \frac{1}{4}|\delta \vec{B}|^2 + \frac{1}{4} \delta E_i^* \partial_\omega (\omega \epsilon_{ij})\delta E_j \, .
\end{eqnarray}
This transport equation can be solved in a manner similar to the case of the isotropic resonance, with the result being given by
\begin{equation}\label{eq:pa_resonant}
    p_{a\rightarrow \gamma} = \frac{\pi}{2} \frac{\omega^2 \gagg^2 B^2}{k_a} \frac{1}{|\omega_p \partial_x \omega_p + \frac{\omega^2-\omega_p^2}{\omega^2 \tan\theta_B} \omega_p^2 \partial_x \theta_B|} \, .
\end{equation}
This result has been confirmed in numerical simulations in~\cite{Gines:2024ekm}. Notice, that as before, this approach relies on the assumption that the background is sufficiently slowly varying that the WKB approximation can be applied. As argued in the introduction, this is not always the case. 

\subsection{Behavior near cut-offs and resonances}  \label{sec:LZAlfven}

In an isotropic plasma, the dispersion relation of the ordinary mode is given by $\omega^2 = k^2 + \omega_p^2$. Here, one can see the appearance of a cut-off in the dispersion relation at $\omega = \omega_p$. Alone, this feature is not particularly interesting, as it merely represents total reflection of any incident wave.

The story becomes slightly more interesting in the context of a magnetized plasma. Here, the momentum of the LO mode and Alfv\'{e}n modes are given by
\begin{eqnarray}
    k^2 = \frac{\omega^2 (\omega^2 - \omega_p^2)}{\omega^2 - \omega_p^2 \cos^2\theta} \, ,
\end{eqnarray}
where the LO mode exists only for $\omega \geq \omega_p$  and the Alfv\'{e}n mode exists only for $\omega \leq \omega_p 
\times |\cos\theta|$. Notice that the expression for $k$ implies the existence of both a cut-off at $\omega = \omega_p$, and a resonance at $\omega = \omega_p |\cos\theta_B|$. These features are nicely illustrated in Fig.~\ref{fig:mode_comp}, where we plot the LO and Alfv\'{e}n modes for various angles as a function of $\omega/\omega_p$. 

Typically, one only expects the axion to mix with the LO mode -- this follows from the fact that the Alfv\'{e}n branch is sub-luminal, $\omega < k$, while the axion is super-luminal, $\omega > k$ (the inherent mismatch in the energy-momentum leads to a huge supression of any mixing).  When the plasma gradient is large, however, and the spatial separation of the resonance and cut-off small, one can tunnel directly from the LO branch to the Alfv\'{e}n branch. In the context of axion physics, this mechanism provides an alternative indirect means for exciting sub-luminal plasma modes. Below, we describe how this phenomena arises in standard electrodynamics.

Starting from the wave equation with an anisotropic dielectric tensor, one can write the differential equations for each electric field component; for a wave traveling in the x-direction, and assuming all gradients of the background media are aligned with this direction (allowing one to reduce the functional dependence $\vec{E}(\vec{x}) = \vec{E}(x)$), this reduces to:
\begin{align}
    E_x &= \frac{\omega_p^2 \cos\theta_B \sin\theta_B}{\omega^2 - \omega_p^2 \cos^2\theta_B} E_y \\ 
    \partial^2_x E_y &= \omega_p^2 \cos\theta_B \sin\theta E_x + (\omega_p^2 \sin^2\theta_B - \omega^2) E_y \\ 
     \partial^2_x E_z &= - \omega^2 E_z \, .
\end{align}
Inserting the longitudinal solution into the differential equation for $E_x$ yields
\begin{eqnarray}
     \partial^2_x E_y = \frac{\omega^2 (\omega_p^2 - \omega^2)}{\omega^2 - \omega_p^2 \cos^2\theta_B} E_y \, . 
\end{eqnarray}
Here, one can see the cut-off structure arising at $\omega = \omega_p$, and the resonant structure arising at $\omega = \omega_p \cos\theta_B$. Expanding the plasma frequency near the resonance,  $\omega_p^2 \equiv \Omega \simeq \Omega(x_r) + \partial_x \Omega \big|_{x_r} (x- x_r) $ with $\Omega(x_r) \equiv \omega^2 / \cos^2\theta_B$. Plugging in this substitution and expanding $x$ to leading order about $x_r$ yields
\begin{eqnarray}
    \partial^2_x E_y - \frac{\omega^2}{\cos^2\theta_B} \left(-1 + \frac{\delta x_{cr}}{(x-x_r)} \right) E_y = 0 \, ,
\end{eqnarray}
where we have assumed the cut-off and resonance are sufficiently close that one can write $\delta x_{cr} \equiv (x_c - x_r) \simeq (\omega^2 - \Omega(x_r))/ (\partial_x \Omega \big|_{x_r})$. Re-defining $x \rightarrow \xi \cos\theta_B / \omega + x_r$, one arrives at the known `Budden equation'~\cite{budden1961,swanson2020plasma},
\begin{eqnarray}
  \partial^2_\xi E_y + \left(1 + \frac{-\delta x_{cr} (\frac{\omega}{\cos\theta_B}) }{\xi + i \epsilon} \right) E_y = 0,
\end{eqnarray}
where we have made it clear that the integration of the solution must be done in the complex plane in order to avoid the singular point at the resonance.

We can now look for solutions which respect the asymptotics of the problem of interest: namely an incident wave which is partially reflected, and partially transmitted. 

The solution to this equation yields a Landau-Zener like conversion probability 
\begin{eqnarray}
    P_{c} =  e^{- \pi \eta },
    \label{eq:PAlfvenAnalytic}
\end{eqnarray}
where 
\begin{eqnarray}\label{eq:buddeneta}
    \eta = \frac{ \omega^2(\sec^2\theta_B-1)}{2\,|\partial_x \omega_p|_{x_r}}
= \frac{\omega^2\tan^2\theta_B}{2\,|\partial_x\omega_p|_{x_r}}
\end{eqnarray}
and when moving to Eq.~\ref{eq:buddeneta} we have assumed that $\eta \ll 1$. Roughly speaking, this result shows a non-negligible excitation amplitude when the separation between the cut-off and the resonance is small with respect to the wavelength of the photon. Note that there appears to be a divergence in Eq.~\ref{eq:buddeneta} at $\theta_B = \pi/2$ -- this corresponds to the limit in which the resonance, located $\omega_p(x_r) = \omega/|\cos\theta_B|$, is infinitely separated from the cut-off, breaking the assumptions in derivation above.

This is a nice result, however it relies on a number of assumptions -- namely, the existence of a stratified medium (allowing the functional dependence on $\vec{E}$ to be reduced to one dimension), the ability to locally expand and solve between $x_r$ and $x_c$, and the assumption that the spatial variation of the amplitude is driven merely by the tunneling effect. In generalized backgrounds, and in the scenario in which the LO mode is being simultaneously excited by the background axion, this approximation is likely to break down. We demonstrate in the sections below how one can numerically reconstruct the Alfv\'{e}n mode which is indirectly excited in this manner, allowing for the tools developed here to study axion-induced mode conversion across a wider range of scenarios.

\subsection{Alternative approach to axion-induced electric field}

We started this section by arguing that energy dissipation could occur via on-shell photon production, or indirectly via the work done by axion-excited electromagnetic fields on the ambient plasma. Above, we have provided various examples to illustrate how these fields on electromagnetic modes can be produced, and under what assumptions these results are valid. The case of indirect energy dissipation represents a scenario which is often highly suppressed. There are two reasons for this: first, the harmonic oscillations in the axion field imply there is often a high degree of cancellation between different phases, and second, the presence of a dense medium with $\omega_p \gg \omega$ can heavily suppress the effective amplitude of the induced electric field. The former effect can be partly evaded if there is net energy dissipation on timescales much less than the oscillation timescale of the axion. The latter effect can also be partly evaded, as recently discussed in~\cite{Caputo:2023cpv}, if there exist small plasma under-densities, as can often arise in astrophysical settings. Here, we briefly discuss how this effect can arise, and demonstrate in the later sections numerically how the electric field behaves in various regimes.

One should effectively think of this problem as being similar to that of an axion haloscope -- namely, one is hoping to solve the vacuum wave equation in Eq.~\ref{eq:wave_vac} with boundary conditions. Let us focus on the case of an isotropic plasma which fills most of the space, except a small cubic region of characteristic scale-size $L$. We will take an axion plane wave solution propagating in the x-direction, and a perpendicular magnetic field in y-direction. We will always work in the small coupling limit, such that the external plasma can be treated as local. Outside the vacuum region, the dense plasma drives the electric field to zero, producing an effective cavity. Is it the imposition of these boundary conditions which are what differentiate this solution from that of Eqs.~\ref{eq:vac_1}-\ref{eq:vac_3}. 

In order to derive solutions for the axion-induced electric field, we begin by noting that the homogeneous solution inside the cavity (and respecting the boundary conditions, which force $E_\perp = 0$ on the boundary) supports a discrete set of modes, $\vec{E}(\vec{x}) = \sum_n c_n \vec{\epsilon}_n(\vec{x})$, where the modes are given by
\begin{align}
\epsilon_x^{(mnp)}(\mathbf{x})
&=
-\frac{mn}{m^2+p^2}\,E_0^{(mnp)} \\
& \times \cos\!\left(\frac{m\pi x}{L}\right)
\sin\!\left(\frac{n\pi y}{L}\right)
\sin\!\left(\frac{p\pi z}{L}\right) \, \nonumber
\\[10pt]
\epsilon_y^{(mnp)}(\mathbf{x})
&=
E_0^{(mnp)}
\sin\!\left(\frac{m\pi x}{L}\right)
\cos\!\left(\frac{n\pi y}{L}\right)
\sin\!\left(\frac{p\pi z}{L}\right),
\\
\epsilon_z^{(mnp)}(\mathbf{x})
&=
-\frac{pn}{m^2+p^2}\,E_0^{(mnp)} \\ &\times 
\sin\!\left(\frac{m\pi x}{L}\right)
\sin\!\left(\frac{n\pi y}{L}\right)
\cos\!\left(\frac{p\pi z}{L}\right) \nonumber \, .
\end{align}
where $m,n,p$ are integers, and $E_0^{(mnp)}$ a free amplitude. Plugging this solution into the source-less wave equation, one can see that each eigenstate has an energy $\omega_{mnp}^2 = \frac{\pi^2}{L^2}(m^2 + n^2 + p^2)$. The axion-induced wave equation can be solved for by plugging in this solution, projecting onto a given state, and exploiting the orthogonality to find the weight coefficients:
\begin{equation}
    c_{mnp} = \frac{\omega^2}{\omega_{mnp}^2 - \omega^2} \frac{-\gagg a B}{E_0^{(mnp)}} \frac{64}{mnp \pi^3} \frac{m^2 + p^2}{m^2 + n^2 + p^2} \, .
\end{equation}
Notice in the limit that $\omega \rightarrow 0$, only the lowest order state contributes, and the induced electric field amplitude scales, relative to the vacuum field, as $|\vec{E}| / |\vec{E}^{\rm vac}| \propto \omega^2 / \omega_{111}^2 \propto (m_a \times L)^2$. Here, one can see that the parametric suppression of $(m_a / \omega_p)^2$ was traded for a suppression of $(m_a \times L)^2$, which may imply that the naive in-medium suppression is not nearly as severe as one would have otherwise thought. 

The above example is oversimplified. The suppression at the boundary will rarely ever be perfect, implying that the electric field will leak across the boundary for realistic plasma densities. In addition, in anisotropic media, the boundary conditions are modified, since the plasma does not guarantee a suppression for all electric field components (recall that the plasma response becomes one dimensional).  We return to comment on the impact of such variations below.

\subsection{Radiative energy losses from inhomogeneous axions}

Before continuing, we comment on one final scenario in which energy can be lost by an inhomogeneous axion field, and which is not explicitly included in the examples above, but has been studied previously in~\cite{Caputo:2023cpv}.

Starting from Eqns.~\ref{eq:nabladotEGauss}-\ref{eq:maxEB}, one can derive wave equations for the electric and magnetic fields excited by an axion in a large background $\vec{B}$ field, assuming no background charges; the result is given by
\begin{eqnarray}
    (\partial_t^2 - \nabla^2)\delta \vec{E} = \nabla \rho_{ae} + \partial_t \vec{J}_{ae} \\
    (\partial_t^2 - \nabla^2)\delta \vec{B} = - \nabla \times \vec{J}_{ae} \, 
\end{eqnarray}
with $\rho_{ae}$ and $\vec{J}_{ae}$ being the axion charge and current density. Using Green's functions, assuming an axion field profile $\propto e^{-i\omega t}$, and working in the far-field limit $|\vec{x} - \vec{x}^\prime| / |\vec{x}| \ll 1$, one finds
\begin{align}
    \delta \vec{E} &= \frac{e^{-i \omega t + i \omega \vec{x}}}{4\pi |\vec{x}|}\int d^3 \vec{x}^\prime e^{- i \omega \vec{x} \cdot \vec{x}^\prime / |\vec{x}|} \times \nonumber \\ & \left(\nabla \rho_{a,x}(\vec{x}^\prime) - i \omega \vec{J}_{a,x}(\vec{x}^\prime) \right) \\
    \delta \vec{B} &= -\frac{e^{-i \omega t + i \omega \vec{x}}}{4\pi |\vec{x}|}\int d^3 \vec{x}^\prime e^{- i \omega \vec{x} \cdot \vec{x}^\prime / |\vec{x}|} \left( \nabla \times \vec{J}_{a,x}(\vec{x}^\prime) \right) \, ,
\end{align}
where we have redefined the charge and current densities so as to factor out the harmonic oscillation, $\vec{J}_{ae} = \vec{J}_{a,x} e^{-i \omega t}$ and $\rho_{ae} = \rho_{a,x} e^{-i \omega t}$.

Let us now restrict our attention to a localized, spherical, clump of axions which extend over a scale $L \sim 1/k$. Assuming that the background magnetic field is sufficiently uniform over this region, and that $k \ll \omega$, one finds
\begin{eqnarray}
    \delta \vec{E} &\simeq& \frac{e^{-i \omega t + i \omega \vec{x}}}{4\pi |\vec{x}| \omega^2 k}  i \omega \vec{J}_{a,x} \\
    \delta \vec{B} &=& \frac{e^{-i \omega t + i \omega \vec{x}}}{4\pi |\vec{x}| \omega^2 k} i \vec{k} \times \vec{J}_{a,x} \, ,
\end{eqnarray}
which is equivalent to the expression derived in~\cite{Caputo:2023cpv} in the limit that $k \sim \omega$.  One can immediately see from this equation that there exists a non-zero Poynting flux, $\left< \vec{E} \times \vec{B} \right>$, so long as $\vec{k}$ has some component non-parallel to $\vec{B}$. In general, the radiated luminosity $L$ will scale as 
\begin{eqnarray}
    \left< L\right>_{\gamma} \sim 4\pi |\vec{x}|^2 \delta \vec{E} \times \delta \vec{B} \nonumber \\
    \propto \frac{1}{\omega \, k} \gagg^2 B_0^2 a^2 
\end{eqnarray}
where in the second line we have factored out the functional dependencies. This radiative process is usually highly inefficient in most contexts of interest, unless very high field configurations can be generated. We will not discuss this possibility further.

\section{Simulations and Results}\label{sec:results}

\subsection{Numerical Setup}\label{subsec:numerics_sims}

We start by giving a brief outline of the time- and frequency-domain simulation codes used in this work. More details are provided in Appendix~\ref{sec:Simulations}. Note that while in the previous sections the computations have been carried out considering fields that propagate along the $x$ axis, in our numerical simulations propagation is along the $y$ axis.

\subsubsection{Time Domain}
We evolve the system directly in the time domain by solving the coupled evolution equations as an initial-value problem on a numerical grid. The field equations are reduced to a set of first-order evolution equations using the formalism of the 3+1 decomposition, and then discretized on a 2D spatial domain, assuming homogeneity along the third direction. Time integration is performed with the method of lines: spatial derivatives are computed using fourth-order finite-difference stencils, while the time evolution is carried out with a sixth-order Runge–Kutta integrator.

We found that evolving a large uniform background magnetic field can introduce numerical errors that accumulate over time and may compromise stability. To mitigate this issue, we split the magnetic field into a fixed background component and a dynamical foreground perturbation. Only the foreground part is evolved numerically, while the total field (background plus perturbation) is used when evaluating the source terms in the evolution equations.

We employ periodic boundary conditions, making sure to set the grid dimension in such a way that, taking also into account the geometry of the problem, signals reaching the boundaries do not arrive to the region of interest during the simulation time.

We are interested in studying the scattering of axion wave packet on a strongly magnetized plasma. We therefore construct initial data by first setting a plasma profile characterized by a rectangular barrier (in the $x$ and $y$ directions) on top of a small background plasma density. Then we add an axion wave packet that has a gaussian profile along the $y$ direction, and extends all over the grid along the $x$ direction. Such wave packet is placed in a region sufficiently far from the barrier, where the plasma density assumes the background value, and it is constructed as a vacuum propagation eigenstate. In this way the packet can propagate across the domain without significant mode splitting before interacting with the plasma structure.
This setup allows us to simulate the fully nonlinear dynamics of the fields and the plasma with sufficient accuracy. Additional details of the numerical method, the magnetic-field splitting, and the initialization procedure are provided in Appendix~\ref{sec:EvolutionEquations1DAxion}.

\subsubsection{Frequency Domain}
In the frequency domain, one directly solves the field equation, Eq.~\ref{eq:waveeq_ani}, on a pre-defined grid. This is done in two dimensions, with no need to solve for the motion of the plasma, since this is implemented directly via the dielectric tensor. In that sense, the frequency domain simulations allow for the computation of properties across a wider dynamical range that the time-domain simulations; this comes at the cost of losing dynamical information, and the ability to deal with scenarios for which non-trivial time-domain dynamics arise (such as in one of the examples which will be provided below).

We adopt an approach which closely follows that of~\cite{Gines:2024ekm}. Namely, we impose a perfectly matched layer (PML) around the boundary of the simulation domain, which serves to damp waves and suppress spurious reflections. We impose a magnetic field profile across a subset of the domain, which is constant in a rectangular region, and which is suppressed via a  $\sin^2$ function along the boundaries, where the suppression takes place over a handful of grid points  -- this procedure is necessary in order to avoid having sharp features of the magnetic field help source electromagnetic fields. We adopt monochromatic axion plane wave solutions, with each simulation taking a different initial axion velocity, with values ranging from $v_a \sim 0.1$ to $v \sim 0.7$; in each case, we ensure that the axion wavelength is always resolved with at least $\gtrsim 8$ grid points. Details on the specific functional forms and values of each simulations are included in Appendix~\ref{sec:fdomset}.

\subsection{Breakdown of the WKB approximation}\label{subsec:wkb}

Having outlined the procedure of the time- and frequency-domain simulations, we now turn our attention toward their application. We begin by studying the most straightforward scenario: the case in which a rapidly varying plasma leads to a breakdown in the WKB approximation in the vicinity of the resonance. The intuition and qualitative understanding of this limit is relatively easy to understand (although, may be somewhat counterintuitive for those familiar with resonant dark photon mixing), and thus serves as a natural sanity check for ensuring the convergence and consistency of the simulation results.

\subsubsection{Time domain}

We perform 3 sets of simulations in which we test the transition from the regime where the plasma is slowly-varying, to the regime where plasma varies in a step-like behavior, and non-resonant conversion becomes dominant. While here we will report some quantities using international units, our code uses rationalized Heaviside units, with $c = \hbar = 1$, so that the electron mass is set to be $m_e = 511 \, \keV$, and the elementary charge to $e = \sqrt{4 \pi/137}$. 

In all the runs we use $\gagg = 4 \times 10^{-12} \, \eV^{-1}$ (although this is not particularly relevant, as we present re-normalized quantities), and $B_0 = 10^{-5} \, T$, along the x-axis (note that this corresponds to a cyclotron frequency $\omega_c \sim 10^{-9}$ eV). The plasma density on the background is set to $n_{\rm{bkg}} = 10^{-6} \, \cm^{-3}$, yielding a plasma frequency $\omega_{p, \rm{bkg}} = 3.70 \times 10^{-14} \, \eV$. The mass of the axion is set to $m_a = 10^{-10} \, \eV$, and the central wave number of the packet is $k_0 = 2.06 \times 10^{-10} \, \eV$, which, on the background, yields a frequency $\omega_+ \approx \sqrt{m_a^2 + k_0^2} = 2.29 \times 10^{-10} \, \eV$ (note that this configuration yields $\omega_c \gg \omega, \omega_p$, corresponding to the magnetized limit), and a group velocity $v_g \approx k_0 / \omega_{+} \approx 0.9$. In each set we use a different value of the plasma density at the top of the barrier, which correspond to different mixing angles, and allows us to perform multiple tests of the transition towards the regime of non-resonant conversion. Specifically, we select $n_{\rm{max}} = \{10, 15, 20\} \, \cm^{-3}$, corresponding to $\omega_{p, \rm{max}} = \{1.171, 1.435, 1.657\} \times 10^{-10} \, \eV$. Note that $\omega_{p, \rm{max}}$ is always smaller than $\omega_+$, so that both the photon and the axion propagate inside the barrier after the conversion. However, for smaller values of $n_{\rm max}$ the plasma frequency is closer to $m_a$, causing the wave packets of the photon and the axion to have similar group velocity and taking longer to separate. In each set of simulations we vary the parameter $W$ in \ref{eq:InitialPlasmaDensity}, which sets the steepness of the boundaries of the barrier, in the range $0.01 \, \km^{-1} \le W \le 10 \, \km^{-1}$. The parameter $x_0$ is set to $x_0 = 0 \, \km$, while $y_0$ to $y_0 = l_y / 2$, in such a way that the boundary where the conversion takes place is at $y = 0 \, \km$. The transverse width of the barrier is $l_x = 5000 \, \km$, while the longitudinal is varied in $1250 \, \km \le l_y \le 6000 \, \km$. This is done in order to optimize the computational cost, as the size of the plasma barrier along $y$ also affects the extension of the grid that has be used; we aim to minimize $l_y$ based on the parameter $W$ and the space the two wave packets take to separate, which depends on $n_{\rm max}$. The amplitude of the wave packet is set to $\AA = 1$, its width to $\sigma_k = 0.05 \, k_0$, and its initial position is varied in the range $-2500 \, \km \le y_0 \le -400 \, \km$, with the purpose to place the packet sufficiently far from the left boundary of the plasma barrier, while trying to minimize the time to reach it.

Let us now focus on the numerical grid. The base grid step is $\Delta x = \Delta y = 5.6 \times 10^8 \, eV^{-1}$, but is decreased up to $\Delta x = \Delta y = 1.2 \times 10^{8} \, \eV^{-1}$ for large values of $W$, in order to resolve the boundaries of the plasma barrier. The time step $\Delta t$ is set to $\Delta t = CFL \, \Delta z/v_p$, where $CFL$ is the Courant–Friedrichs–Lewy factor, and is set to $0.1$, while $v_p$ is the largest phase velocity in our setup, which is the phase velocity of the photon inside the barrier, estimated as $v_p \approx \omega / k_{p} \approx \omega_+ / \sqrt{\omega_+^2 - \omega_{p, \rm{max}}^2}$. The total time of integration $T$ is large enough that at the end of simulation the conversion process has completed and the two wave packet have separated. Concretely, $T$ is of the order of $T \approx 10^{12} \div 10^{13} \, \eV^{-1}$, which allows the axion wave packet to travel for a distance $L = v_gT \approx \OO{10^{3} \, \km}$. The position of the left boundary of the computational domain along the y axis, $y_{-\infty}$, is varied across the simulations to reduce the computational cost, keeping it at a distance from the initial position of the wave packet of at least $L$. The right boundary $y_{+\infty}$ is instead placed sufficiently far from the right boundary of the plasma barrier, $y = l_y$, varying its value on the basis of the steepness parameter $W$. Typically, the distance between $y_{+\infty}$ and $y = l_y$ is $10^3 \, \km$ when $W \le 1.3 \, \km^{-1}$, but can decrease up to $50 \, \km$ for larger values of $W$. Lastly, since we wish to consider a setup that is homogeneous with respect to both the transverse directions, we set the numerical grid to extend for only 2 grid point along the x axis.

\begin{figure}
    \centering
    \includegraphics[width=\columnwidth]{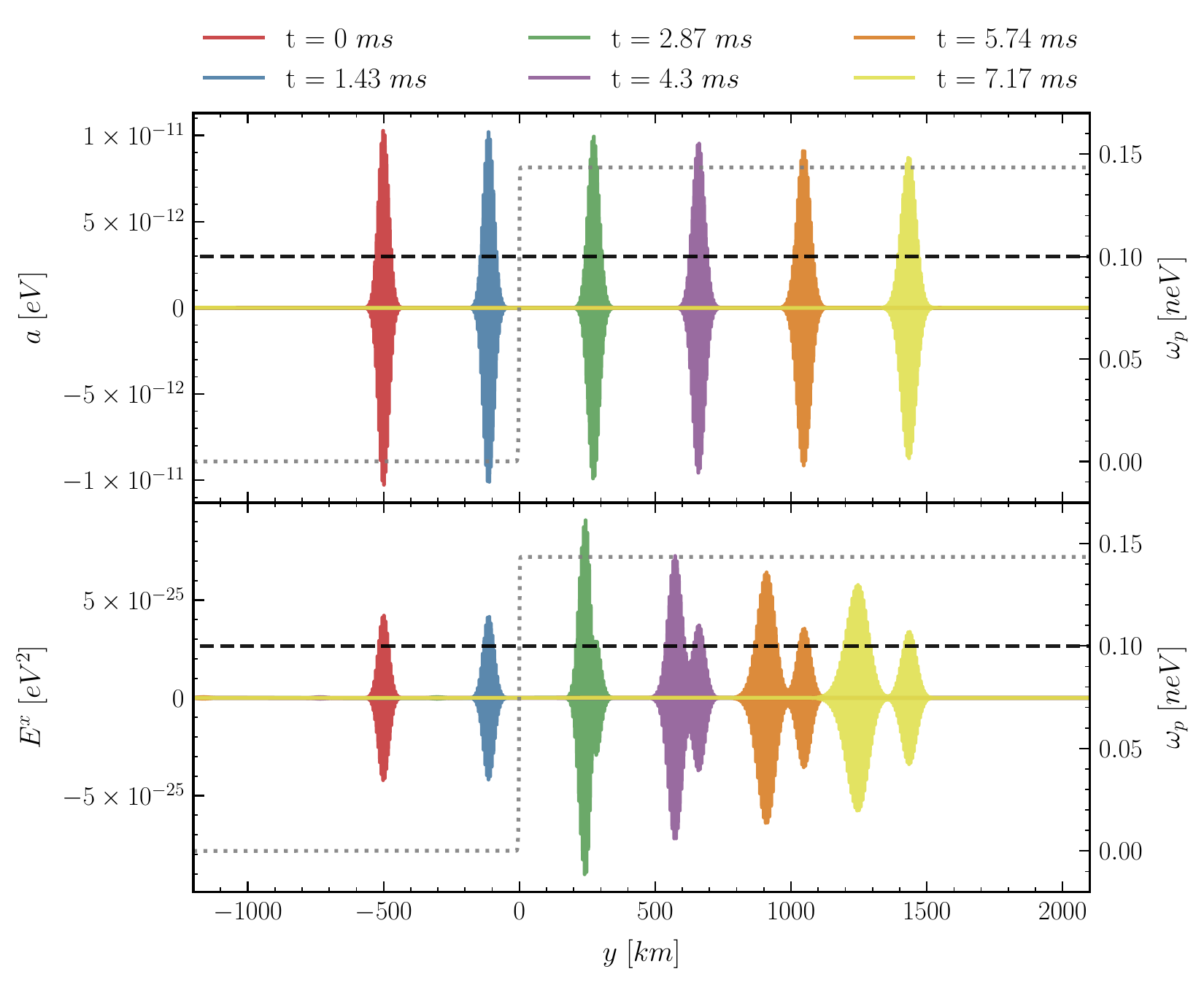}
    \caption{Snapshots of the evolution for the simulation with $n_{\rm{max}} =  15 \, \cm^{-3}$, and $W = 1.3 \, \km^{-1}$ in the region close to the boundary of the barrier. The upper panel shows the profile of the axion, while the lower panel the profile of the component of the electric field that is parallel to the background magnetic field, $E^y$. Different colors refer to different snapshots. The horizontal dashed line indicates the value $m_a$, while the gray dotted line denotes the profile of the plasma frequency, extracted at the beginning of the simulation. The simulation starts with a wave packet of the axion in the unmixed basis outside the barrier. When it enters the plasma it undergoes a conversion process. A photon wave packet is produced, which propagates towards the right together with the axion. Since they posses different group velocities, they gradually separate, with the photon traveling slower. An additional wave packet of the electromagnetic field propagating toward the left is also produced in the conversion, but its amplitude is substantially smaller, and it is barely visible in the bottom panel.}
    \label{fig:LZTestsSnapshots}
\end{figure}

The general dynamical behavior can be observed in Fig.~\ref{fig:LZTestsSnapshots}, where we plot some snapshots of the simulation with $W = 1.3 \, \km^{-1}$ and $n_{\rm{max}} = 15 \, \cm^{-3}$. Here the upper panel shows the profile of the axion, and the lower panel the profile of $E^x$, \ie the component of the electric field that is parallel to the background magnetic field. Different colors denote different snapshots, and the gray dotted line is the profile of the plasma density, extracted at $t = 0 \, \ms$. The horizontal black dashed line denotes instead the value of the axion mass, $m_a$. As we can see the simulation starts with a wave packet of the axion in the unmixed basis, so that it possesses a component of the electromagnetic field. After the packet crosses the boundary of the barrier it undergoes a conversion process, and three wave packets emerge: a left-moving photon wave packet, a right-moving photon and a right-moving axion in the unmixed basis. The former has a small amplitude and it is barely visible in the plot. Instead the wave packets propagating toward the right are initially overlapped and then gradually separate, due to the fact that they have different group velocities, with the photon being slower, as $\omega_{p, \rm{max}} > m_a$. The component of the axion field in the propagation eigenstate corresponding to the photon is not visible, since it is too small compared to the scales shown.

We extract the conversion probability as
\begin{equation}
    p_{a\rightarrow \gamma} = \frac{\overline{\EE}_i - \overline{\EE}_f}{\overline{\EE}_i} = 1 - \frac{\overline{\EE}_f}{\overline{\EE}_i},
    \label{eq:ProbabilityFromSimulationDef}
\end{equation}
where $\overline{\EE}_{i}$ and $\overline{\EE}_f$ are the energies per unit surface of the axion, extracted at times sufficiently before and after the conversion process, respectively. Note that we typically re-normalize the conversion probability by an amplitude $p_0$, since the choice of $\gagg$ (and to some degree, also $B_0$) in our simulations is arbitrary.  The axion energy density is given by
\begin{equation}
    \rho_a = \frac{1}{2} \Bigl[ \Pi^2 + \Theta_x^2 + \Theta_y^2 + m_a^2 a^2\Bigr],
    \label{eq:AxionEnergySimulations}
\end{equation}
and we estimate $\overline \EE$ evaluating $\rho_a$ on a one-dimensional slice along y, and integrating it from the initial position of the axion wave packet to the right boundary of the plasma:
\begin{equation}
    \overline \EE = \frac{1}{2} \int_{y_0}^{l_y} dy \, \biggl[ \Pi^2 + \Theta_x^2 + \Theta_y^2 + m_a^2 a^2\biggr].
    \label{eq:AxionEnergyDensityDef}
\end{equation}
The use of the energy per unit surface in the computation of $p_{a\rightarrow \gamma}$, instead of the energy $\EE = \int_V d^3x \, \rho_a$ is justified by the fact that we are considering a system that is homogeneous along the $xz$-plane, so that the surface terms cancel out when computing the ratio in Eq.~\ref{eq:ProbabilityFromSimulationDef}. The decision to start the integration at $y_0$ allows us instead to get rid of left-propagating components of the axion, which might appear due to the approximations in constructing the initial data.

\begin{figure*}
    \centering
    \includegraphics[width=\columnwidth]{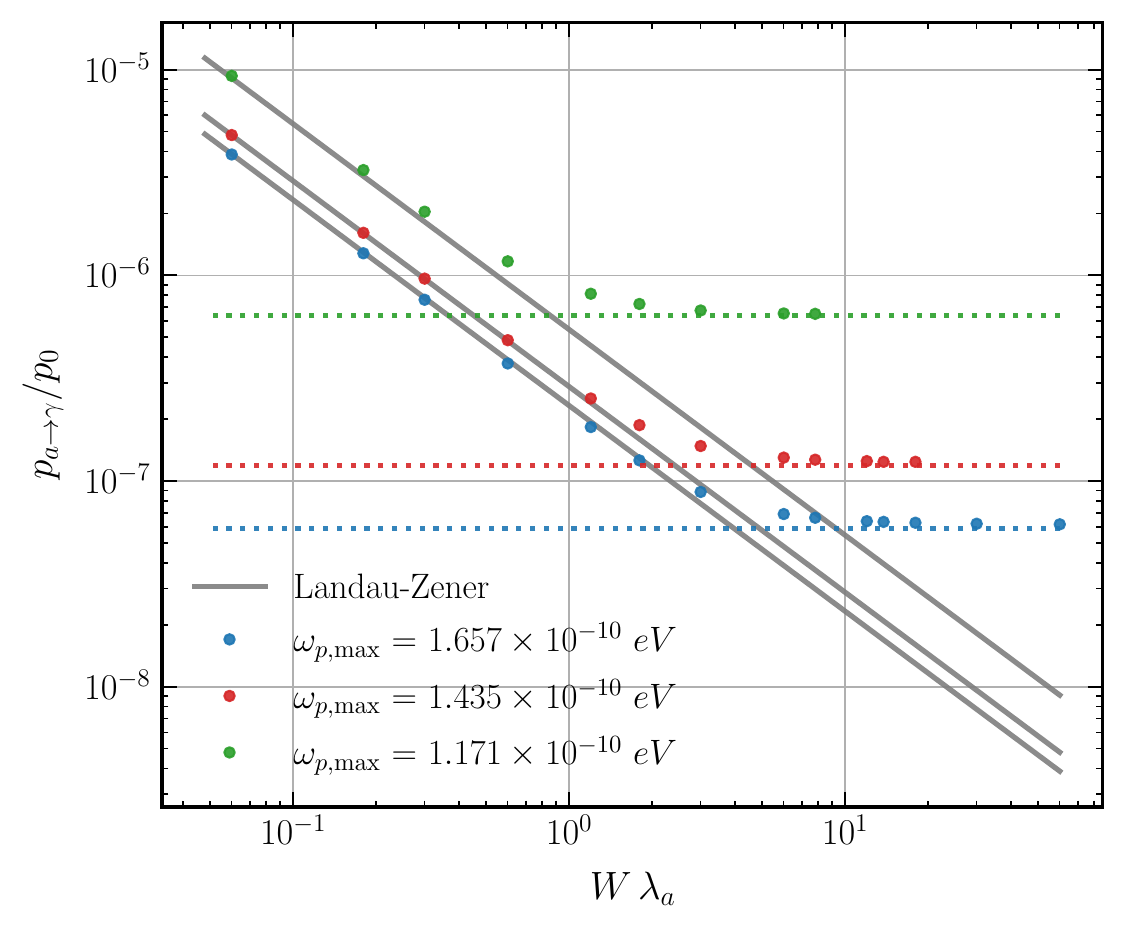}
    \includegraphics[width=\columnwidth]{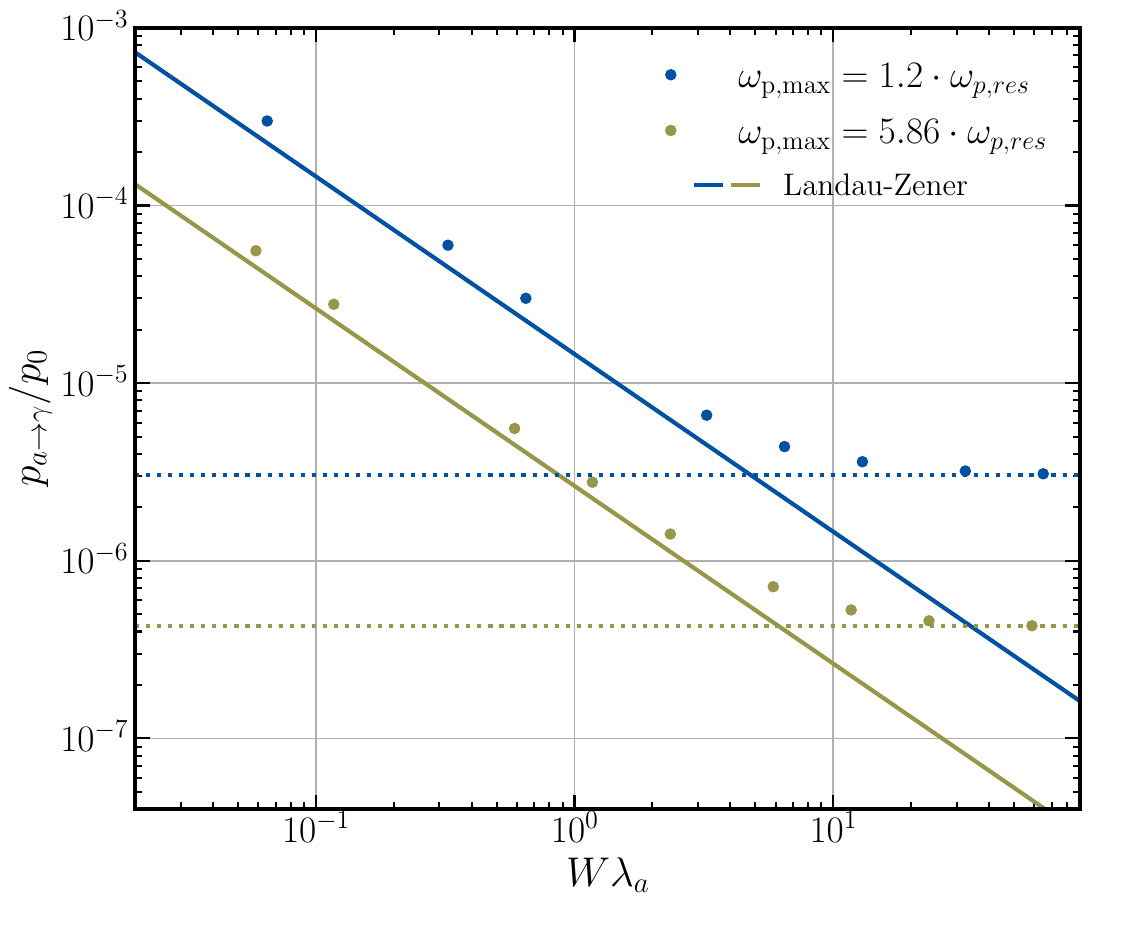}
    \caption{ Normalized conversion probability for an axion traversing a magnetized medium as a function of the inverse spatial gradient scale of the background plasma $W$, normalized in units of the axion de Broglie wavelength, for the time-domain (left) and frequency-domain (right) simulations. In the time-domain simulations, we adopt an axion mass $m_a = 10^{-10} \, \eV$, a group velocity $v_g \approx 0.9$, a coupling constant is $\gagg = 4 \times 10^{-12} \, \eV^{-1}$ and a background magnetic field $B_0 = 10^{-5} \, T$, and we run the simulations for three values of the maximum height of the plasma boundary $\omega_{p, {\rm max}}$. In the frequency domain simulation, we instead show results for an axion mass $m_a = 10^{-5} \, \eV$, an axion velocity of $v = 0.7$ and a background magnetic field strength $B_0 = 10 \, T$. In both cases, we compare the simulation results (dots) with the analytic expression for the resonant conversion probability given in Eq.~\ref{eq:pa_resonant}, and with the non-resonant conversion probability (neglecting the sharp boundaries). Both figures show excellent agreement with the limiting cases, with deviations from these asymptotic results occurring near $W \lambda_a \sim \mathcal{O}(1)$ values. This result provides confidence that both time-domain and frequency-domain simulations are capable of accurately resolving resonant and non-resonant dynamics in a regime where the WKB approximation is strongly violated.  } \label{fig:ConversionProbabilityVaryingW}
\end{figure*}

The behavior of the conversion probability we obtained from our simulations is shown in the left panel of Fig.~\ref{fig:ConversionProbabilityVaryingW}, where each color identifies a set with the corresponding value of the plasma frequency at the top of the barrier. The x-axis displays $W \lambda_a$, which estimates the steepness of the boundary of the barrier compared to the variations of the fields, as the ratio between the wavelength of the axion at the beginning of the simulation, $\lambda_a$, and the scale over which the plasma varies at the boundary, $\approx W^{-1}$. The gray solid lines represent the conversion probability estimated with the Landau-Zener approximation, Eq.~\ref{eq:LandauZener}, computed setting the position of the right boundary far to the right, so that only the left boundary matters. The dotted lines instead represent the estimates obtained assuming non-resonant conversion, as discussed in Sec.~\ref{sec:NonResonantConversion}. As we can see for small values of $W \lambda_a$, when the slowly-varying plasma approximation holds, the packet undergoes resonant conversion, and the probability is correctly estimated by the Landau-Zener formula. For large values of $W \lambda_a$, instead, the non-resonant process dominates, and the conversion probability approaches a constant value, obtained by re-projecting the initial state in the two new propagation eigenstates.

\subsubsection{Frequency domain}

In the frequency domain, we perform simulations using an axion mass $m_a = 10^{-5}\,\mathrm{eV}$\footnote{Note that the time and frequency domain simulations are performed using different mass axions and different scales purely for historical reasons -- our results are always presented in terms of renormalized quantities, and thus the physical scale is of no relevance to the physics of interest. } and with various orientations of the magnetic field, including $\theta_B = 90^\circ$ and $\theta_B = 45^\circ$, with the former corresponding to the limit of an isotropic plasma. The background magnetic field $\vec{B}_0$ is implemented over a large smooth sub-domain of the simulation with constant magnitude and length $50\,\lambda_a$, and a $\sin^2$ smoothing region is imposed on the boundary of this region, extending over $25\,\lambda_a$ on each side. We begin by imposing an axion plane wave solution with a velocity $v = 0.7$ (corresponding to $\omega \simeq 1.4\,m_a$), and with an initial velocity perpendicular to the plasma barrier. The computational domain has dimensions $16 \times 16\,\mathrm{m}$, corresponding to approximately $126.6 \times 126.6\,\lambda_a$. The plasma frequency varies only along the axion propagation direction (the $y$-coordinate) and is given by 
\begin{equation}
\omega_p(y) = \omega_{\mathrm{p,bkg}} + \left(\omega_{\mathrm{p,bkg}} - \omega_{\mathrm{p,max}}\right)\,(\delta_1 + \delta_2 - 1),
\end{equation}
where $\delta_1 = \left[1 + e^{W_1 (y - y_1)}\right]^{-1}$ and $\delta_2 = \left[1 + e^{W_2 (y - y_2)}\right]^{-1}$ are sigmoid functions. The parameter $y_1$ is chosen such that the first level crossing ($\omega_p = m_a$) occurs at the center of the simulation domain, and the barrier gradient is controlled by $W_1 = -W_2 \equiv W$. We consider two values of the maximum plasma frequency, $\omega_{\mathrm{p,max}} = \{1.2,\,5.86\}\,\omega_{\mathrm{p,res}}$; in the case of the former, the sourced photon can propagate freely through the barrier, while for the latter there is total reflection.

\begin{figure}
    \centering
    \includegraphics[width=\columnwidth]{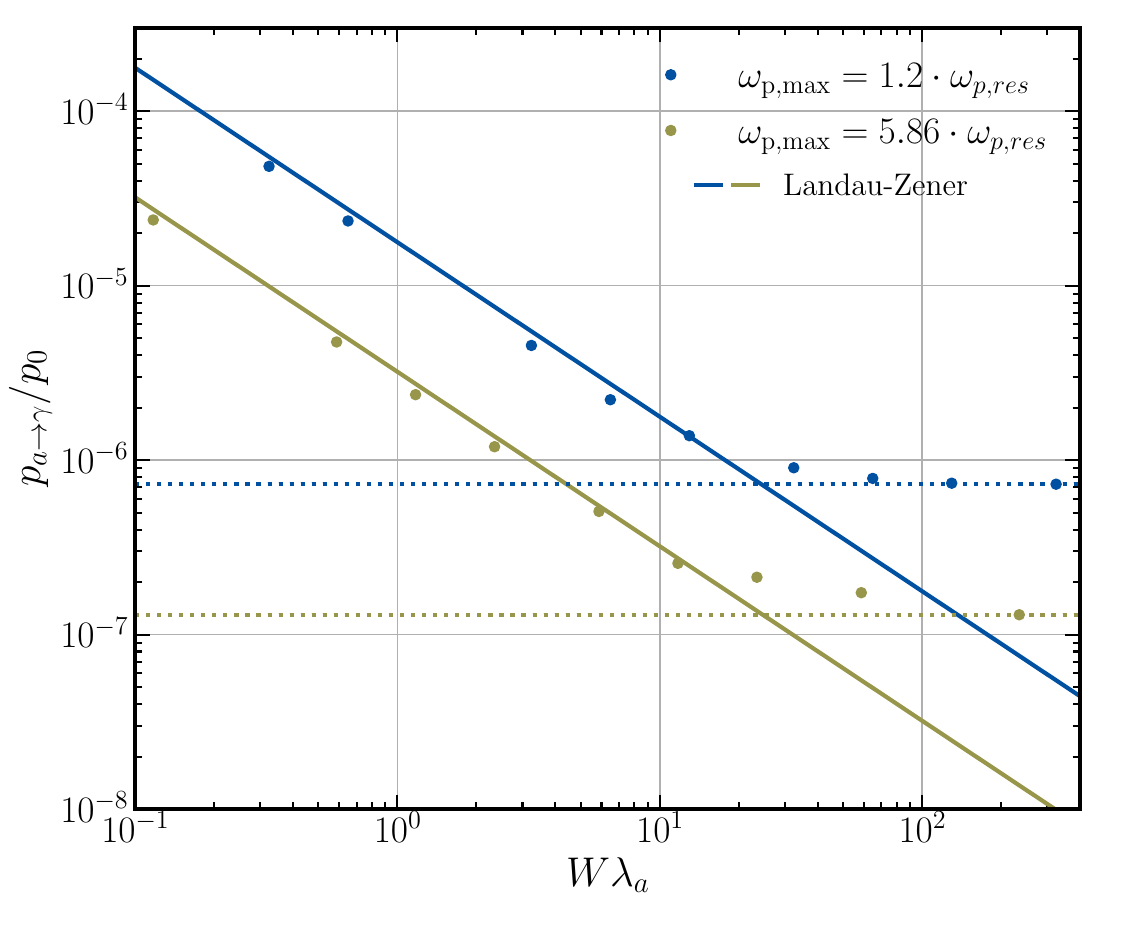}
    \caption{Same as the right panel of Fig.\ref{fig:ConversionProbabilityVaryingW}, but taking $\theta_B = 45^\circ$. } \label{fig:ConversionProbabilityVaryingWTheta45}
\end{figure}

As in the previous section, we calculate the numerical conversion probability for varying gradients of the plasma frequency. The results are shown in the right panel of Fig.~\ref{fig:ConversionProbabilityVaryingW} for the isotropic plasma limit (for direct comparison with the time domain simulation), and in Fig.~\ref{fig:ConversionProbabilityVaryingWTheta45} with $\theta_B = 45^\circ$. Results are shown in both cases for the two different amplitudes of the plasma barrier, and in both cases $y_2$ is set to a large enough value to fall outside the simulation box, so as to simulate a single-sided barrier. As observed in the time-domain simulations, one begins to see deviations from the analytic predictions of the resonant conversion probability when the spatial variation of the plasma is fast with respect to the de Broglie wavelength of the axion. 

\begin{figure*}
    \centering
    \includegraphics[width=\columnwidth]{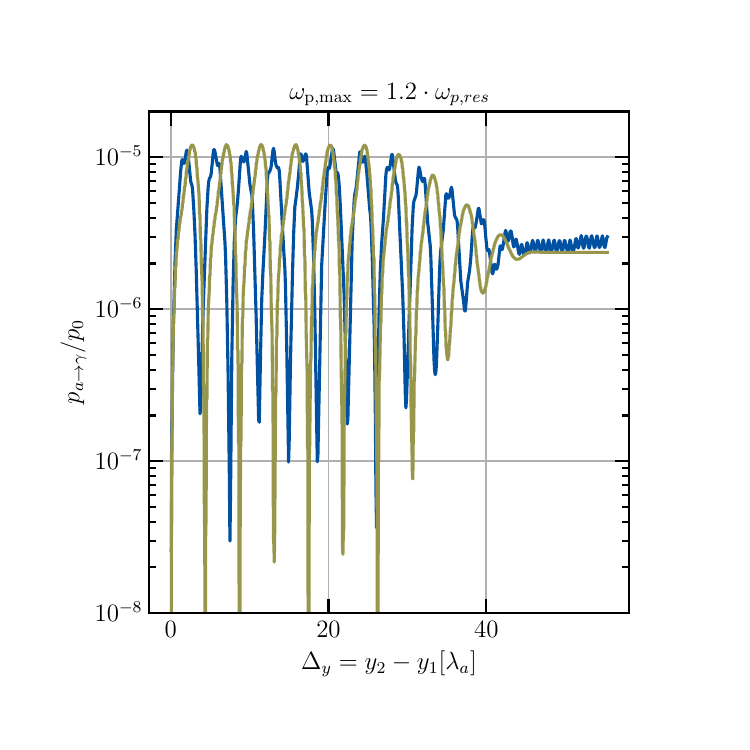}
    \includegraphics[width=\columnwidth]{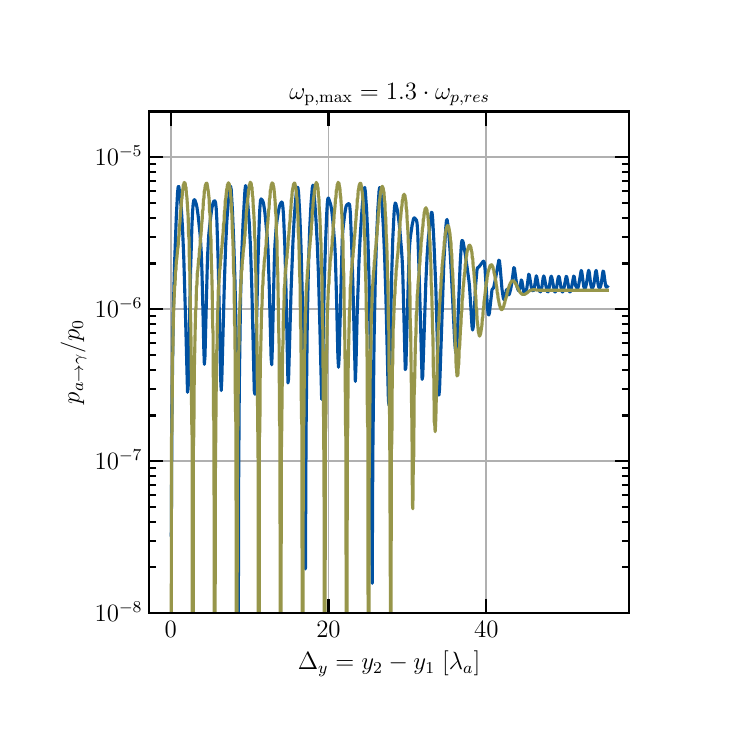}
    \caption{Analytical ($\it{green}$) and simulated  ($\it{blue}$) conversion probabilities in the frequency domain for two different amplitudes of the plasma barrier, $\omega_{\rm{p,max}}$, as a function of the width of the plasma barrier $\Delta_y$ (defined in units of the axion De Broglie wavelength $\lambda_a$). Here, we can clearly see the relative trade-off between resonant and non-resonant conversion, with the former corresponding to the limit of large $\Delta_y$, and the latter corresponding to the limit in which $\Delta_y \rightarrow 0$. Note that finite resolution effects naturally lead to an unavoidable phase shift in the oscillatory nature of the non-resonant conversion, see~\cite{Gines:2024ekm} for a discussion.}
    \label{fig:NonResonantTheoSimComparison}
\end{figure*}

In order to better understand the behavior at large plasma gradient, we can plot the conversion probability as a function of the width of the plasma barrier (for a fixed value of $W \lambda_a= 23$), parameterized via $\Delta_y = y_2 - y_1$. Varying $\Delta_y$ between 0 and  $\mathcal{O}(10) \times \lambda_a$ allows one to smoothly control the strength of the non-resonant conversion probability -- this result of this is shown in Fig.~\ref{fig:NonResonantTheoSimComparison} (for two heights of the plasma barrier, $1.2 \cdot \omega_{\rm{p,res}}$ and $1.3 \cdot \omega_{\rm{p,res}}$), and are compared with the theoretical expectation, which is computed directly using Green's functions. Note that the slight phase shift observed between the theoretical and analytic results is a consequence of finite spatial resolution effects, see~\cite{Gines:2024ekm}.

\subsection{Sub-luminal Mode Excitation}\label{sec:modeexcite}

Having accurately captured the quantitative behavior of the mixing of the axion and the LO mode in the regime of strongly varying plasma gradients, we now turn our attention to the more interesting problem of what happens when axions are exciting plasma modes which are simultaneously mixing with other modes. As mentioned in Sec.~\ref{sec:axelec}, this mixing behavior is expected when the spatial separation between a cut-off and a resonance in plasma mode dispersion relations is small -- the WKB approximation is naturally violated in the vicinity of the cut-off / resonance, and thus it is possible to generate tunneling from super-luminal to sub-luminal plasma modes. This opens a new possibility for the interactions of axions in non-trivial backgrounds.  

Since the Alfv\'en mode is directly generated, and exists exclusively, in a localized region of high plasma density, this problem is most naturally studied in the time domain (as one can more clearly differentiate axion induced electric fields from on-shell plasma modes). Below, we outline our simulation procedure and our analysis.

\subsubsection{Results}

In order to check if the Alfv\'en mode is produced, we perform a set of simulations in which we vary the angle $\theta_B$ between $10^\circ$ and $90^\circ$. The axion mass is set to $m_a = 5 \times 10^{-11} \, \eV$, and the central wave number of the packet to $k_0 = 9.8 \times 10^{-11} \, \eV$, so that its frequency is, up to $\OO{\gagg^2}$ terms, $\omega \approx \sqrt{m_a^2 + k_0^2} = 1.1 \times 10^{-10} \, \eV$, and the group velocity is $v_g \approx k_0 / \omega \approx 0.891$. The width of the wave packet is set to $\sigma_k = 0.05 k_0$, its the central position to $z_0 = 0 \, \km$, and its amplitude to $\AA = 1$. The intensity of the background magnetic field is chosen in such a way that the cyclotron frequency is $10$ times larger than the axion frequency $B_0 = m_e \omega_c/ e = 10 m_e \sqrt{m_a^2 + k_0^2} / e$, while the coupling constant is set to $\gagg = 4 \times 10^{-12} \, \eV^{-1}$. The plasma barrier is centered in $(x_0, y_0) = (0, 825) \, \km$, with extensions $(l_x, l_y) = (5000, 650) \, \km$. The value of $l_x$ is significantly larger than the extension of the grid in the $x$ direction, so that the corners of the plasma barrier are outside the domain of integration; this simplifies considerably the dynamics, making the identification of the Alfv\'en mode easier. The parameter $W$ appearing in Eq.~\ref{eq:InitialPlasmaDensity} is set to $W = 0.1 \, \km^{-1}$, and the background plasma density to $n_{\rm bkg} = 10^{-6} \, \cm^{-3}$, which results in a plasma frequency $\omega_{p, \rm bkg} = 3.7 \times 10^{-14} \, \eV$. The parameter $n_{\rm max}$ is set to $n_{\rm max} = 20 \, \cm^{-3}$ (corresponding to $\omega_{p, \rm max} = 1.66 \times 10^{-10} \, \eV$) for the simulations with $\theta_B > 20^\circ$. For lower angles, the group velocity (given by $v_g = \partial_k \omega$) of the Alfv\'en mode starts getting closer to that of the axion-induced electric field, and therefore the two wave packets take longer to separate. In order to keep the computational cost constant, in the cases with $\theta_B \le 20^\circ$ we counteract the increase in the group velocity of the Alfv\'en by gradually decreasing $n_{\rm max}$.

The numerical grid extends in $[-650 \, \km, 650 \, \km] \times [-500 \, \km, 1400 \, \km]$, with grid steps $\Delta x = \Delta y = 1.5 \times 10^9 \, \eV^{-1} = 296 \, \m$. The time step is set to $\Delta t = CFL \, \frac{\Delta x}{v_{p, \rm a}}$, where $v_{p, \rm a} = \sqrt{m_a^2 + k_0^2} / k_0$ is an estimate of the phase velocity of the axion, and the $CFL$ factor is set to $0.2$ . Lastly, the total integration time is set to $T = 4.49 \, \ms$, which allows the axion wave packet to travel $1200 \, \km$.

After performing the set of simulations we obtain that the Alfv\'en mode is not produced in the cases with $\theta_B \ge 50^\circ$, in agreement with theoretical expectations (this follows from the fact that the resonance occurs at $\omega = \omega_p |\cos\theta_B|$, and for larger $\theta_B$ the right hand side is always less than $\omega$). For $\theta_B \le 40^\circ$, one instead sees the appearance of a new wave packet inside the plasma barrier, being sourced only after the axion encounters the resonance point. We also check that the systems maintains homogeneity along the transverse directions by comparing the profiles of the fields at $t = 4.39 \, \ms$ on two slices along the y-axis, and we obtain that they coincide up to machine precision for all the simulations. Taking advantage of this, we repeat the simulations with $\theta_B \le 40^\circ$ using a grid that extends for only two grid points along the x axis, and increasing the spatial resolution by a factor $3$. In some selected cases, we also consider longer integration times, to be able to better characterize the Alfv\'en mode. In the following paragraphs we will only refer to data extracted from this second set of runs.

\begin{figure}
    \centering
    \includegraphics[width=\columnwidth]{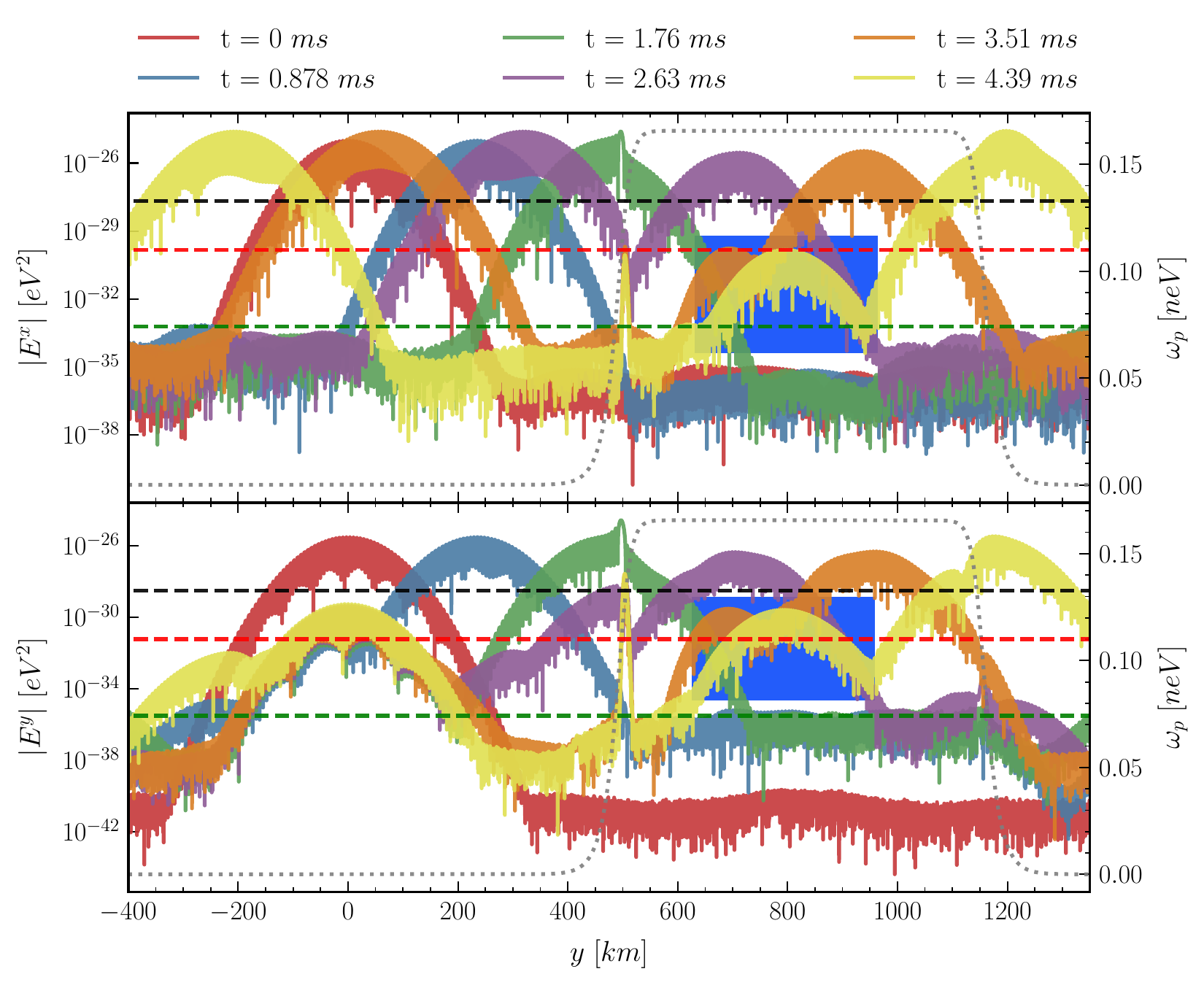}
    \caption{Snapshots of the absolute values of $E^x$ and $E^y$ at different time steps for the simulation with $\theta_B = 34^\circ$. Solid lines of different colors denote the profile of the electric field at different time steps; the gray dotted line denotes the plasma frequency; the red dashed line represents the axion frequency estimated as $\omega_a = \sqrt{m_a^2 + k_0^2}$; the green dashed lines represents the resonant frequency $\omega = \frac{m_a \omega_a}{\sqrt{m_a^2 \cos^2 \theta_B + \omega_a^2 \sin^2 \theta_B}}$; the black dashed line denotes the Alfvén mode resonance $\omega = \omega_a / \cos \theta_B$. At the beginning of the simulation the electric field is only given by the axion-induced component of the wave packet. When the axion enters the barrier the conversion takes place, and a LO mode is produced and propagates towards the left, as it cannot penetrate the barrier. The axion induced electric field component continues to propagate toward the right, but at later time a smaller wave packet forms inside the barrier, which is the Alfv\'en mode; this packet is highlighted with a blue rectangle for the snapshot at $t = 4.39 \, \ms$ (yellow lines).
    }
    \label{fig:Theta34TimeSnapshots}
\end{figure}

In order to show how we identify the Alfv\'en mode, in Fig.~\ref{fig:Theta34TimeSnapshots} we plot some snapshots of the simulation with $\theta_B = 34^\circ$. The $x$ and $y$ components of the absolute values of the electric field are represented by with solid lines in the upper and lower panel, respectively, using different colors to identify the time steps. The dotted line denotes the profile of the plasma frequency, extracted at the initial step. The red horizontal dashed line indicates an estimate of the axion frequency, $\omega_a = \sqrt{m_a^2 + k_0^2}$, while the green and black dashed lines represent the resonant frequencies of the axion, $\omega \approx \frac{m_a \omega_a}{\sqrt{m_a^2 \cos^2 \theta_B + \omega_a^2 \sin^2 \theta_B}}$, and the Alfvén mode, $\omega \approx \omega_a / \cos \theta_B$, respectively. The simulation is initialized with an approximation of a wave packet of the propagation eigenstate corresponding to the axion, therefore it contains a component of the electromagnetic field, that is visible in the red curve. Since the background magnetic field has a component along the propagation direction, then $E^y$ is non-vanishing. When the axion reaches the plasma barrier, the conversion takes place and a photon propagates toward the left. This is a LO mode, which cannot penetrate the barrier, as $\omega < \omega_{p, \rm max}$. Instead the axion continues to propagate, and this is reflected in the electric field it induces. At later times, a new wave packet appears behind the axion and inside the barrier, which is our candidate for the Alfv\'en mode, and is highlighted with a blue box for the last time snapshots shown. We also notice a component of $E^y$ that remains at the initial position and does not propagate. As we will show in Appendix~\ref{app:convergence}, this is not induced by numerical error; however, we do not exclude it is only related to the approximations in constructing initial data.

\begin{figure}
    \centering
    \includegraphics[width=\columnwidth]{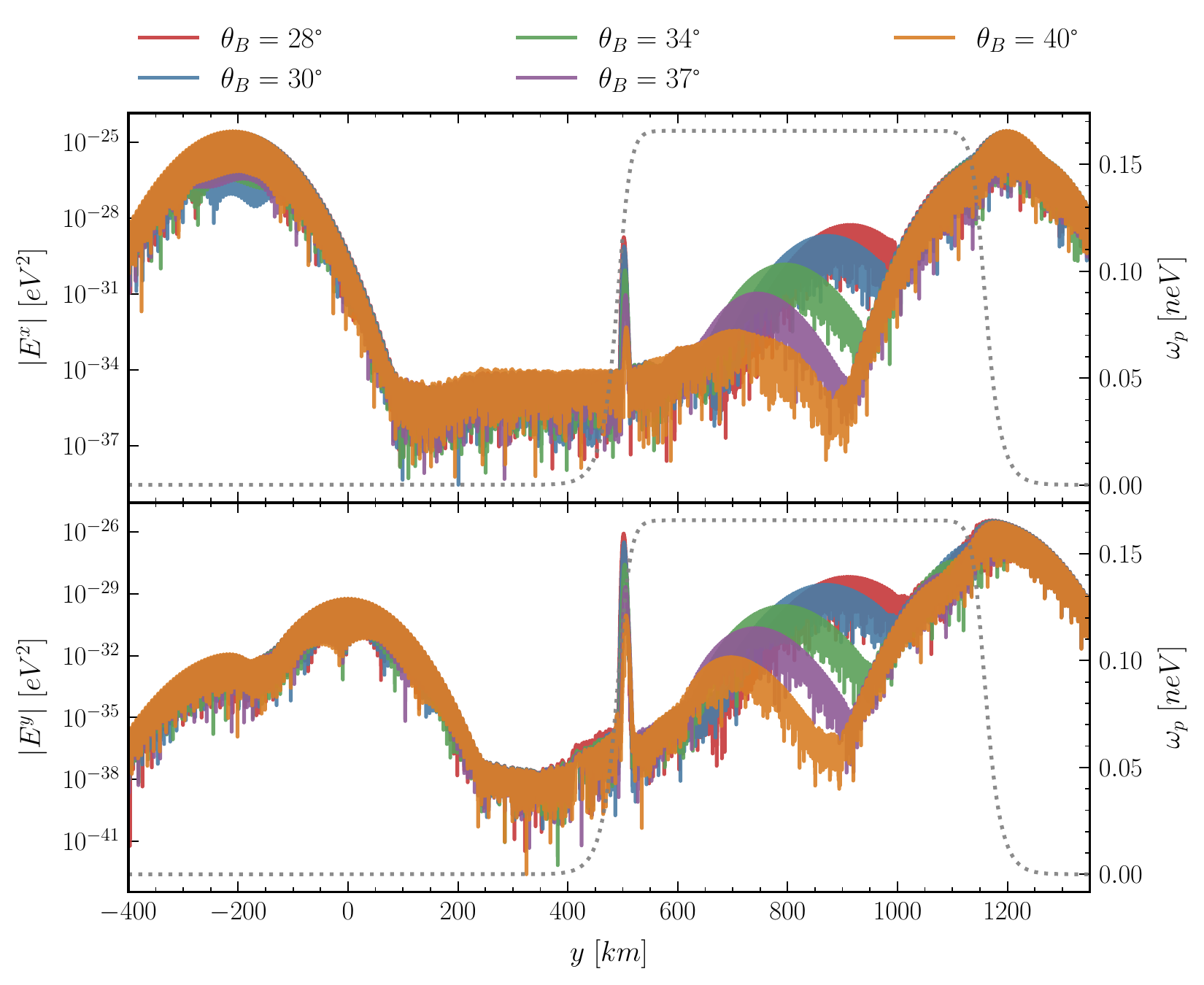}
    \caption{Snapshots of the absolute values of $E^x$ and $E^y$ at $t = 4.39 \, \ms$ for the set of simulations in which we vary the angle $\theta_B$ and we keep $n_{\rm max}$ constant. The gray dashed line represents the profile of the plasma frequency. We can see that as $\theta_B$ decreases the Alfv\'en mode becomes faster, consistently with the dispersion relation, and its amplitude increases.}
    \label{fig:VaryingThetaSnapshots}
\end{figure}

In order to show that the wave packet we identified inside the barrier is indeed an Alfv\'en mode, we first display in Fig.~\ref{fig:VaryingThetaSnapshots} the snapshots of $E^x$ and $E^y$ at $t = 4.39 \, \ms$ for different runs with fixed values of $\omega_{p, \rm max}$. As we can see in simulations with lower $\theta_B$ the wave packet propagating inside the barrier has traveled for longer distances; this means that the group velocity increase as $\theta_B$ decreases, in agreement with theoretical expectations for the Alfv\'en mode. 

As a further check, we compare the numerical values of the wave number, the group velocity, and the ratio between the amplitudes of the components of the mode against analytical expectations. To estimate such quantities from the results of the numerical simulations we proceed in the following way. For each run, we extract the data in two time steps, $t_1$ and $t_2$, and we slice them along the $y$-axis. Then, on both the slices, we identify the region where the Alfv\'en candidate is present. In such regions, we extract the local maxima of $\lvert E^y \rvert$, and we interpolate them with a spline. We then identify the maximum of the interpolant and we extract the wave length $\lambda_A$ from the locations of two non-consecutive zeros of $E^y$. The wave vector is then estimated as $k_A = 2 \pi ~ / ~ \lambda_A$. Note that due to dispersion the value of $k_A$ obtained with this procedure can vary depending on the extraction position. The group velocity is instead computed as $v_g = \left( y_A(t_2) - y_A(t_1) \right) ~ / ~ \left( t_2 - t_1 \right)$, where $y_A(t_{1,2})$ are the location of the maxima extracted with the interpolation procedure. Since such values are extracted from an interpolation of the local maxima, which are separated by half a wave length, we decided to associate an uncertainty $\frac{\sqrt{2}}{t_2 - t_1} \frac{\lambda_A}{2}$ to the estimate. Lastly, we estimate the ratio $\RR$ between the amplitudes of the longitudinal and transverse component of the mode, as the ratio between the RMS of $E^y$ and that of $E^x$ in the range we identify the mode to be in at time $t_2$. The results are shown in Table~\ref{tab:AlfvenModeParameters}, together with the theoretical estimates for the parameters we analyzed, coming from the dispersion relation~\ref{eq:AlfvenDispersionRelation} and the polarization vector~\ref{eq:AlfenPolarizationVector}, as well as the relative discrepancies we obtained for the wave number and the ratio $\RR$. 

\begin{table*}
    \centering
    \caption{Parameters of the Alfv\'en mode extracted from the simulations, and comparison with the expected values. $k$ is the wave number, $\RR$ denotes the ratio between the amplitude of $E^y$ and the amplitude of $E^x$, while $v_g$ is the group velocity.}
    \begin{ruledtabular}
        \begin{tabular}{ddddddddddd}
            \multicolumn{1}{c}{$\theta_B ~ [^\circ]$} & \multicolumn{1}{c}{$\omega_{\rm{p,max}} ~ [neV]$} & \multicolumn{1}{c}{$k_{\rm ext} ~ [neV]$} & \multicolumn{1}{c}{$k_{\rm th} ~ [neV]$} & \multicolumn{1}{c}{$\Delta k / k_{\rm th}$} & \multicolumn{1}{c}{$\RR_{\rm ext}$} & \multicolumn{1}{c}{$\RR_{\rm th}$} & \multicolumn{1}{c}{$\Delta \RR / \RR_{\rm th}$} & \multicolumn{1}{c}{$v_{g, \rm ext}$} & \multicolumn{1}{c}{$\Delta v_{g, \rm ext}$} & \multicolumn{1}{c}{$v_{g, \rm th}$} \\
            \colrule
            40 & 0.166 & 0.185 & 0.215 & -0.14 & 2.50 & 3.38 & -0.26 & 0.26 & 0.09 & 0.158 \\
            37 & 0.166 & 0.172 & 0.185 & -0.071 & 2.11 & 2.44 & -0.14 & 0.31 & 0.10 & 0.242 \\
            34 & 0.166 & 0.162 & 0.166 & -0.021 & 1.77 & 1.88 & -0.06 & 0.39 & 0.10 & 0.332 \\
            30 & 0.166 & 0.151 & 0.148 & 0.018 & 1.41 & 1.40 & 0.01 & 0.48 & 0.11 & 0.454 \\
            28 & 0.166 & 0.146 & 0.141 & 0.033 & 1.26 & 1.22 & 0.03 & 0.54 & 0.11 & 0.514 \\
            20 & 0.138 & 0.134 & 0.133 & 0.0079 & 1.27 & 1.29 & -0.02 & 0.51 & 0.12 & 0.454 \\
            15 & 0.128 & 0.128 & 0.127 & 0.0045 & 1.23 & 1.26 & -0.03 & 0.53 & 0.13 & 0.446 \\
            12.5 & 0.124 & 0.125 & 0.125 & 0.0013 & 1.20 & 1.28 & -0.06 & 0.56 & 0.13 & 0.431 \\
            10 & 0.121 & 0.122 & 0.121 & 0.0093 & 2.26 & 1.17 & 0.92 & 0.44 & 0.14 & 0.462
        \end{tabular}
    \end{ruledtabular}
    \label{tab:AlfvenModeParameters}
\end{table*}

The agreement between the parameters extracted from the simulation and theoretical expectations for the Alfvén mode is remarkably good, with a relative discrepancy on $\RR$ and $k$ getting at the percent level for $15^\circ \lesssim \theta_B \lesssim 30^\circ$. The simulations with $\theta_B = 10^\circ$ and $\theta_B = 40^\circ$ are the hardest to extract the parameters of the mode from, and show worse agreement. Nevertheless, given the low values assumed in general by the relative discrepancies, we conclude that the wave packet propagating inside the barrier is indeed the Alfvén mode.

The amplitudes for such mode we obtained in our simulations appears to be orders of magnitude smaller than that of the axion-induced electric field, or of the LO mode. In order to better characterize the conversion process we decided to run an additional set of simulations in which we keep the angle $\theta_B$ fixed to $30^\circ$, but we vary the parameter governing the steepness of the plasma barrier,  $W$. In this set we keep $n_{\rm max} = 20 \, \cm^{-3}$, and we use a barrier centered in $(x_0, y_0) = (0, 1275) \, \km$, with extensions $(l_x, l_y) = (5000, 1150) \, \km$. All the other physical parameters are assigned in the same way as in the previous set. As for the grid, we consider a numerical domain that extends in $[-1200 \, \km, 2500 \, \km]$ along the $y$ axis, and that contains only 2 grid points on the $x$ axis, owing to the symmetry of the system. The CFL factor is set to $CFL = 0.2$ as before. For the smallest value of the steepness of the barrier, $W = 0.05 \, \km^{-1}$, we set the grid steps to $\Delta x = \Delta y = 5.6 \times 10^8 \, \eV^{-1} = 110 \, \m$. As the parameter $W$ increases we adjust the grid spacing and extension to be able to resolve well the fields, especially at the point where conversion takes place, and produce a clear wave packet of the Alfvén mode.

\begin{figure}
    \centering
    \includegraphics[width=\columnwidth]{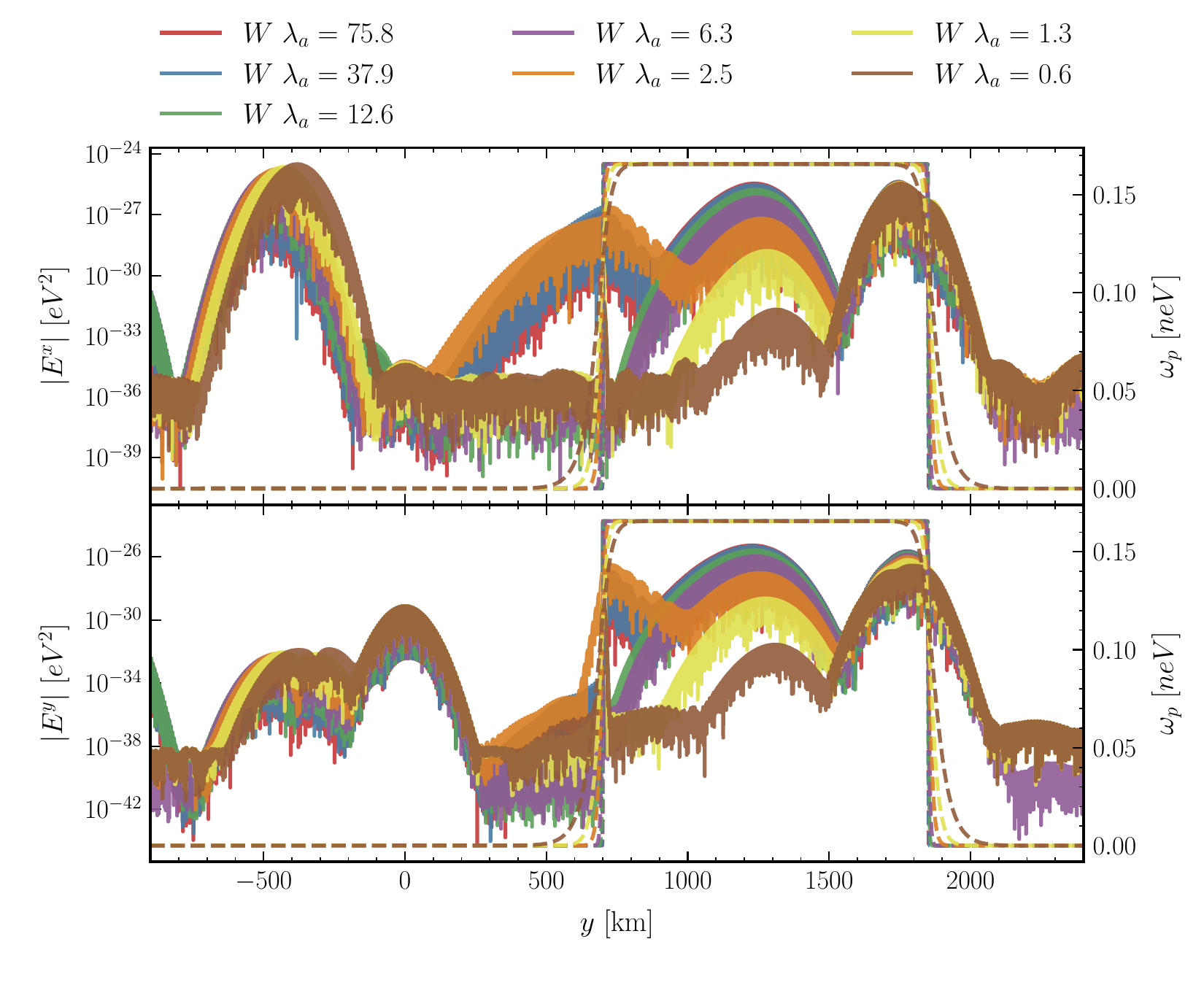}
    \caption{Snapshots of the the absolute values of the $x$ and $y$ components of the electric field at $t = 6.56 \, \ms$ for a representative set of simulations with $\theta_B = 30^\circ$ and different values of the steepness of the plasma barrier. For each color, solid lines represent the profiles of the electric field, while dashed lines represent the plasma frequency profiles. As we can see the amplitude of the Alfvén mode increases with $W$, and can even become comparable to the that of the LO mode.}
    \label{fig:FixedThetaSnapshots}
\end{figure}

In Fig.~\ref{fig:FixedThetaSnapshots} we show the profiles of the absolute values of $E^x$ and $E^y$ at $t = 6.56 \, \ms$ for some selected simulations of the set. Different colors refer to different values of $W$, which is expressed in terms of the axion wave length $\lambda_a = 2\pi / k_0$. Solid lines denote the profile of the electric field, while dashed line the profile of the plasma frequency.

\begin{figure}
    \centering
    \includegraphics[width=\columnwidth]{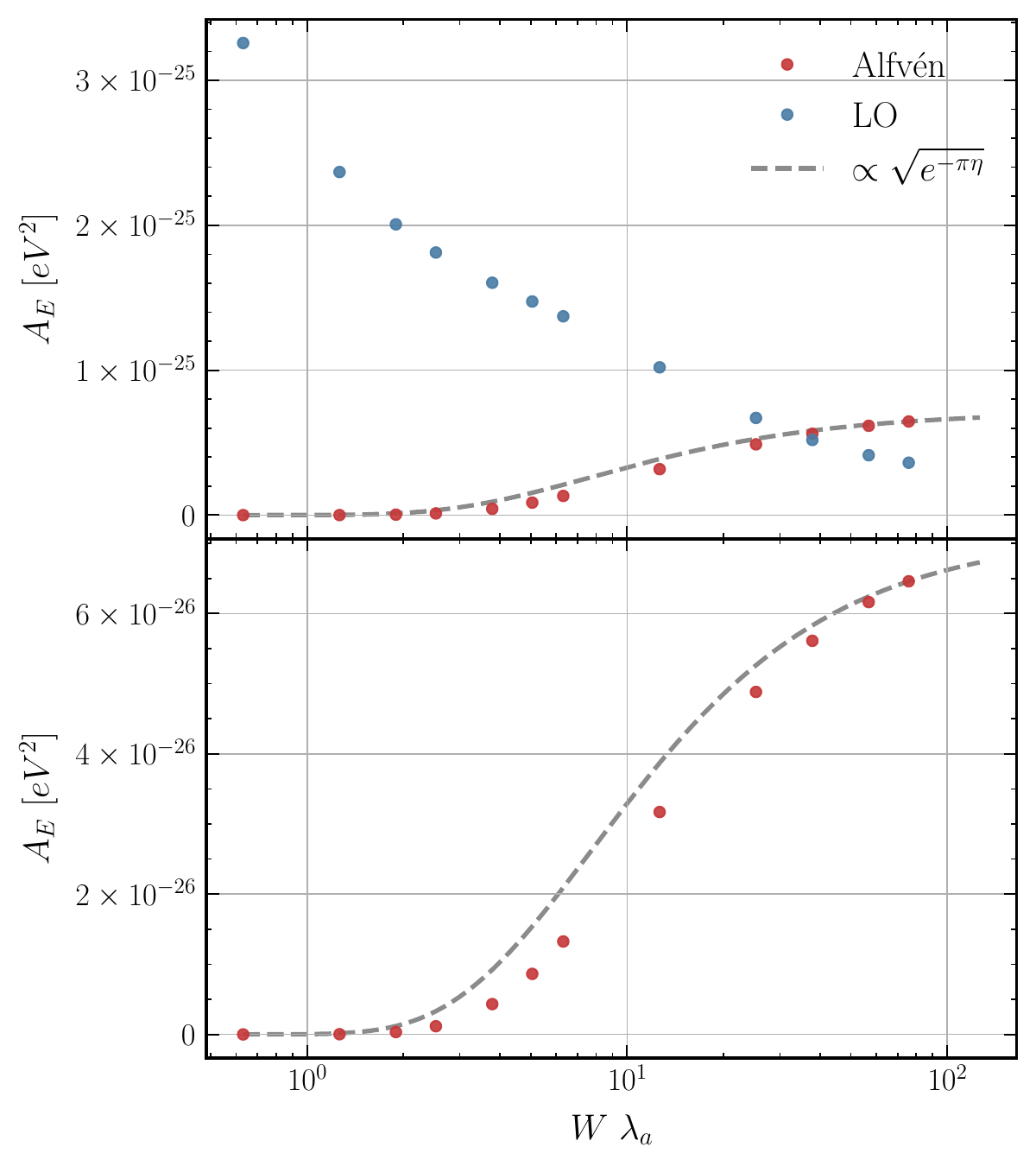}
    \caption{Comparison between the amplitudes of the Alfv\'en mode (red) and the LO mode (blue) for different values of the ratio between the axion wave length, $\lambda_a$, and the typical length scale over which the plasma density varies at the boundary of the barrier, $W^{-1}$. The bottom panel shows a zoom-in on the behavior of the amplitude of the Alfv\'en mode. The amplitude is estimated as the value of $A_E = \sqrt{(E^x)^2 + (E^y)^2}$ at the peak of the wave packets. We can clearly see that for large values of $W \, \lambda_a$ the amplitude of the Alfv\'en mode becomes larger than the amplitude of the LO mode. This is a key result of this work, which clearly demonstrates the indirect excitation of a sub-luminal plasma with an amplitude comparable to that of the super-luminal modes (\ie those typically studied in the context of axion-photon mixing). The gray dashed line represents the expected behavior of the amplitude of the Alfvén mode under the assumption of a fixed amplitude incident LO mode (an assumption which is not intrinsically expected to hold), obtained as the square root of the conversion probability in Eq.~\ref{eq:PAlfvenAnalytic}. The gray line has been normalized to the observed value at large gradient, and appears to reproduce the approximate functional scaling at lower values of $W$.  }
    \label{fig:AlfvenLOAmplitudes}
\end{figure}

As the scale over which the plasma density varies, $W^{-1}$, decreases, the amplitude of the Alfv\'en mode inside the plasma barrier increases, eventually becoming larger than the amplitude of the LO mode. For large values of $W$ it starts saturating, and the Alfv\'en mode for $W ~ \lambda_a = 75.8$ is barely visible behind the line corresponding to the case $W ~ \lambda_a = 37.9$. To provide a more quantitative characterization of the effect, we estimate the amplitudes of the modes as the value of $\sqrt{(E^x)^2 + (E^y)^2}$ extracted at the peak of the wave packets. Such positions are identified as the points where $E^y$ assumes its maximum value in $1100 \, \km \le y \le 1400 \, \km$ for the Alfv\'en mode, and as the point where $E^x$ assume its maximum value in $-600 \, \km \le y \le -250 \, \km$ for the LO mode. The amplitudes we obtain are shown in Fig.~\ref{fig:AlfvenLOAmplitudes} in terms of $W ~ \lambda_a$, which can be also understood as the ratio between the axion wave length and the length scale over which the plasma density varies at the boundary of the barrier. Red and blue points denote the amplitudes of the Alfvén and LO mode, respectively. As we can see the amplitude of the Alfvén mode increases with $W ~ \lambda_a$ while that of the LO mode decreases. At $W ~ \lambda \approx 30$ they cross each other, with the amplitude of the Alfvén mode becoming larger than that of the LO mode. Additionally, we compared our results with the behavior of the conversion probability computed in Sec.~\ref{sec:LZAlfven}. Since this quantity is expected to scale with the square of the amplitude, we include in Fig.~\ref{fig:AlfvenLOAmplitudes} the square root of the conversion probability in Eq.~\ref{eq:PAlfvenAnalytic}, appropriately rescaled to cross the rightmost red point (gray dashed line). As one can see, the analytic estimate reproduces the qualitative behavior of the amplitude of the Alfvén mode we obtain from our numerical simulations; it should be emphasized that this comes at something of a surprise, as the former is computed assuming a constant amplitude incident LO wave, whereas the efficiency with which the LO mode is sourced is itself a function of the plasma gradient. 

\begin{figure}
    \centering
    \includegraphics[width=\columnwidth]{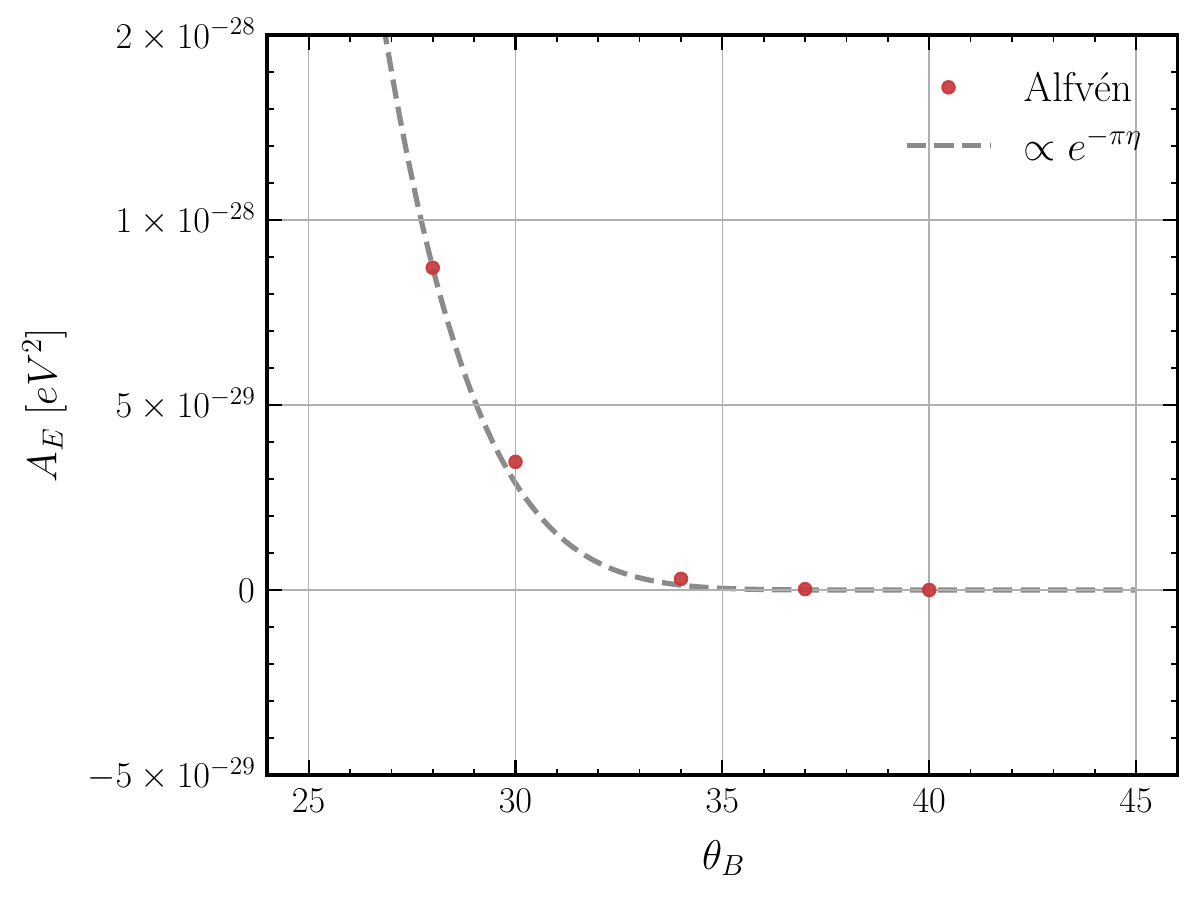}
    \caption{Amplitude of the Alfvén mode at $t = 4.39 \, \ms$ for the five simulations show in Fig.~\ref{fig:VaryingThetaSnapshots}, shown as a function of $\theta_B$. The gray dashed line denotes the expected behavior of the Alfvén mode amplitude assuming a constant incident LO mode, computed as the square root of the conversion probability from Eq.~\ref{eq:PAlfvenAnalytic}, and rescaled by a constant to intersect the amplitude for the case $\theta_B = 28^\circ$. The analytic behavior is in good agreement with data from the simulations.}
    \label{fig:AlfvenLOAmplitudesTheta}
\end{figure}

To assess whether Eq.~\ref{eq:PAlfvenAnalytic} also accurately captures the dependence of the conversion probability on $\theta_B$, we extract the amplitude of the Alfvén mode at $t = 4.39 \, \ms$ from the five simulations shown in Fig.~\ref{fig:VaryingThetaSnapshots}. Since the group velocity depends on $\theta_B$, the wave packets are in different locations, and we adjust the interval over which we search for the maximum of $E^y$ across the different simulations. The results are shown in Fig.~\ref{fig:AlfvenLOAmplitudesTheta}, where we plot the amplitude of the Alfv\'en mode as a function of $\theta_B$. As can be seen, the analytical prediction reproduces well the behavior observed in our numerical simulations.

Let us highlight briefly that the electric field appears to have a narrow and sharp feature situated directly near the resonance itself. This peculiar feature appeared across the mode-mixing simulations, and slightly shifts in location and amplitude depending on the simulation parameters. Zooming into this region, we have identified a number of interesting features. First, we find it is typically located in between the cut-off and the resonance, with a slight biasing toward the location of resonance. Next, we find that that its properties are evolving in time; in particular, we see that the perpendicular component of the electric field appears to decay (while the amplitude of the longitudinal feature remains nearly invariant), and the wavelength appears to decrease with time until it is no longer well-resolved. We investigated whether increasing numerical resolution had any impact on the reconstruction or properties of the system (both the peak standing wave feature, as well as the reflected and Alfv\'en wave), but these tests were found to always yield similar results (and there appeared to be no correlation with loss of overall accuracy -- see Appendix~\ref{app:convergence}).

Note that the properties of this feature resemble that of a high-$k$ mode standing wave (namely, a static longitudinal wave with support at wavelengths below the spatial resolution of the simulation); the question is how did this standing wave arise, and why does it remain. As mentioned in the proceeding sections, near sharp boundaries and resonances, the WKB approximation is strongly violated -- the strong breaking of translational invariance induced by the changing background can lead to violation of momentum conservation in the wave equation, allowing the transfer of energy to high momentum modes. This was not directly seen in the context of the mode-mixing probability (\ie the Budden equation), as this equation only captures the asymptotic solutions; the local electric field solutions sitting in the region between the cut-off and resonance are, in general, more complex~\cite{swanson2003plasma}. This opens the question of whether the localized standing wave is a physical feature that can arise in genuine systems, or if it is a consequence of unphysical assumptions that enter based on how the simulation was constructed.  

\begin{figure}
    \centering
    \includegraphics[width=\columnwidth]{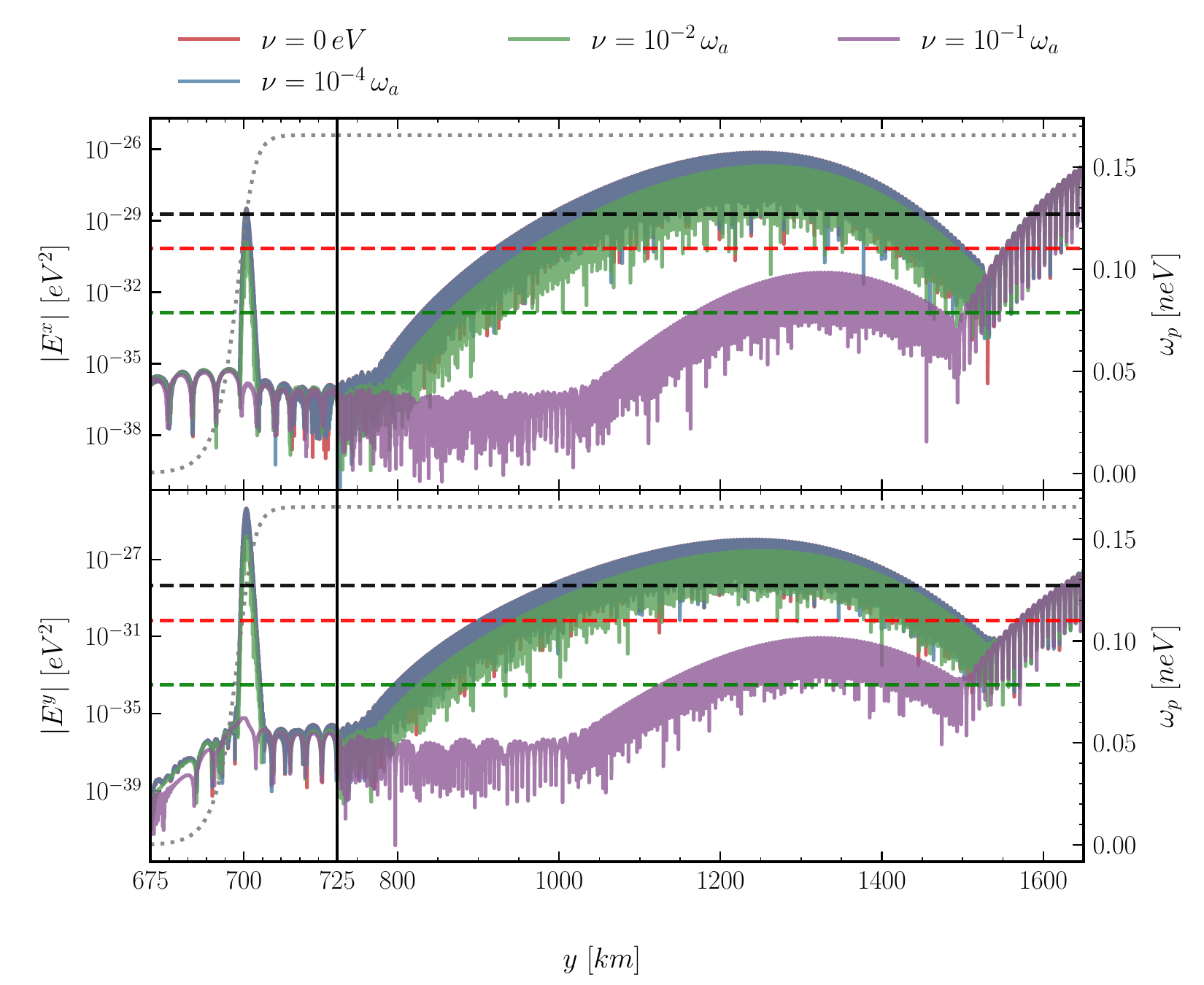}
    \caption{Absolute value of $E^x$ and $E^y$ at $t = 6.56 \, \ms$ for simulations with $W ~ \lambda_a = 6.3$ and different levels of energy damping in plasma, encoded by the parameter $\nu$. The left panels show the area of the spike at the edge of the plasma barrier, while the right panels show the area where the Alfvén mode is present. The red, green and black dashed lines represent the frequency of the wave packet, the axion resonant frequency, and the Alfvén mode resonant frequency, respectively, as in Fig.~\ref{fig:Theta34TimeSnapshots}. For $\nu = 10^{-4} \, \omega_a$ the effect of the damping term is extremely weak, and the red curve is barely visible behind the blue one. As $\nu$ increases the effect become stronger, and both the spike and the Alfvén mode get attenuated.}
    \label{fig:spike_diss}
\end{figure}

Let us note that our fiducial simulations do not include the possibility of energy dissipation within the plasma itself (the plasma is non-thermal, and electron-ion scattering processes are neglected). This raises the possibility that the energy being transferred into the local electric field solutions remains artificially trapped, with no means of dissipating. In order to test this possibility, we modify the equations of motion for the electrons to include an additional damping term, $\partial_t \UU^i \propto -\nu \UU^i$, which ensures there exists an energy damping mechanism. We re-run the simulation with $W ~ \lambda_a = 6.3$ in the set shown in Fig.~\ref{fig:FixedThetaSnapshots} for various values of $\nu$. The result is shown in Fig.~\ref{fig:spike_diss}, where we see that even a small amount of energy dissipation leads to a very strong suppression of the standing wave (the Alfv\'en mode also experiences some suppression, as one would expect, but on a scale which is orders of magnitude less than the standing wave feature). This is indeed indicative of these standing wave features being an artificial feature arising from a slightly simplified treatment of the background plasma.

\subsection{Electric field induced in small-scale under-densities}\label{sec:inhomo}

Here, we return to the question of how axion-induced electric fields can evade the strong parametric plasma suppression induced by high density plasmas. As mentioned in the preceding sections, this can happen when there exists small localized regions of plasma under-density (as for example occurs in the vacuum gaps on the polar caps of a neutron stars or black holes). We choose to work in the frequency domain, as we require a large separation of scales, which is most naturally achieved in this setup. We do this by directly imposing the spatial structure of the plasma frequency in the dielectric tensor -- note that this is only a valid approximation in the limit that the electron motion is small relative to the boundaries imposed on the under-dense region itself. Effectively, this imposes a constraint on the amplitude of the axion field and the coupling constant. Moving beyond this approximation would require time-domain simulations with appropriate boundary conditions. Since the weakly coupled limit is typically most natural for the axion, we restrict our attention to this scenario.

\begin{figure*}
	\centering
        \includegraphics[width=\linewidth] {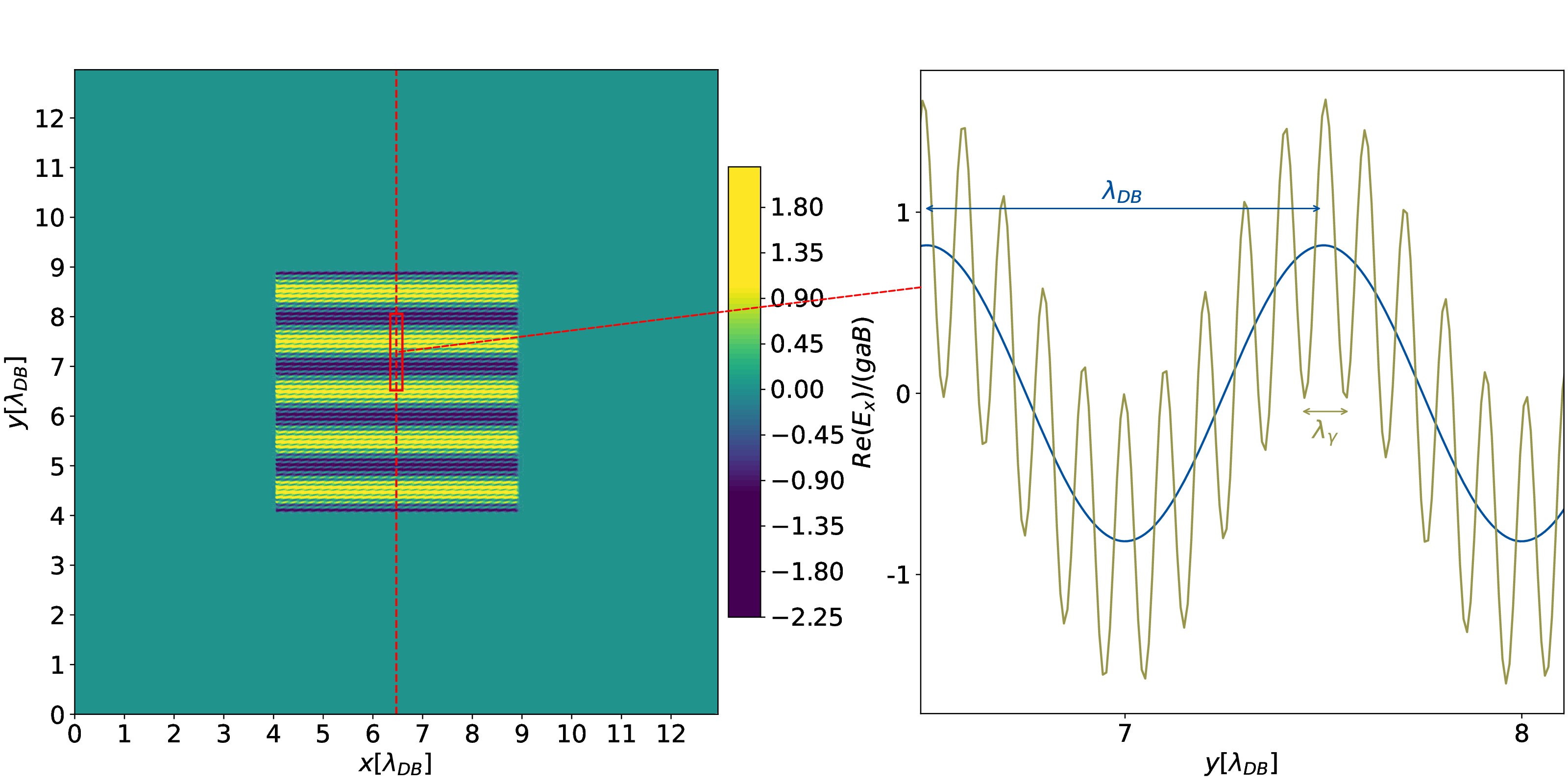}
        \caption{{\it Left:} Normalized electric field inside and outside a square, steep plasma under-density of width $5 \sim \lambda_a$. The field outside the under-density is strongly suppressed by a large plasma frequency. {\it Right:} Detail of a longitudinal section along the axion propagation direction showing the electric field as the superposition of the non-propagating axion-induced electric field and propagating on-shell photons excited by axion-photon conversions through interaction with the surrounding plasma. 
        }
	\label{fig:excitedphotons}
\end{figure*}

\begin{figure*}
	\centering
        \includegraphics[width=\linewidth] {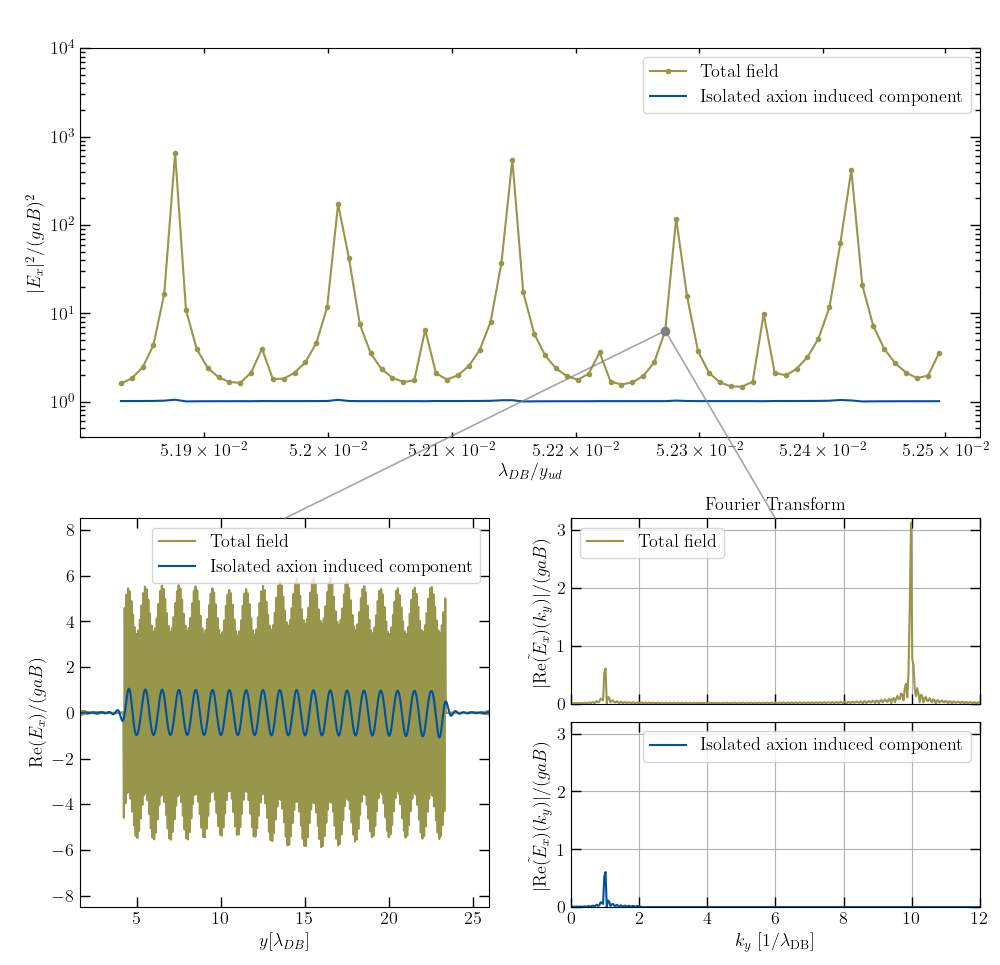}
        \caption{{\it Top:} Total normalized squared transverse electric field (green) and the isolated axion component (blue) for varying length of the under-density. As explained in the text the isolation is performed on the frequency domain by filtering out the high frequency component. {\it Bottom left :} The real component of the total transverse field and its axion induced component  along the middle of the simulation box in the propagation direction for a particular length of the under-density. {\it Bottom right:} The Fourier transform of the field and its isolated axion induced component.}
	\label{fig:ft_analysis}
\end{figure*}

An illustration of the simulation setup is depicted in in Fig. \ref{fig:excitedphotons}, where a large plasma density, with an amplitude six orders of magnitude large than that of the axion energy, permeates most of the simulation domain, except a small region in the center. The magnetic field is instead taken to be uniform across the simulation domain. In the central vacuum region, one can see the excitation of the electric field, whose amplitude has been normalized to the vacuum expectation, $E \sim \gagg a B$. The boundary transition regions are taken to be extremely sharp, such that the plasma frequency transitions from vacuum to $10^6 \times m_a$ over a distance of $10 \, \cm$, with a  spatial resolution in the simulation mesh at the level of $40\, \mm$ (we have verified explicitly that this is sufficient to fully resolve the features of the electric field away from these boundaries). The right panel shows a one-dimensional projection along the red region highlighted in left panel, where one can see that the signal is comprised of two distinct components, one corresponding to the axion induced electric field (shown in blue, with a wavelength set by the Debye wavelength of the axion) and one corresponding to an excited on-shell photon (shown in yellow, with a wavelength set by the axion frequency). The second component is a genuine electromagnetic wave that satisfies the local photon dispersion relation of the medium, that is to say it is a solution of the homogeneous wave equation, not just the directly forced particular solution. Ultimately, it appears because of the presence of boundaries, which act like places where the driven axion field can scatter into true propagating photon modes. In general, the simulations resolve only the sum of these two contributions. Since we are primarily interested in the axion induced electric field, we must separate these contributions in post-processing. This is done by Fourier transforming the signal region, filtering out the high frequency component, and going back to real space; an illustration is provided in the bottom panel of Fig.~\ref{fig:ft_analysis}. 

In the top panel of Fig.~\ref{fig:ft_analysis} we show the amplitude of the electric field component aligned with $\vec{B}$ to the expectation of the vacuum limit. The green curve shows the full reconstructed electric field as a function of axion velocity (plotted as the axion De Broglie wavelength $\lambda_{\rm DB}$ normalized to the size of the under-density $y_{\rm ud}$), while the blue curve shows the isolated axion-induced electric field. Here, one can see that in the limit where the axion wavelength is small relative to the scale of the under-density, the electric field is dominated by on-shell photon production (which in this case is resonantly enhanced due to the fact that the fixed background setup is effectively creating a cavity-like experiment, similar to traditional axion haloscopes). After removing the on-shell contribution, one indeed recovers the expected amplitude.

\begin{figure*}
	\centering
        \includegraphics[width=\linewidth] {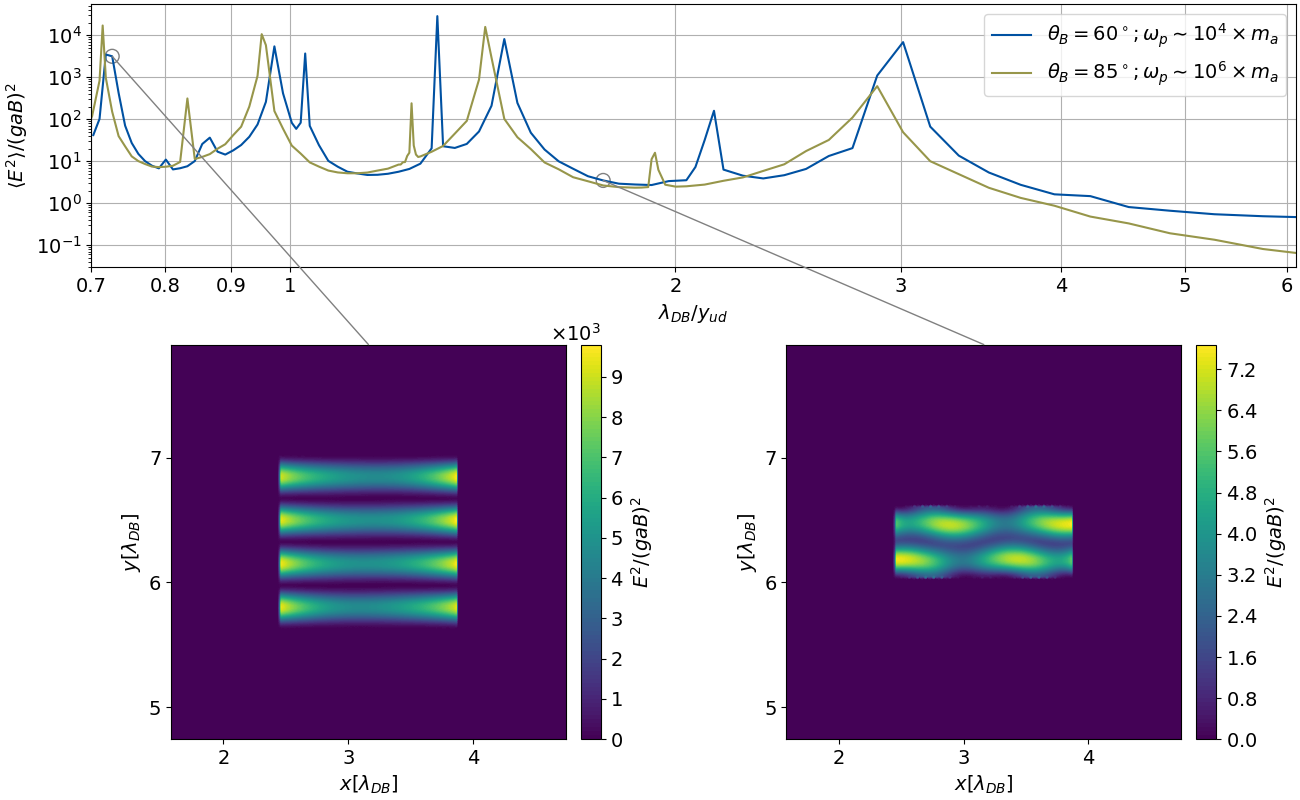}
        \caption{{\it Top:} Averaged (squared) electric field inside the under-density, normalized by $g_{a\gamma\gamma} a B$, for a rectangular under-density with varying vertical length scale, $y_{ud}$, along the propagation direction. Results are shown for two values of $\theta_B$, and two different heights of the plasma barrier. {\it Bottom:} 2-D representations of the field for two different lengths of the under-density. Between the two points there is a difference of 4 orders of magnitude due to the geometrical resonance caused by the cavity effects.}
	\label{fig:square_underdensity_varying_length}
\end{figure*}

The regime where $\lambda_{\rm DB} \gtrsim y_{\rm ud}$ is shown instead in Fig.~\ref{fig:square_underdensity_varying_length}\footnote{Note that we plot the full reconstructed electric field in this figure, rather than isolating the two components and plotting only the axion-induced field. This is because in the limit of very small under-densities, it becomes difficult to accurately separate the two modes -- there is a non-zero overlap in  Fourier space which makes the division obscured.}. Here, we also show two-dimensional projections of the electric field amplitude for two fixed points in the simulation, one corresponding to an on-resonance transition (left) which occurs when the scale of the under-density is a fixed harmonic of the axion wavelength, and one in which the photon excitation is relatively inefficient (right). The interest here is in understanding the scaling behavior at large $\lambda_{\rm DB}$, and on the transition region near the boundary (where the scales are comparable). Here, we include two calculations, one performed with $\theta_B = 85^\circ$ and with a plasma frequency outside the domain $\omega_p \sim 10^6 \times m_a$, and one with $\theta_B = 60^\circ$ and $\omega_p \sim 10^4 \times m_a$ (note that it becomes difficult to reduce the plasma threshold further, as the electric field leaks out and is not fully damped by the PML, with the latter effect arising from the fact that we are studying the large De Broglie wavelengths). Decreasing the angle and reducing the plasma frequency alters the boundary conditions, shifting them away from the plasma-induced cavity structure discussed in Sec.~\ref{sec:axelec}. Here, one can see residual resonant features extend to $\lambda_{\rm DB} \sim 3 \times y_{\rm ud}$, followed by a slow-fall off at larger values of the De Broglie wavelength. We have confirmed that in the isotropic limit, there is a primary resonant feature that appears as $\lambda_{\rm DB} \simeq y_{\rm ud}$, with smaller resonances separated at half wavelength distances; here, it is the anisotropy that shifts the resonant structure and distributes power more evenly across the difference resonances. We can see that the anisotropic plasma also reduces the suppression arising for large $\lambda_{\rm DB}$, tending toward the vacuum limit for small angles.

\begin{figure*}
	\centering
        \includegraphics[width=\linewidth] {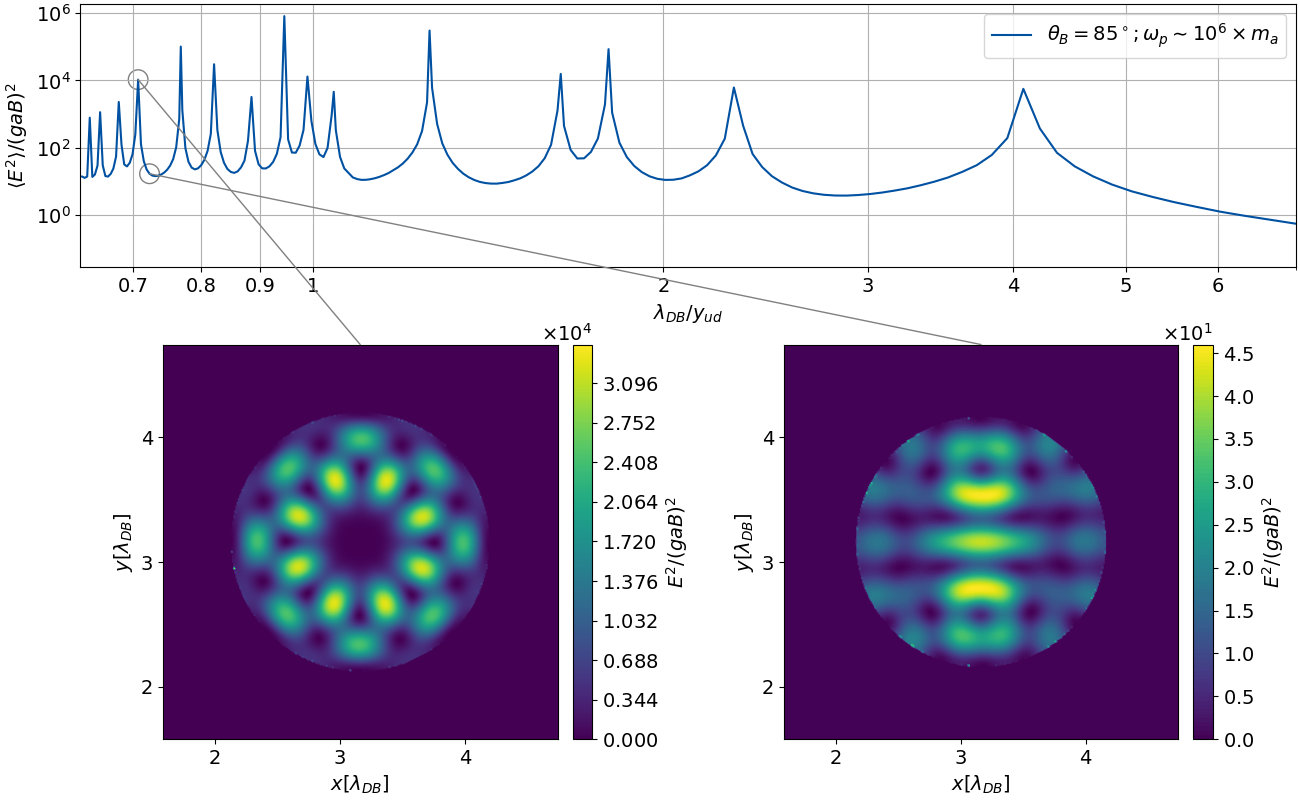}
        \caption{Same as Fig.~\ref{fig:square_underdensity_varying_length} for an under-density with circular geometry.}
	\label{fig:circle_underdensity_varying_diameter}
\end{figure*}

To highlight the role of the plasma geometry in Fig.~\ref{fig:square_underdensity_varying_length}, we repeat these simulations using a spherical rather than square under-density (and using $\theta_B = 85^\circ$ and with a plasma frequency outside the domain $\omega_p \sim 10^6 \times m_a$). The result is shown in Fig.~\ref{fig:circle_underdensity_varying_diameter}. Here, one can see the resonant structure of the induced electric field varies significantly, with far more resonances than in the square under-density scenario, but the general structure remains the same.

\section{Conclusions}

The goal of this work is to study energy transfer and loss from non-relativistic and semi-relativistic axion fields traversing inhomogeneous and strongly varying magnetized plasmas. This encompasses scenarios which naturally occur in a variety of astrophysical settings (see discussion in introduction), and for which analytic approximations often break down. This has been done using both time- and frequency-domain simulation codes, based on the ones developed in~\cite{Corelli:2024lvc} and~\cite{Gines:2024ekm}, that solve for the induced electric fields and the excited electromagnetic modes in highly magnetized plasmas. These techniques provide complimentary approaches, and yield consistent results across a variety of test scenarios. 

Arguably the most exciting aspect of this numerical study is the realization that axions can, in some circumstances, indirectly excite Alfv\'en modes (and more broadly, sub-luminal plasma modes) -- and with an efficiency that is comparable to that of ordinary modes (or, more precisely, LO modes in a magnetized plasma). This can occur in strongly varying backgrounds, where the axion is resonantly mixing with the LO  (occurring near $\omega_p \simeq m_a$) at a location which is very close to the cut-off in the LO mode dispersion relation (occurring at $\omega = \omega_p$) and the resonance of the Alfv\'en mode dispersion relation (occurring at $\omega = \omega_p |cos\theta_B|$). When the plasma gradient is sufficiently large, the WKB approximation is strongly violated, and all of three regimes spatially collapse into a narrow region -- this allows for a simultaneous sourcing of LO modes, alongside a tunneling of the LO modes into Alfv\'en modes. 

In addition to the aforementioned, we have investigated how strong parametric suppression of the axion-induced electric field can be evaded in the event that there exists small localized regions of vacuum (which are natural expectations arising in a variety of systems). For an isotropic background with high plasma barriers, one effectively trades a suppression scaling as $(m_a / \omega_p)^2$ for one which scales like $(L \times m_a)^2$, where $L$ is the characteristic scale of the under-density. We have investigated how this geometric suppression operates across different plasma scales and in anisotropic plasmas (where the boundary conditions are modified), showing that the geometric suppression can also be largely mitigated since the anisotropy of the plasma strongly modifies the boundary conditions themselves.  For low axion masses and high-density plasmas, this trade-off opens new avenues for more efficient energy transfer and dissipation in astrophysical systems.

Let us conclude by highlighting the potential relevance of these results in physical systems which may be of interest. First, let us note that traditional analytic approaches tend to break down when there are strong background gradients. Such features are natural near compact objects, which host spatially varying electromagnetic fields (which drives spatial variations in the plasma density) and current sheets (\ie narrow, dense, localized regions of high current carrying plasma). Moreover, it should be highlighted that axion dark matter resonantly sources electromagnetic modes at background densities very near the cut-off of the dispersion relation of the ordinary mode (this follows from the fact that axion dark matter is non-relativistic, or perhaps semi-relativistic near compact objects) -- when large background variations are present, this implies the axion induced resonance is occurring while energy is tunneled into the sub-luminal Alfv\'en modes. We have demonstrated here that this process can, in some scenarios, be quite efficient. This opens the question of whether there exists novel channels for energy dissipation in slowly-growing high-density axion field configurations, such as \eg `axion clouds'~\cite{Noordhuis_2023_C,Witte:2024akb} around neutron stars or black holes~\cite{Arvanitaki:2009fg}. This mode mixing process also opens a new possibility of energy transfer through plasma barriers. For example, previous work on radio wave excitation from axions near neutron stars have assumed low-energy photons undergo perfect reflection off the dense plasma~\cite{Witte:2021arp,McDonald:2023shx,Battye:2021xvt,Tjemsland:2023vvc,Witte:2022cjj} -- this mixing may offer an avenue for partially isotropizing highly anisotropic emission, as \eg occurs when axion mini-clusters or axion stars encounter neutron stars~\cite{Witte:2022cjj,Buckley:2020fmh,Nurmi:2021xds,Bai:2021nrs,Kouvaris:2022guf}. These are merely representative examples of how the results of these simulations may be applied in astrophysics -- we leave a more extensive investigation of such effects to future work.

%%%%%%%%%%%%%
%%%%%%%%%%%%%%

\section{Acknowledgments}%
%%%%%%%%%%%%%%%%%%%%%%%

AC is supported by an ERC STG grant (``AstroDarkLS'', grant No. 101117510). AC acknowledges the Weizmann Institute of Science for hospitality at different stages of this project and the support from the Benoziyo Endowment Fund for the Advancement of
Science. SJW acknowledges support from a Royal Society University Research Fellowship (URF-R1-231065). This work is also
supported by the Deutsche Forschungsgemeinschaft under Germany’s Excellence Strategy—EXC 2121 “Quantum Universe”—390833306. This article/publication is based upon work from COST Action COSMIC WISPers CA21106, supported by COST (European Cooperation in Science and Technology). EU acknowledges the financial support from the MCIU with funding from the European Union NextGenerationEU (PRTR-C17.I01) and Generalitat Valenciana (ASFAE/2022/020). This work has been supported by the Spanish grant PID2023-148162NB-C22 and by the European SE project ASYMMETRY (HORIZON-MSCA-2021-SE-01/101086085-ASYMMETRY) and well as by the Generalitat Valenciana grants PROMETEO/2019/083 and CIPROM/2022/69. s. E.C. acknowledges financial support
provided under the European Union’s H2020 ERC Advanced Grant “Black holes: gravitational engines of discovery” grant agreement no. Gravitas–101052587. E.C. acknowledges support from the
Villum Investigator program supported by the VILLUM
Foundation (grant no. VIL37766) and the DNRF Chair
program (grant no. DNRF162) by the Danish National
Research Foundation.
FC's research is supported by the European Research Council (ERC) Horizon Synergy Grant “Making Sense of the Unexpected in the Gravitational-Wave Sky” grant agreement no. GWSky–101167314.
Some numerical simulations have been performed at the Vera cluster supported by the Italian Ministry for Research and by Sapienza University of Rome. Some numerical simulations were performed at the cluster “Baltasar-Sete-Sois”
and supported by the H2020 ERC Advanced Grant “Black holes: gravitational engines of
discovery” grant agreement no. Gravitas–101052587.

\appendix

\section{Simulations}\label{sec:Simulations}

In this Appendix, we provide more details on the algorithms used for our time- and frequency-domain simulations.

\subsection{Time domain simulations}  \label{sec:EvolutionEquations1DAxion}

Working in the time domain, we can obtain the system of evolution equations and constraints with a 3+1 decomposition. In a nutshell, the spacetime is foliated with a set of spacelike hypersurfaces, and the coordinates are such that $t$ is constant on the hypersurfaces. Then, through a series of projections onto the hypersurfaces and onto the unit vector normal to them, it is possible to derive a set of evolution equations and constraints. Working in this formulation, we will use both Greek and Latin letters for tensor indices. The former indicate spacetime components, and run in $\{0, 1, 2, 3\}$, while the latter indicate spatial component, and run in $\{1, 2, 3\}$. 

The Ansatz for the line element is
\begin{equation}
    ds^2 = -(\alpha^2 + \beta_i \beta^i) \, dt^2 + 2 \beta_i \, dt \, dx^i + \gamma_{ij} \, dx^i \, dx^j,
    \label{eq:3+1metric}
\end{equation}
where $\alpha$ is the lapse, $\beta^i$ the shift vector, and $\gamma_{ij}$ the spatial 3-metric. The unit vector normal to the hypersurfaces can be expressed as $n^\mu = \left(\frac{1}{\alpha}, -\frac{\beta^1}{\alpha}, -\frac{\beta^2}{\alpha}, -\frac{\beta^3}{\alpha}\right)$. This vector is timelike, and can be thought as the 4-velocity of an observer, which is called \textit{eulerian observer}. 

Before obtaining the equations, it is necessary to introduce some definitions. In particular, the electric and the magnetic fields are defined as \cite{Alcubierre:2009ij}
\begin{equation}
    E^\mu = -n_\nu F^{\nu\mu}, \qquad B^\mu = - n_\nu \tilde{F}^{\nu\mu},
    \label{eq:EBdefinitions}
\end{equation}
which allows us to rewrite the electromagnetic tensor as
\begin{equation}
    F^{\mu\nu} = n^\mu E^\nu - n^\nu E^\mu + \spatial{\epsilon}^{\mu\nu\sigma} B_\sigma,
    \label{eq:ElectromagneticTensor3+1}
\end{equation}
where $\spatial{\epsilon}^{\mu\nu\sigma} = n_\lambda \epsilon^{\lambda\mu\nu\sigma}$.

For the plasma fluid, we project the 4-velocity $u^\mu$ onto $n^\mu$ and onto the hypersurfaces, and define the quantities
\begin{equation}
    \Gamma = -n_\mu u^\mu, \quad \spatial{u}^\mu = \tensor{h}{^\mu_\nu} u^\nu,
\end{equation}
where $\tensor{h}{^\mu_\nu}$ is the projector onto the hypersurfaces. However, we will not work directly with $\spatial{u}^\mu$, but rather with $\UU^\mu$, which is defined via the relation $\spatial{u}^\mu = \Gamma \UU^\mu$. Also the electron $n_e$ is not directly used, but we work with $\nel = \Gamma n_e$.

With these definitions, the evolution equations for the plasma fluid can be expressed as
\begin{align}
    \tder \Gamma &= \beta \partial_i \Gamma - \alpha \UU^i \partial_i \Gamma + \alpha \Gamma K_{ij}\UU^i \UU^j \notag \\
                 &- \Gamma \UU^i \partial_i \alpha + \frac{e}{m_e} \alpha E^i \UU_i, \label{eq:GammaEvolCov} \\
    \tder \UU^i  &= \beta^j \partial_j \UU^i - \UU^j \partial_j \beta^i - D^i \alpha - \alpha \UU^i K_{jl} \UU^j \UU^l \notag \\
                 &+ \frac{\alpha}{\Gamma} \frac{e}{m_e} \Bigl( -\UU^i E^j \UU_j + E^i + \spatial{\epsilon}^{ijl} B_l \UU_j \Bigr) \notag \\
                 &+ 2 \alpha \tensor{K}{^i_j} \UU^j + \UU^i \UU^j \partial_j \alpha - \alpha \UU^j D_j\UU^i \label{eq:UUEvol} \\
    \tder \nel   &= \beta^i \partial_i \nel + \alpha K \nel - \alpha \UU^i \partial_i \nel - \alpha \nel \nabla_\mu \UU^\mu, \label{eq:nelEvol}
\end{align}
where $D_i$ is the covariant derivative with respect to the 3-metric $\gamma_{ij}$, $K_{\mu\nu} = - \tensor{h}{^\lambda_\mu}\tensor{h}{^\sigma_\nu} \nabla_\sigma n_\lambda$ is the extrinsic curvature, and $K$ is its trace. Furthermore we have a constraint coming from the normalization of the fluid 4-velocity:
\begin{equation}
    \Gamma^2 \bigl( 1 - \UU_i \UU^i \bigr) = 1.
    \label{eq:FluidVelocityNormalization}
\end{equation}

Let us now focus on the equations for the axion. Here we need to define the conjugate momentum
\begin{equation}
    \Pi = n^\mu \nabla_\mu a = -\frac{1}{\alpha} \tder a + \frac{\beta^i}{\alpha} D_i a.
    \label{eq:AxionMomentumDef}
\end{equation}
Then the evolution equations for $a$ read
\begin{align}
    \tder a &= -\alpha \Pi + \beta^i \partial_i a, \label{eq:AxionEvolution} \\
    \tder \Pi &= \beta^i \partial_i \Pi + \Pi K \alpha - \alpha D^2 a - \gamma^{ij} (D_i \alpha) (D_j a) \notag \\
              &+ m_a^2 a \alpha - \alpha \gagg E_i B^i. \label{eq:AxionMomentumEvolution}
\end{align}

Finally, to write the equations for the electromagnetic field, we need to decompose the current defining the charge density $\rho = -n_\mu J^\mu$ and the 3-current $\spatial{J}^\mu = \tensor{h}{^\mu_\nu} J^\nu$. In our case the current is composed by two terms, accounting for ions and electrons: $J^\mu = J_{\rm{(ions)}}^\mu + J_{\rm{(e)}}^\mu$. Ions are assumed to be at rest (with respect to the eulerian observer), while electrons follow the velocity field $u^\mu$; thus we have
\begin{align}
    J_{\rm{(ions)}}^\mu &= \rho_{\rm{(ions)}} n^\mu, \label{eq:Jions} \\
    J_{\rm{(e)}}^\mu &= -e n_e u^\mu = - e \nel (n^\mu + \UU^\mu). \label{eq:Jelectrons}
\end{align}
With these definitions we obtain two evolution equations:
\begin{align}
    \tder E^i &= \mathcal{L}_\beta E^i + \alpha K E^i + \Bigl[ \vec{D} \times (\alpha \vec{B}) \Bigr]^i - \alpha e \nel \UU^i \notag \\
              &+ \alpha \gagg \Pi B^i + \alpha \gagg \Bigl[ \vec{E} \times \vec{D} a \Bigr]^i,  \label{eq:EEvol} \\
    \tder B^i &= \mathcal{L}_\beta B^i + \alpha K B^i - \Bigl[ \vec{D} \times (\alpha \vec{E}) \Bigr]^i,  \label{eq:BEvol}
\end{align}
where the arrow indicates a 3-vector, while $\mathcal{L}_\beta$ is the Lie derivative with respect to the shift vector. We also obtain two constraints, which play the role of the Gauss law, and the absence of magnetic monopoles,
\begin{align}
    D_i E^i &= -\rho_{\rm{(ions)}} + e \nel - \gagg B^i D_i a, \label{eq:Gauss} \\
    D_i B^i &= 0. \label{eq:MagneticGauss}
\end{align}

The equations we derived hold for a generic background metric. Nevertheless here we intend only to use a minkowskian metric $g_{\mu\nu} = \eta_{\mu\nu} = \diag\{-1, 1, 1, 1\}$, which results in $\alpha = 1$ and $\beta^i = 0$, as well as $K_{ij} = 0$ and $K = 0$ due to flatness. Furthermore, we restrict to systems that are homogeneuous with respect to the $z$ axis. This reduces the geometry of the simulation domain from 3+1 to 2+1 and removes all the terms containing derivatives with respect to $z$ from the field equations, cutting considerably the computational cost.

To arrive to the final set of equations we used in our code, one final step has to be performed. Indeed, the system now contains second order spatial derivatives of the axion field $a$. To reduce it to first order, we introduce the auxiliary variables $\Theta_x = \xder a$ and $\Theta_y = \yder a$ \footnote{Note that, thanks to the assumption of homogeneity on the $z$ axis, it is not necessary to introduce the auxiliary variable for $\partial_z a$, as it vanishes automatically.}. Their evolution equations will then be obtained as
\begin{equation}
    \tder \Theta_x = \tder \xder a = \xder \tder a = - \xder \Pi,
\end{equation}
and the same for $\Theta_y$.

After these operations, we can now write the full set of evolution equations. For the axion we have
\begin{align}
    \tder a &= - \Pi, \label{eq:AxionEvol} \\
    \tder \Theta_x &=  - \xder \Pi, \label{eq:ThetaxEvol} \\
    \tder \Theta_y &=  - \yder \Pi, \label{eq:ThetayEvol} \\
    \tder \Pi &= -\xder \Theta_x - \yder \Theta_y + m_a^2 a \notag \\
              &- \gagg (E^x B^x + E^y B^y + E^z B^z); \label{eq:AxionMomentumEvol}
\end{align}
for the electromagnetic field we have
\begin{align}
    \tder E^x &= \yder B^z - e \nel \UU^x + \gagg \Pi B^x - \gagg E^z \Theta_y, \label{eq:ExEvol} \\
    \tder E^y &= - \xder B^z - e \nel \UU^y + \gagg \Pi B^y + \gagg E^z \Theta_x, \label{eq:EyEvol} \\
    \tder E^z &= \xder B^y - \yder B^x - e \nel \UU^z + \gagg \Pi B^z \notag \\
              &+ \gagg \bigl( E^x \Theta_y - E^y \Theta_x \bigr), \label{eq:EzEvol} \\
    \tder B^x &= - \yder E^z, \label{eq:BxEvol} \\
    \tder B^y &= \xder E^z, \label{eq:ByEvol} \\
    \tder B^z &= \yder E^x - \xder E^y; \label{eq:BzEvol}
\end{align}
for the plasma we have
\begin{align}
    \tder \Gamma &= -\UU^x \xder \Gamma - \UU^y \yder \Gamma \notag \\
                 &+ \frac{e}{m_e}\bigl(E^x \UU^x + E^y \UU^y + E^z \UU^z \bigr), \label{eq:GammaEvol} \\
    \tder \UU^x &= \frac{1}{\Gamma} \frac{e}{m_e} \Bigl[- \UU^x \bigl(E^x \UU^x + E^y \UU^y + E^z \UU^z \bigr) + E^x \notag \\
                &+ \UU^y B^z - \UU^z B^y \Bigr] - \UU^x \xder \UU^x - \UU^y \yder \UU^x, \label{eq:UxEvol} \\
    \tder \UU^y &= \frac{1}{\Gamma} \frac{e}{m_e} \Bigl[ -\UU^y \bigl(E^x \UU^x + E^y \UU^y + E^z \UU^z \bigr) + E^y \notag \\
                &+ \UU^z B^x - \UU^x B^z \Bigr] - \UU^x \xder \UU^y - \UU^y \yder \UU^y, \label{eq:UyEvol} \\
    \tder \UU^z &= \frac{1}{\Gamma} \frac{e}{m_e} \Bigl[ -\UU^z \bigl(E^x \UU^x + E^y \UU^y + E^z \UU^z \bigr) + E^z \notag \\
                &+ \UU^x B^y - \UU^y B^x \Bigr] - \UU^x \xder \UU^z - \UU^y \yder \UU^z, \label{eq:UzEvol} \\
    \tder \nel &= - \UU^x \xder \nel - \UU^y \yder \nel - \nel \bigl(\xder \UU^x + \yder \UU^y \bigr). \label{eq:NelEvol}
\end{align}
The constraints are
\begin{align}
    \Theta_x &= \xder a, \label{eq:ThetaxConstraint} \\
    \Theta_y &= \yder a, \label{eq:ThetayConstraint} \\
    \xder E^x + \yder E^y &= -\rho_{\rm{(ions)}} + e \nel \notag \\
                          &- \gagg \bigl(B^x \Theta_x + B^y \Theta_y \bigr), \label{eq:GaussConstraint} \\
    \xder B^x + \yder B^y &= 0, \label{eq:MagneticConstraint} \\
    1 &= \Gamma^2 \Bigl[ 1 - \bigl(\UU^x\bigl)^2 - \bigl(\UU^y\bigl)^2 - \bigl(\UU^z\bigl)^2 \Bigr]. \label{eq:PlasmaConstraint}
\end{align}

Note that, since we are using a minkowskian metric, we have $V^i = V_i$ for every 3-vector $\vec{V}$; therefore we used upper or lower indices as convenient.

\subsubsection{Numerical setup}

Numerical evolution is performed with the method of lines. Spatial derivatives appearing in the right-hand sides of the evolution equations are computed with the finite difference method, using centered stencils that guarantee fourth order of accuracy. Time integration is instead performed with the sixth-order accurate version of the Runge-Kutta algorithm. We use periodic boundary conditions, which allow us to easily perform evolution without introducing numerical error in the simulation, provided that the initial profile of the plasma at the boundaries matches (to avoid discontinuities). However, we shall always use grid extensions large enough that during the simulations signals reaching the boundaries cannot come back the relevant regions where interactions happen, in order to avoid artificial interferences.

In the configuration we are interested in simulating we should include a constant background magnetic field $\vec B_0$, in order to allow for conversion to take place. The background magnetic field cannot be localized in a specific region of the space, as at the boundaries it would start propagating due to the term $\vec \partial \times \vec B$ appearing in the right hand side of the evolution equation for $\vec E$. We therefore decided to extend the support of $\vec B_0$ to the entire spatial domain.  

In principle the inclusion of the background magnetic field could be done at the initialization stage, adding $\vec B_0$ to the initial profile of the magnetic field, and then evolving the system with the equations derived in Sec.~\ref{sec:EvolutionEquations1DAxion}. However this can introduce numerical errors in the simulation, and we instead proceeded by separating the background from the foreground component, expressing $\vec B$ as $\vec B = \vec B_0 + \vec B_{\rm{fg}}$. Since $\tder \vec B_0 = 0$ and $\partial_i \vec B_0 = 0$, we evolve only the foreground component $\vec B_{\rm{fg}}$ using Eqs.~\ref{eq:BxEvol}-\ref{eq:BzEvol}, and the other fields by evaluating $\partial_i \vec B = \partial_i \vec B_{\rm{fg}}$ and $\vec B = \vec B_0 + \vec B_{\rm{fg}}$ when they appear in the right hand sides. It is worth underlying that this procedure does not involve a linear expansion of the magnetic field with respect to its background values, but only a redefinition of the evolution variable, preserving the fully nonlinear character of the system.

\subsubsection{Initialization procedure}

\begin{figure}
    \centering
    \includegraphics[width=\linewidth]{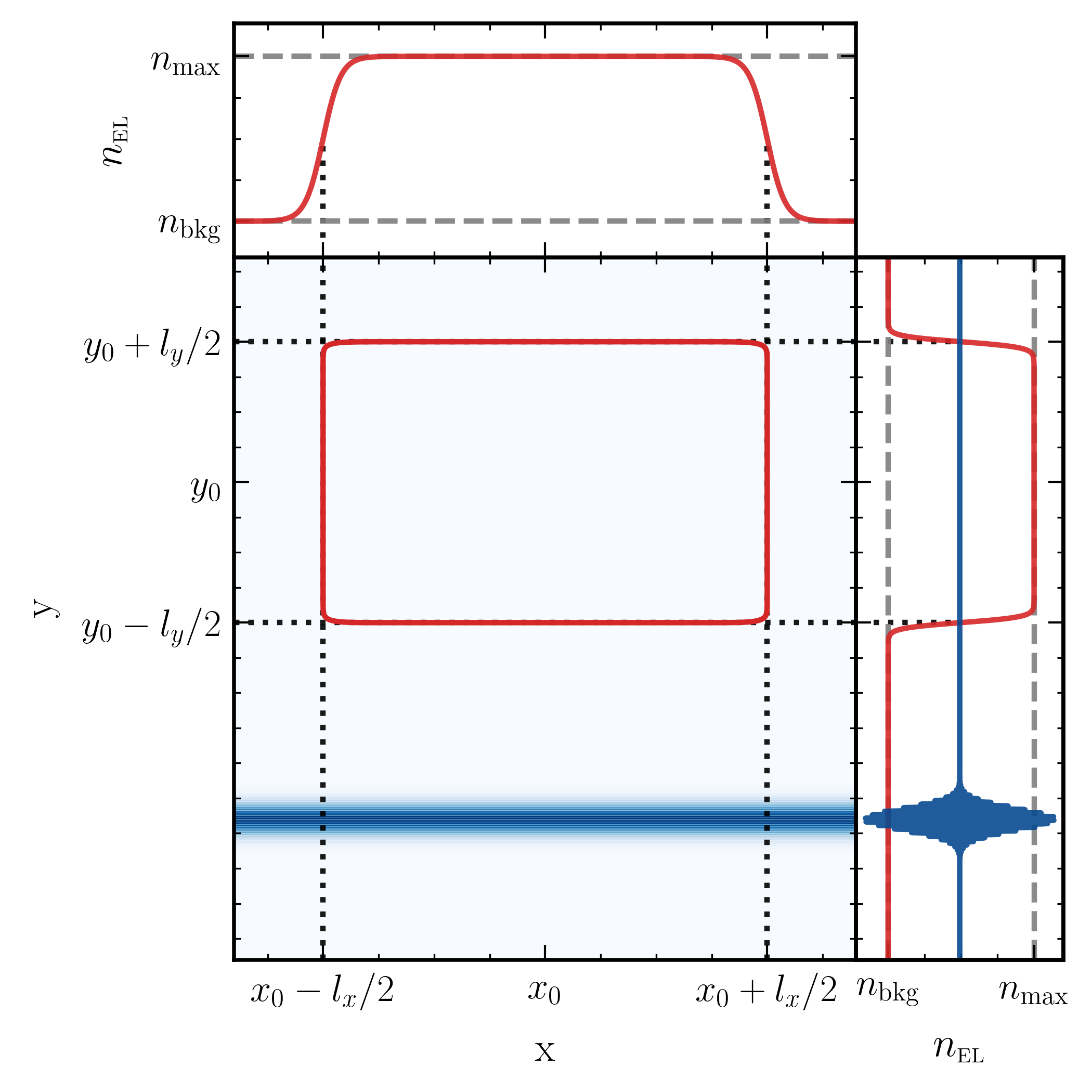}
    \caption{Illustration of 2D plasma frequency profile, and initialized wave-packet (blue), adopted in the time-domain simulation. The rectangular region highlighted in red in central panel roughly highlights the resonant boundary. }
    \label{fig:id_profile}
\end{figure}

The scenario we are interesting in simulating is a wave packet of the axion (in the unmixed basis) scattering on a plasma barrier. Let us start by describing the initialization of the plasma. The initial 4-velocity of electrons is taken to be $u^\mu = (1, \vec 0)$, so that $\Gamma(x, y, t = 0) = 1$ and $\vec \UU(x, y, t = 0) = 0$. The initial plasma density profile $n_0 = \nel(x, y, t=0)$ is that of a rectangular barrier, given by
\begin{align}
    n_0 &= n_{\rm bkg} + (n_{\rm max} - n_{\rm bkg}) \notag \\
        &\times \Bigl[\sigma\bigl(x - x_0; W, l_x/2\bigr) + \sigma\bigl(- (x - x_0); W, l_x/2\bigr) - 1\Bigr] \notag \\
        &\times \Bigl[\sigma\bigl(y - y_0; W, l_y/2\bigr) + \sigma\bigl(- (y - y_0); W, l_y/2\bigr) - 1\Bigr]
    \label{eq:InitialPlasmaDensity}
\end{align}
where $n_{\rm{bkg}}$ is the plasma density at the background and $n_{\rm{max}}$ is its value at the top of the barrier; $(x_0, y_0)$ is the locations of the center of the rectangle, $l_x$ and $l_y$ its width across the $x$ and $y$, respectively; the boundaries of the barrier are modeled with the sigmoid functions $\sigma(\xi; W, \xi_0) = \bigl(1+e^{-W(\xi - \xi_0)}\bigr)^{-1}$, with $W$ encoding their steepness.
A schematic representation of this plasma density profile can be found in Fig.~\ref{fig:id_profile}. 
To complete the initialization of the plasma, we set the charge density of ions to $\rho_{\rm{(ions)}} = e \nel(x, y, t = 0) = e n_0$, in such a way that the total charge density $\rho$ is initially vanishing. Note that $\rho_{\rm{(ions)}}$ is taken to be constant throughout the simulation, as ions are assumed to be at rest. Furthermore it only enters in the Gauss law for the electric field, Eq.~\ref{eq:GaussConstraint}, so that it is only relevant for constructing the initial profile, and for evaluating the convergence of the code.

Let us now focus on the the profiles of the axion, the electric field and the foreground component of the magnetic field. We wish to initialize them with a propagation eigenstate representing an axion wave packet which starts outside the barrier and moves toward it. We consider a plane wave and take $y$ as the propagation direction. The wave packet initial profile will therefore depend only on $y$, and extend across all the grid along the $x$ axis, reaching the boundaries. This is not a problem for boundary conditions, as an homogeneity with respect to $x$ is compatible with periodicity on the same axis. In other words, such initial profile will not generate any signal propagating from the boundary, neither physical nor unphysical, and the only signals arriving to the boundary are those coming from the interaction of the axion and electromagnetic fields with the plasma.

The requirement of a propagation eigenstate allows us to avoid an initial splitting of the packet on two components traveling with different velocities, gaining more control on the initial data, and cleaning the process from additional interactions of the packets between each other and with the plasma barrier. Furthermore, since the wave packet will be placed outside the barrier, \ie in a region where $\nel \approx n_{\rm bkg}$, and since we would like to consider frequency components well above the background plasma frequency, we will set $\omega_p = 0$. 

Let us start by defining the reference frame we use in our computation. The background magnetic field can have component along all the three axes, $\vec B = (B_0^x, B_0^y, B_0^z)$, so we define a set of spatial coordinates $(\mathring x, \mathring y, \mathring z)$ such that $\vec B_0$ is in the $\mathring{x}\mathring{y}$ plane. In practice we have $\mathring{y} = y$, and the unit vectors pointing to the $\mathring{x}$ and $\mathring{y}$ directions, in the original frame, are
\begin{align}
    \hat{e}_{\mathring{x}} &= \left( \frac{B_0^x}{B_{0, T}}, 0, \frac{B_0^z}{B_{0, T}} \right), \label{eq:TransverseBUnitVector} \\
    \hat{e}_{\mathring{y}} &= (0, 1, 0), \label{eq:LongitudinalBUnitVector}
\end{align}
where $B_{0, T} = \sqrt{\bigl(B_0^x\bigr)^2 + \bigl(B_0^z\bigr)^2}$.

In this frame, we express the 4-potential of the electromagnetic field and the axion in terms of their Fourier transforms, as
\begin{align}
    A^\mu(t, \mathring{y}) &= \int_{\mathbb{R}} d\omega \, \tilde{A}^\mu(\omega; \mathring{y}) e^{i \omega t} \notag \\
    a(t, \mathring{y}) &= \int_{\mathbb{R}} d\omega \, \tilde{a}(\omega; \mathring{y}) e^{i \omega t} = -i \int_{\mathbb{R}} d\omega \, \bar{a}(\omega; \mathring{y}) e^{i \omega t},
    \label{eq:InitialWavepacketFrequencyDecomp}
\end{align}
where $\bar{a}(\omega; \mathring{y}) = i ~ \tilde{a}(\omega; \mathring{y})$. The field equation for $\tilde{A}^{\mathring{z}}$ reads
\begin{equation}
    \left(\omega^2 + \partial_{\mathring{y}}^2 \right) \tilde{A}^{\mathring{z}} = 0,
    \label{eq:LinearizedAzEqVacuum}
\end{equation}
which is solved by $\tilde{A}^{\mathring{z}} = 0$; the equations for $\tilde{A}^{\mathring{x}}$ and $a$ instead read
\begin{equation}
    \left(\omega^2 + \partial_{\mathring{y}}^2 \right) 
    \begin{pmatrix}
        \tilde{A}^{\mathring{x}} \\ \bar{a}
    \end{pmatrix}
    =
    \begin{pmatrix}
        0 & -\omega \gagg B_0^{\mathring{x}} \\
        -\omega \gagg B_0^{\mathring{x}} & m_a^2
    \end{pmatrix}
    \begin{pmatrix}
        \tilde{A}^{\mathring{x}} \\ \bar{a}
    \end{pmatrix}.
    \label{eq:LinearizedATaVacuum}
\end{equation}
The eigenvalues of the mass matrix are then 
\begin{equation}
    \mu_{\pm}^2(\omega) = \frac{m_a^2 \pm \sqrt{m_a^4 + 4 \omega^2 \gagg^2 \left( B_0^{\mathring{x}} \right)^2}}{2},
    \label{eq:MassEingevaluesVacuum}
\end{equation}
and the rotation angle to the unmixed basis is
\begin{equation}
    \theta(\omega) = \frac{1}{2}\arctan{\left( \frac{2 \gagg B_0^{\mathring{x}} \omega}{m_a^2} \right)}.
    \label{eq:RotationAngleVacuum}
\end{equation}
The eigenvalue of the mass matrix corresponding to the axion is $\mu_{+}^2(\omega)$, so that a solution representing an axion wave packet can be written as
\begin{equation}
    \tilde \Psi_D(\omega; \mathring{y}) = \PsiVector{0}{1} \int_{\mathbb{R}} dk \, \hat \Sigma(\omega, k) e^{-i k \mathring{y}} \delta \Bigl[\omega^2 - k^2 - \mu_+^2(\omega)\Bigr],
\end{equation}
where $\hat \Sigma(\omega, k)$ represents the spectrum of the wave packet, and $\delta$ is the Dirac delta function. Solving the condition $\omega^2 - k^2 - \mu_+^2(\omega) = 0$, gives the dispersion relation for the axion, which is
\begin{align}
    \omega_+(k) &= \frac{1}{\sqrt{2}} \biggl\{ 2 k^2 + m_a^2 +\gagg^2 \bigl( B_0^{\mathring{x}} \bigr)^2 + \Bigl[m_a^4 \notag \\
                  &+ \gagg^4 \bigl( B_0^{\mathring{x}} \bigr)^4 + 2(2 k^2 + m_a^2) \gagg^2 \bigl( B_0^{\mathring{x}} \bigr)^2 \Bigr]^{1/2} \biggr\}^{1/2}.
    \label{eq:AxionDispersionRelationVacuum}
\end{align}
Up to some factors to be reabsorbed in $\hat \Sigma$, $\tilde \Psi_D(\omega; \mathring{y})$ can be rewritten as
\begin{align}
    \tilde \Psi_D(\omega; \mathring{y}) &= \PsiVector{0}{1} \int_{\mathbb{R}} dk \, \hat \Sigma(\omega, k) e^{-i k \mathring{y}} \notag \\
                             &\times \Bigl[\delta \bigl(\omega - \omega_+(k) \bigr) + \delta \bigl(\omega + \omega_+(k) \bigr)\Bigr].
\end{align}
Going back to the mixed basis we have
\begin{align}
    \tilde \Psi(\omega; \mathring{z}) &= \PsiVector{-\sin\theta(\omega)}{\cos\theta(\omega)} \int_{\mathbb{R}} dk \, \hat \Sigma(\omega, k) e^{-i k \mathring{y}} \notag \\
                             &\times \Bigl[\delta \bigl(\omega - \omega_+(k) \bigr) + \delta \bigl(\omega + \omega_+(k) \bigr)\Bigr].
\end{align}
We can now transform back to $A_\parallel(t, \mathring{z})$ and $a(t, \mathring{z})$; imposing the solutions to be real we get
\begin{align}
    A^{\mathring{x}}(t, \mathring{y}) &= - \int_\mathbb{R} dk \, \sin \theta(\omega_+(k)) \Bigl[\hat \Sigma(k) e^{i (\omega_+(k) t - k\mathring{y})} + c.c. \Bigr], \\
    a(t, \mathring{y}) &=  - \int_\mathbb{R} dk \, \cos \theta(\omega_+(k)) \Bigl[ i \hat \Sigma(k) e^{i (\omega_+(k) t - k\mathring{y})} + c.c. \Bigr],
\end{align}
where $c.c.$ denotes the complex conjugate, and $\hat \Sigma(k) = \hat \Sigma(\omega_+(k), k)$. Reabsorbing a factor $2$ in $\hat \Sigma(k)$, we finally arrive at
\begin{align}
    A^{\mathring{x}}(t, \mathring{y}) &= \Re \biggl\{ - \int_\mathbb{R} dk \, \sin \theta(\omega_+(k)) \hat \Sigma(k) e^{i (\omega_+(k) t - k\mathring{y})} \biggr\}, \label{eq:PhotonInitialWavePacket} \\
    a(t, \mathring{y}) &= \Re\biggl\{ - i \int_\mathbb{R} dk \, \cos \theta(\omega_+(k)) \hat \Sigma(k) e^{i (\omega_+(k) t - k\mathring{y})} \biggr\},
    \label{eq:AxionInitialWavePacket}
\end{align}
where $\Re$ denotes the real part. In our code, $\hat \Sigma(k)$ is set to be a gaussian centered around some wave number $k_0$:
\begin{equation}
    \hat \Sigma(k) = \AA ~ e^{-\frac{(k - k_0)^2}{2 \sigma_k^2}},
    \label{eq:WavePacketSigma}
\end{equation}
were $\mathcal{A}$ is an amplitude parameter, and $\sigma_k$ is the width of the packet in terms of the wave number. Since $\mathring{y} = y$, the initial profile of the axion is determined by computing $a$, $\Pi$, $\Theta_x$ and $\Theta_y$ from Eq.~\ref{eq:AxionInitialWavePacket}, with 
\begin{align}
    \Theta_x(t = 0, y) &= 0 \notag \\
    \Theta_y(t = 0, y) &= \yder a(t, y)|_{t = 0} \notag \\
    \Pi(t = 0, y) &= -\tder a(t, y)|_{t = 0}.
\end{align}
As for the photon, we have that the full 3-vector potential can be expressed as $\vec A = A^{\mathring{x}} \hat{e}_{\mathring{x}} + A^{\mathring{y}}\hat{e}_{\mathring{y}}$. The electric field is therefore given by
\begin{equation}
    \vec E = -\tder \vec A - \vec \partial \varphi = 
    \begin{pmatrix}
        - \frac{B_0^x}{B_{0, T}} \tder A^{\mathring{x}} \\
        - \tder A^{\mathring{y}} - \yder \varphi \\
        - \frac{B_0^z}{B_{0, T}} \tder A^{\mathring{x}}
    \end{pmatrix} ,
    \label{eq:InitialElectricFieldNoy}
\end{equation}
where $\varphi = - n_\mu A^\mu = A^t$ is the electromagnetic scalar potential. In Eq.~\ref{eq:InitialElectricFieldNoy} we can determine the $x$ and $z$ components of $\vec E$ using Eq.~\ref{eq:PhotonInitialWavePacket}, but we have not computed $A^{\mathring{y}}$ and $\varphi$. However, we can avoid doing it by using the Gauss law which, due to the assumptions of a vanishing plasma density and the homogeneity of the fields with respect to $x$ and $z$, reduces to
\begin{equation}
    \yder E^y = - \gagg B_0^y \yder a.
    \label{eq:GaussLawVacuum}
\end{equation}
Integrating on both sides we get an expression for $E^y$. The electric field can thus be expressed as
\begin{equation}
    \vec E(t = 0, y) = 
    \begin{pmatrix}
        - \frac{B_0^x}{B_{0, T}} \tder A^{\mathring{x}} (t, y)|_{t = 0} \\
        - \gagg B_0^y a (t, y)|_{t = 0} \\
        - \frac{B_0^z}{B_{0, T}} \tder A^{\mathring{x}} (t, y)|_{t = 0}
    \end{pmatrix} ,
    \label{eq:InitialElectricField}
\end{equation}
For the foreground component of the magnetic field we have
\begin{equation}
    \vec B_{\rm{fg}}(t = 0, z) = \vec \partial \times \vec A = 
    \begin{pmatrix}
        \frac{B_0^z}{B_{0, T}} \yder A^{\mathring{x}}|_{t = 0} \\
        0 \\
        -\frac{B_0^x}{B_{0, T}} \yder A^{\mathring{x}}|_{t = 0}
    \end{pmatrix} ,
    \label{eq:InitialMagneticField}
\end{equation}
which can be computed using Eq.~\ref{eq:PhotonInitialWavePacket}. 

The profile we constructed so far describes a wave packet of the axion centered at $y = 0$. As a final step, in order to center it in a generic position $y = y_0$, we shift it by replacing $y \to y - y_0$.

The evaluation of the Fourier integrals in these profiles is performed numerically. We sample the integrand in a region $[-k_{\rm{max}}, k_{\rm{max}}[$, using a spacing $\Delta k$, we apply the FFT algorithm on the sampled data, and we extract the profile in $y$ by means of a cubic Lagrange interpolation. The FFT is performed with the Cooley–Tukey algorithm, so that the number of samples used should be a power of 2. However, we set minimum requirements for the sampling spacing and extension in $k$, in such a way that the spacing in $y$ of the result will be smaller than the grid step used in the simulation, and the extension will be large enough to include the grid.

\subsection{Frequency domain simulations}\label{sec:fdomset}

In the frequency domain, one is effectively attempting to directly solve the axion-induced wave equation given in Eq.~\ref{eq:waveeq_ani}, with the background axion field, background magnetic field and background plasma distribution specified as inputs. This approach has the advantage that one can resolve much smaller scales than the time-domain simulations for fixed computational cost, allowing us, e.g., to push to increasingly non-relativistic velocities and smaller scale inhomogeneities. This benefit, however, comes at a cost -- one naturally loses the ability to study causal or dynamical effects of the system (including back-reaction on the axion field, the dynamical process of excitation, etc.), and in some cases the presence of multiple field contributions can obscure the physics of interest. Below, we outline the procedure adopted to simulate in the frequency domain -- this approach largely follows that of \cite{Gines:2024ekm}.

Following \cite{Gines:2024ekm}, we define a uniform background magnetic field $\vec{B}_0$ across a majority of the spatial domain, with a $\sin^2$ smoothing function at the boundaries of the domain which serve to minimize spurious photon excitation induced from spatial variations of the background, $|\vec{B}| = \vec{B}_0 \cdot \rm{sin}^2(A_y y + B_y) \cdot \rm{sin}^2(A_x x + B_x)$. The parameters $A_{x,y}$ and $B_{x,y}$ are chosen such that the uniform magnetic field vanishes at the edges of the simulation domain, with a sufficiently gradual ramp to suppress unphysical mode excitation. 

The fiducial field strength is arbitrarily set to $|\vec{B}_0| = 1950\;\rm{eV}^2$ -- in presenting our results, however, we normalize out the magnetic field, meaning the precise value of the background field strength is irrelevant. We define background axion plane waves propagating along the $y$-axis with fixed energy $\omega = \sqrt{m_a^2 + k_a^2}$, and adopt for two representative velocities of $v=0.1$ and $v=0.7$. Note that the benefit of the frequency domain simulations is that it allows for the resolution of mildly non-relativistic velocities, which in general require a larger separation of scales.

Similarly to the time-domain evolution code described in the previous section, we consider a cold electron plasma with number density $n_e$ and plasma frequency $\omega_p$. As discussed in the introduction, in a strong background magnetic field electron motion is effectively restricted to the direction of the field. Consequently, an electromagnetic wave propagating through a magnetized medium can only induce plasma oscillations parallel to the magnetic field. The plasma response is therefore anisotropic, exhibiting a preferred spatial direction. This effect is captured directly in the definition of the dielectric tensor: 
\begin{equation}\label{eq:dielectrictensor}
    \boldsymbol\epsilon = \left(\begin{array}{cccc}
    1 - \frac{w_p^2}{w^2} \sin^2 \theta_B & -\frac{w_p^2}{w^2} \sin \theta_B \cos \theta_B & 0 \\
    -\frac{w_p^2}{w^2} \sin \theta_B \cos \theta_B & 1 - \frac{w_p^2}{w^2} \cos^2 \theta_B  & 0 \\
    0 & 0 & 1
    \end{array}\right) \, .
\end{equation}

In this work, we adopt two different angles of $\vec{B}$ with respect to the propagation direction of the axions, $\theta_{B}$. As a fiducial test scenario, we adopt the limiting case of $\theta_{B}=90^{\circ}$, which corresponds to the limit of an isotropic medium. We then study deviations from this limit by looking at small shifts away from perpendicular magnetic fields, taking instead $\theta_{B}=45^{\circ}$.

As in \cite{Gines:2024ekm}, we impose a PML (perfectly matched layer) around the boundary of the simulation domain, which serves to absorb incident waves and suppress spurious reflections -- for each case of interest, we adopt the size and resolution to ensure both the axion and photon wavelengths are damped over the PML.  For example, for an axion mass of $10^{-5}$ eV and a velocity of $0.7$ ($\lambda_{\rm{DB}} \sim 0.13\;\rm{m}$, $\lambda_\gamma$ in vacuum $\sim 0.09\;\rm{m}$), we adopt a $16 \times 16\;\rm{m} \; (\sim 126 \times 126 \;\lambda_a)$ simulation domain with a grid spacing of $1.1\times10^{-2}\;\rm{m}$ ($\sim$ 8 points per $\lambda_\gamma$). Where required to resolve sharp spikes due to spatial resonances, we employed finer grids with spacing down to $4\times10^{-3}\;\rm{m}$. 

In each simulation, we extract the axion-photon conversion probability $p_{a \rightarrow \gamma}$ and the average squared magnitude of the electric field, $|\vec{E}|^2$. Following \cite{Gines:2024ekm}, the conversion probability is computed numerically by integrating the outward Poynting flux over the simulation boundary just inside the enclosing PML region (and normalizing by the incident axion energy flux). The electric field is calculated directly from the simulation output and normalized by the squared amplitude of the non-propagating axion-induced electric field, such that it takes a value of unity in free-space conditions, where no propagating photon modes are excited.

\section{Accuracy and convergence of the time-domain simulation code}  \label{app:convergence}

We assess the convergence properties of our 2+1 time evolution code by checking how the constraint violations on a test simulation scale when numerical resolution is increased. The quantities we monitor are
\begin{align}
    CV_{\rm axion} &= \Theta_y - \yder a, \label{eq:CVAxion} \\
    CV_{\rm Gauss} &= \xder E^x + \yder E^y - e (\nel - n_0) \notag \\
                   &+ \gagg \bigl(B^x \Theta_x + B^y \Theta_y \bigr), \label{eq:CVGauss} \\
    CV_{\rm plasma} &= \Gamma \sqrt{ 1 - \bigl(\UU^x\bigl)^2 - \bigl(\UU^y\bigl)^2 - \bigl(\UU^z\bigl)^2 } - 1, \label{eq:CVPlasma}
\end{align}
which correspond to the violations of the constraints in Eqs.~\ref{eq:ThetayConstraint}, \ref{eq:GaussConstraint} and \ref{eq:PlasmaConstraint}, respectively.

In Ref.~\cite{Corelli:2024lvc} we observed that in the linear regime of interaction the variation of the plasma density in time can be so small to fall below machine precision error, causing the constraint violation not to scale with resolution as expected. Therefore we evaluate convergence on a test setup in which the amplitude of the initial axion wave packet is large enough that the axion-induced electric field can induce a non-negligible backreaction on plasma. In particular we consider a flat plasma profile with density $n_0 = 10^{-3} \, \cm^{-3}$, and a wave packet of propagation eigenstate corresponding to the axion located at $y_0 = -2500 \, \km$, with $k_0 = 9.8 \times 10^{-11} \, \eV$ and $\sigma_k = 0.05 \, k_0$. The mass of the axion is set to $m_a = 10^{-10} \, \eV$, and the frequency is $\omega_a \approx \sqrt{m_a^2 + k_0^2}$. 
We start from the case in which the background magnetic field is orthogonal to the propagation direction, and we set it to $\vec B_0 = (1/\sqrt{2}, 0, 1/\sqrt{2}) \times \, 10^{-5} \, \Tesla$. The coupling constant is $\gagg = 4 \times 10^{-12} \, \eV^{-1}$, while the parameter $\AA$ appearing in Eq.~\ref{eq:WavePacketSigma} is set to $\AA = \frac{0.025}{\epsilon} \, \frac{m_e}{e \, \sigma_k}$, where $\epsilon = \frac{\omega_a B_0 \gagg}{m_a^2}$ is the parameter that encodes the strength of the coupling. In this way the amplitude of the axion in the coordinate space is $\sim \frac{0.025}{\epsilon} \, \frac{m_e}{e}$, and that of the axion-induced electric field is $0.025 \, \omega_a \, \frac{m_e}{e}$, which is close to the threshold where the system enters in the nonlinear regime of interaction between the photon and the plasma~\cite{1970PhFl...13..472K}.

We perform the simulation with two resolutions: $\Delta x_{\rm coarse} = \Delta y_{\rm coarse} = 1.2 \times 10^{9} \, \eV^{-1} = 236 \, \m$ and $\Delta x_{\rm fine} = \Delta y_{\rm fine} = \Delta y_{\rm coarse} ~ / ~ 2$. The grid extends in $[-1.0 \, \km, 1.0 \, \km] \times [- 6000 \, \km, 1000, \, \km]$, so that there are $9$ grid points across the $x$ axis in the run with lower resolution. The $CFL$ factor is $CFL = 0.2$, and the total time of integration is $T = 11.04 \, \ms$.

Since in our numerical integration procedure we use the sixth-order accurate Runge-Kutta method for time integration, and the fourth-order accurate finite differences formulas for computing spatial derivatives, we expect that, at a given time, the constraint violations scale with the fourth power of the spatial step. 
To check whether this is the case, we compute $CV_{\rm axion}$, $CV_{\rm Gauss}$ and $CV_{\rm plasma}$ on the simulation data, evaluating the spatial derivatives with a sixth-order accurate finite difference formula, so that the error introduced by the computation of the constraint violations is subleading with respect to the one expected from the evolution procedure; we then plot the constraint violations, verifying that the profiles in the lower resolution case match the ones obtained by multiplying the profiles in the high resolution case by a factor $\left(\Delta y_{\rm coarse} ~ / ~ \Delta y_{\rm fine} \right)^4 = 16$.

\begin{figure}
    \centering
    \includegraphics[width=\columnwidth]{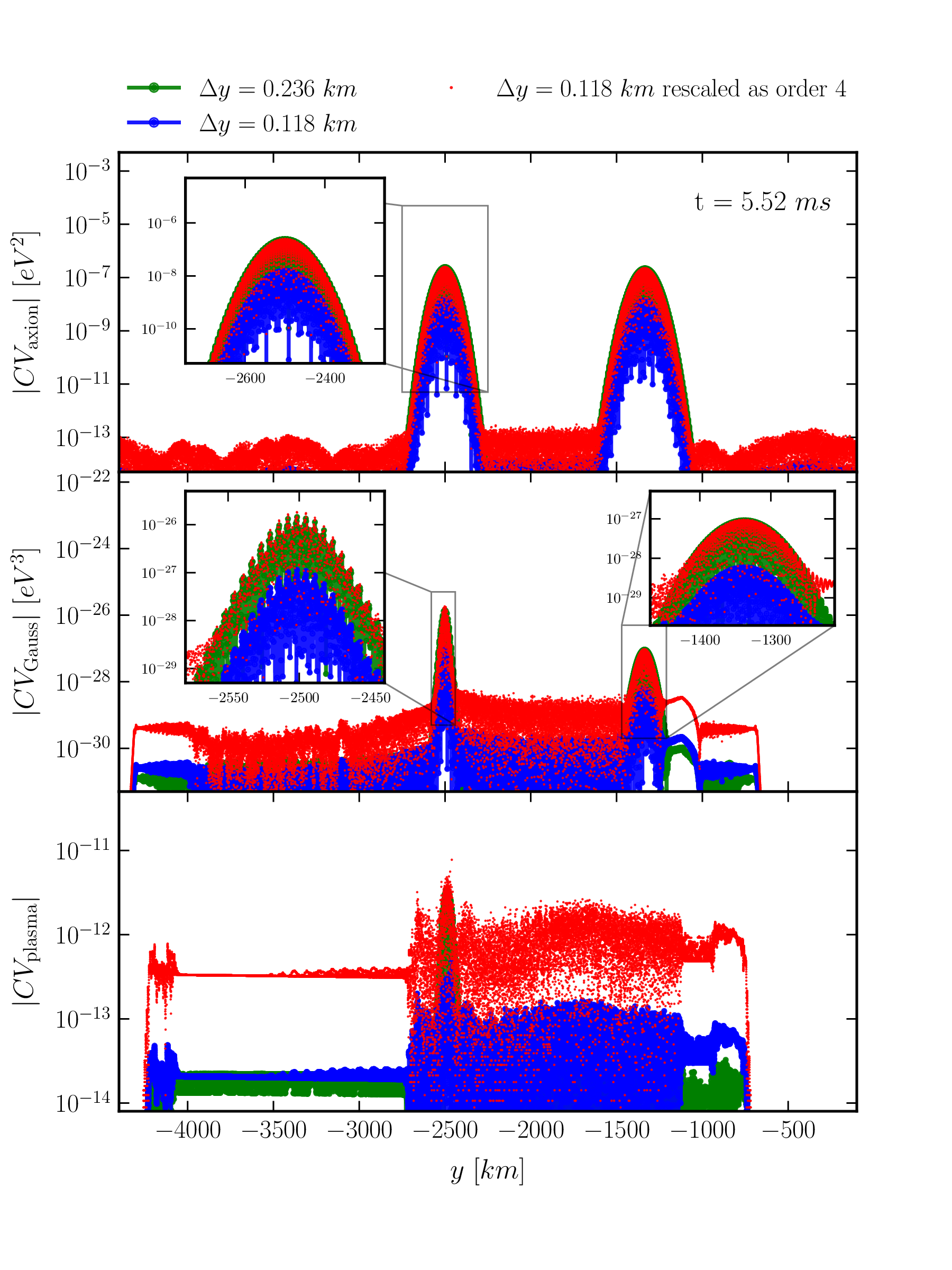}
    \caption{Scaling of the constraint violations at $t = 5.52 \, \ms$ for a simulation in which the amplitude of the initial axion wave packet is large enough that the electromagnetic component of the propagation eigenstate induces a non-negligible back-reaction on the plasma. In the upper, middle and lower panel we consider $CV_{\rm axion}$, $CV_{\rm Gauss}$ and $CV_{\rm plasma}$, respectively. In each panel, the green and blue lines denote the absolute value of the constraint violations corresponding to the simulations with coarser and finer grid, respectively, whereas the red points denote the absolute value of the constraint violations from the simulation with higher resolution, rescaled by a factor $16$. As we can see the code shows good convergence properties.}
    \label{fig:ConvergenceTransverse}
\end{figure}

The results of this test are shown in Fig.~\ref{fig:ConvergenceTransverse}, where the blue and green curves denote the absolute value of the constraint violations extracted from the high and the low resolution simulation, respectively, while the red points denote the constraint violations in the high resolution case rescaled by a factor $16$. Data are extracted at $t = 5.52 \, \ms$, and in the three panels, from top to bottom, we show $CV_{\rm axion}$, $CV_{\rm Gauss}$ and $CV_{\rm plasma}$, respectively. The insets contain a zoom-in of the relevant region where the constraint violations are well above the numerical noise. As we can see the code shows good convergence, with $CV_{\rm Plasma}$ being harder to evaluate as it is dominated by noise. Nevertheless, the evolution of the plasma density enters in the Gauss law, and thus the behavior of $CV_{\rm Gauss}$ provides an additional check on the evolution equations for the plasma.

We observed that at later times the convergence tends slightly to deteriorate; note, however, that the $t = 5.52 \, \ms$ considered here is not too far from the total simulation time used in our production runs.

\begin{figure}
    \centering
    \includegraphics[width=\columnwidth]{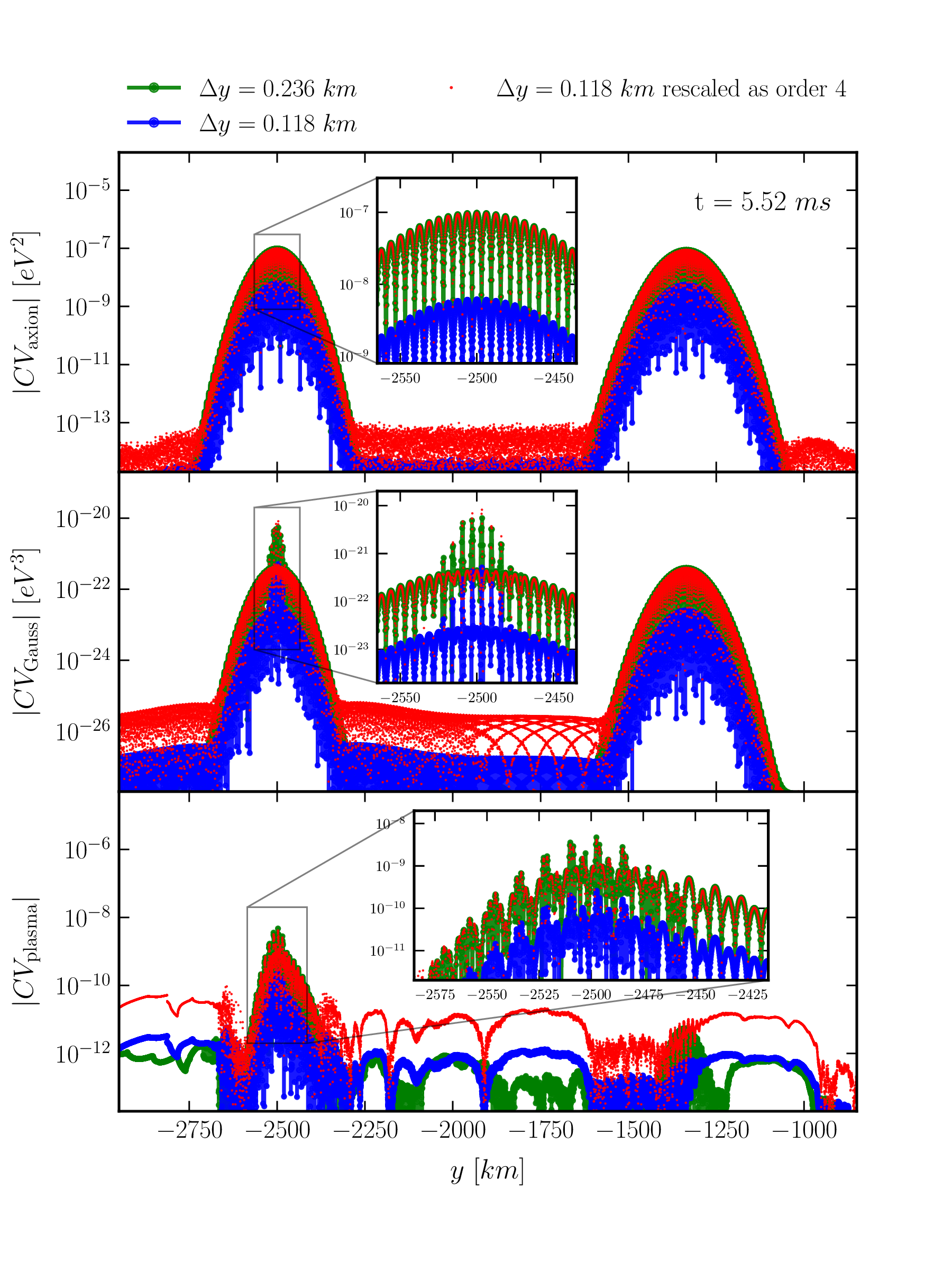}
    \caption{Scaling of the constraint violations at $t = 5.52 \, \ms$ for a configuration with a longitudinal component of the background magnetic field. Conventions are the same as in Fig.~\ref{fig:ConvergenceTransverse}. The code generally satisfies fourth order convergence, in agreement with expectations for the numerical scheme employed. The spikes close to the initial position of the wave packet ($y = -2500 \, \km$) might be due to a large longitudinal component of the electric field, that pushes the system beyond the regime of validity of the fluid approximation for plasma. $CV_{\rm plasma}$ appears to be in general outside the regime of convergence, but shows fourth order scaling around $y = -2500 \, \km$, where it assumes larger values.}
    \label{fig:ConvergenceLongitudinal}
\end{figure}

We now move to asses the convergence on a configuration in which the background magnetic field is not orthogonal to the propagation direction, which is the case of interest for the production of the Alfvén mode. The setup is the same as in the other set, but with $\vec B_0 = (1/\sqrt{3}, 1/\sqrt{3}, 1/\sqrt{3}) \times \, 10^{-5} \, \Tesla$. We observe that the simulation tends to crash, likely due to an excessively large longitudinal component of the electric field, therefore we decrease the amplitude of the axion wave packet, setting it to $\AA = \frac{0.01}{\epsilon} \, \frac{m_e}{e \, \sigma_k}$. The scaling of the constraint violations at $t = 5.52 \, \ms$ is shown in Fig.~\ref{fig:ConvergenceLongitudinal}. $CV_{\rm axion}$ and $CV_{\rm Gauss}$ show two peaks at the initial position of the wave packet and at its position at the extraction time. In both these locations the code shows a fourth order scaling, in agreement with expectations. Around $y = -2500 \, \km$ (the initial position of the wave packet) some spikes appear in the violation of the Gauss constraint. This might be due to a large longitudinal component of the electric field, which pushes the plasma in a regime where the fluid description starts to break down \cite{PhysRev.113.383}. However, this is an inherently nonlinear effect and it is not a concern for the simulations described in the main text. As for $CV_{\rm plasma}$, while it looks generally outside the convergence regime, it satisfies a fourth order scaling at the initial position of the packet, where it assumes larger values.

\begin{figure}
    \centering
    \includegraphics[width=\columnwidth]{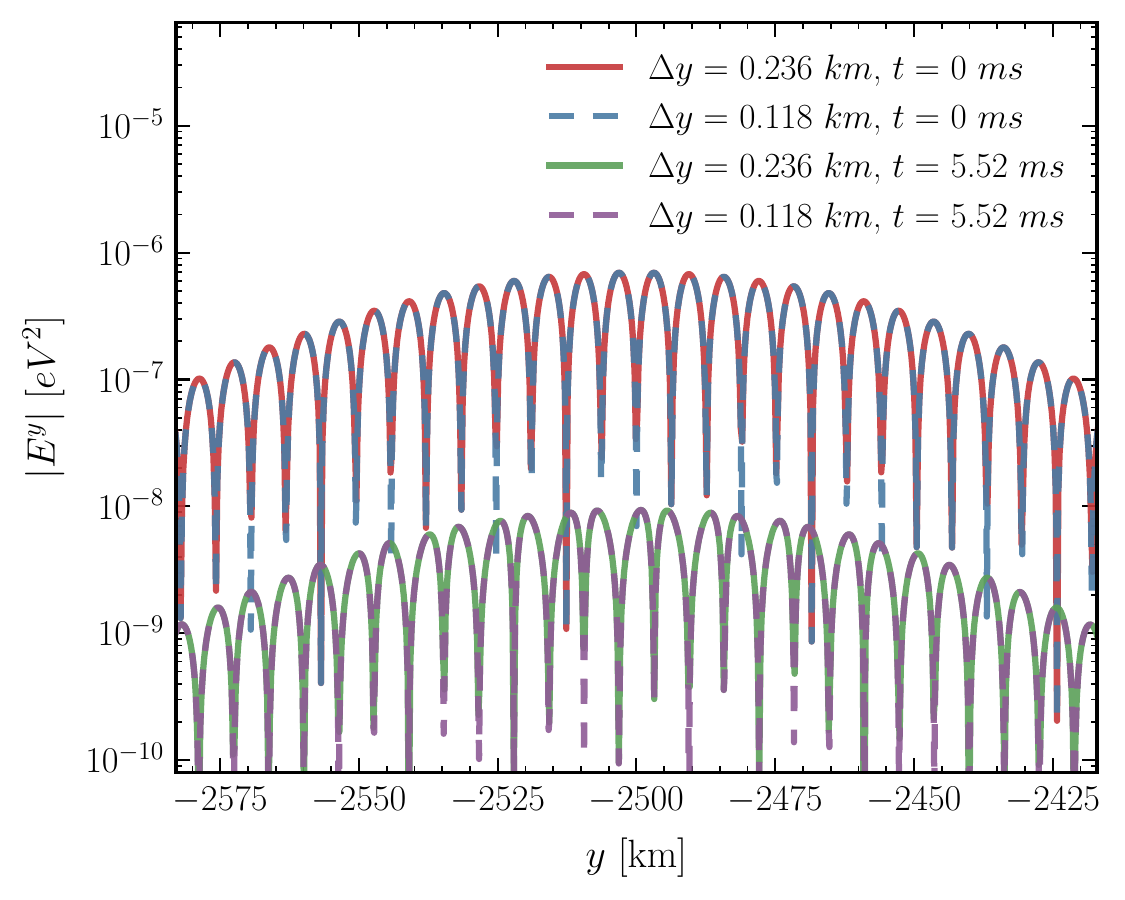}
    \caption{Profile of $E^y$ around the initial position of the wave packet at $t = 0 \, \ms$ and $t = 5.52 \, \ms$. Solid and dashed lines denote data extracted from the simulation at low and high resolution, respectively. As we can see the wave packet contains a static component that remains at rest, and becomes clearly visible after the propagating component has moved away. The perfect matching between the profiles at high and low resolution, together with the scaling of the constraint violation (cf. Fig.~\ref{fig:ConvergenceLongitudinal}) indicates that the appearance of such component is not an artifact of the numerical integration, but a physical effect.}
    \label{fig:StaticWavePacket}
\end{figure}

In these simulations, we also notice a component of $E^y$ that remains at rest in the initial position, analogous to the one appearing in the simulations for the production of the Alfvén mode (see Sec.\ref{sec:modeexcite}). We show such wave packet in Fig.~\ref{fig:StaticWavePacket} where we plot the profile of the electric field at $t = 0 \, \ms$ and at $t = 5.52 \, \ms$ for the two resolutions considered. As we can see the profiles at $t = 5.52 \, \ms$ match perfectly; this, together with the fourth order scaling of the constraint violation in the same region, indicates that such field is not due to numerical error, but is a static wave packet that becomes visible after the propagating component has left its initial position.

\begin{figure}
    \centering
    \includegraphics[width=\columnwidth]{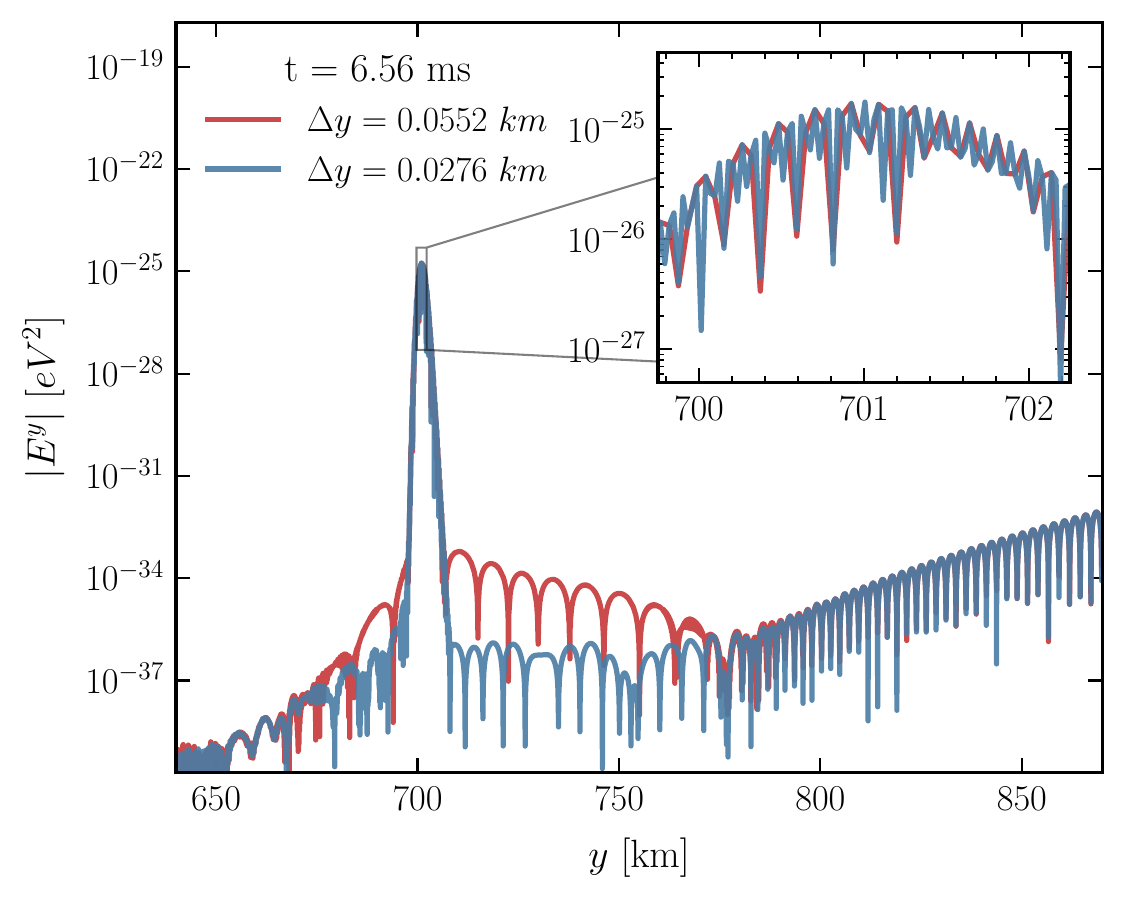} \\
    \includegraphics[width=\columnwidth]{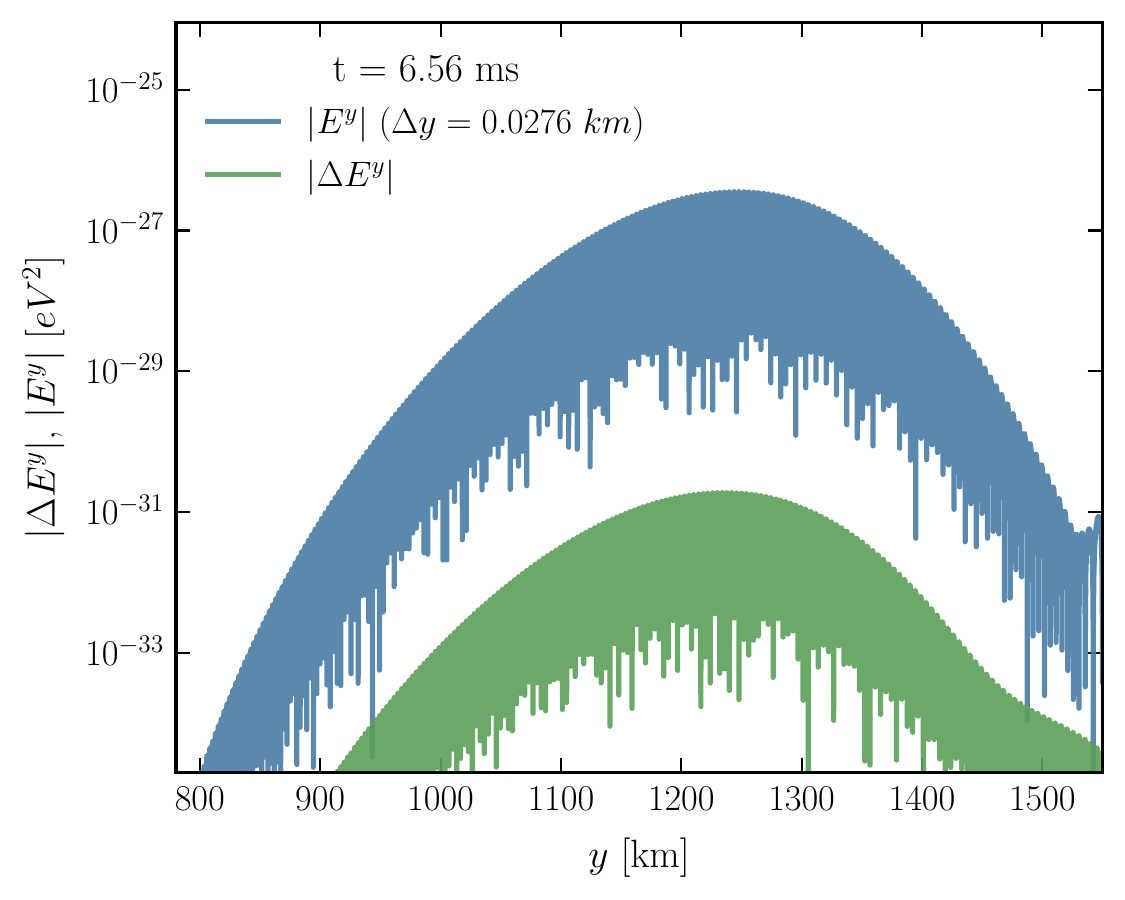}
    \caption{Impact of the spike in the conversion region on the production of the Alfvén mode. Upper panel: absolute value of $E^y$ at $t = 6.56 \, \ms$ in the region close to the spike, extracted from two runs with different resolutions. At least with the choices of $\Delta y$ considered here, the code is not able to describe the spike adequately. However, the component of the electric field emanating from the spike is significantly smaller in the run with higher resolution; this means that such component is unphysical and only due to numerical errors. 
    Lower panel: absolute value of $E^y$ from the simulation with higher resolution (blue), compared with the absolute value of the difference between the profiles of $E^y$ at the two resolutions (green), focusing on the wave packet of the Alfvén mode. As we can see the two profiles are in significant agreement with each other, indicating that the description of the Alfvén mode provided by the code is solid, despite the difficulties in representing the spike.}
    \label{fig:SpikeAnalysis}
\end{figure}

Lastly, we would like to discuss in this Appendix the impact of the spike appearing at the conversion point in the simulations in Sec.~\ref{sec:modeexcite}. We repeated the simulation with $W = 0.3 \, \km^{-1}$ in the set with a fixed angle between the propagation direction and the background magnetic field, doubling the resolution both in space and time, and we compared the profiles of the longitudinal component of the electric field. In particular, in the upper panel of Fig.~\ref{fig:SpikeAnalysis} we show the profile of the absolute value of $E^y$, extracted at $t = 6.56 \, \ms$, focusing on the region close to the spike. Firstly, we can see that the electric field emanating from the spike -- analogous to the one visible in red, blue and orange curves in Fig.\ref{fig:FixedThetaSnapshots} -- is significantly reduced when decreasing the grid step, meaning that it is an artifact of the numerical integration. Secondly, while the intensity of the electric field at the spike is consistent between the two simulations, the rapid oscillatory behavior indicates that, at least with the grid step considered, the code is not able to resolve the behavior of $E^y$. However, using higher resolutions seems to be hardly feasible, as the simulation can take up to $1800$ CPU hr, and the code can supports only single node configurations.

Nevertheless, the profiles of the Alfvén mode obtained with the two resolutions are significantly compatible with each other. This is visible in the lower panel of Fig.~\ref{fig:SpikeAnalysis}, where we show the absolute value of the difference between the profiles of $E^y$ from the two simulations, $\Delta E^y$, together with the absolute value of $E^y$ extracted from the higher resolution run, focusing on the wave packet of the Alfvén mode. As we can see, $\Delta E^y$ is more than three orders of magnitude smaller than $E^y$; this suggests that the production and the characteristics of the Alfvén mode are solid and well described by our code, even in presence of the unphysical electric field emanating from the spike.

\bibliography{biblio}

\end{document}